\documentclass{article}
\usepackage{PRIMEarxiv}
\usepackage{amsmath}

\usepackage[utf8]{inputenc} 
\usepackage[T1]{fontenc}    
\usepackage{hyperref}       
\usepackage{url}            
\usepackage{booktabs}       
\usepackage{amsfonts}       
\usepackage{nicefrac}       
\usepackage{microtype}      
\usepackage{lipsum}
 \usepackage{fancyhdr}       
\usepackage{graphicx}       
\graphicspath{{KAIS/fig}}     

\usepackage{amsthm}
\usepackage[linesnumbered, ruled, vlined]{algorithm2e}

\usepackage[sort&compress, numbers]{natbib}

\usepackage{xspace}
\usepackage{amsthm}

\newcommand{\squishlist}{
 \begin{list}{$\bullet$}
  {  \setlength{\itemsep}{0pt}
     \setlength{\parsep}{3pt}
     \setlength{\topsep}{3pt}
     \setlength{\partopsep}{0pt}
     \setlength{\leftmargin}{2em}
     \setlength{\labelwidth}{1.5em}
     \setlength{\labelsep}{0.5em}
} }
\newcommand{\squishlisttight}{
 \begin{list}{$\bullet$}
  { \setlength{\itemsep}{0pt}
    \setlength{\parsep}{0pt}
    \setlength{\topsep}{0pt}
    \setlength{\partopsep}{0pt}
    \setlength{\leftmargin}{2em}
    \setlength{\labelwidth}{1.5em}
    \setlength{\labelsep}{0.5em}
} }

\newcommand{\squishdesc}{
 \begin{list}{}
  {  \setlength{\itemsep}{0pt}
     \setlength{\parsep}{3pt}
     \setlength{\topsep}{3pt}
     \setlength{\partopsep}{0pt}
     \setlength{\leftmargin}{1em}
     \setlength{\labelwidth}{1.5em}
     \setlength{\labelsep}{0.5em}
} }

\newcommand{\squishend}{
  \end{list}
}

\newcommand{\mpara}[1]{\medskip\noindent{\bf #1}}
\newcommand{\para}[1]{\noindent{\bf #1}}

\newcommand{\xhdr}[1]{\vspace{1mm}\noindent {\bf #1.}}

\newtheorem{theorem}{Theorem}

\newtheorem{lemma}[theorem]{Lemma}

\newcommand{\fpr}[1]{\mathopen{}\left(#1\right)}

\newcommand{\dispfunc}[2]{%
  \ensuremath{%
    \ifthenelse{\equal{\noexpand#2}{}}%
	     {#1}%
		      {{#1}\fpr{#2}}}}

\newcommand{\nnegintegers}{\ensuremath{\mathbb{N}_{+}}\xspace}

\DeclareMathAlphabet{\pazocal}{OMS}{zplm}{m}{n}

\newcommand{\bigO}{\ensuremath{\mathcal{O}}\xspace}
\newcommand{\np}{\ensuremath{\mathbf{NP}}\xspace}
\newcommand{\p}{\ensuremath{\mathbf{P}}\xspace}
\newcommand{\wone}{\ensuremath{\mathbf{W}[1]}\xspace}

\newcommand{\change}[1]{{{#1}}}

\newcommand{\GF}{\ensuremath{\text{GF}}\xspace}

\newcommand{\ktemppath}{\ensuremath{k\text{\sc{-Temp}\-Path}}\xspace}

\newcommand{\pathmotif}{\ensuremath{\text{\sc Path\-Motif}}\xspace}
\newcommand{\rainbowpath}{\ensuremath{\text{\sc Rain\-bow\-Path}}\xspace}

\newcommand{\colorfulpath}{\ensuremath{\text{\sc Color\-ful\-Path}}\xspace}

\newcommand{\restlesspath}{\ensuremath{\text{\sc Restless\-Path}}\xspace}

\newcommand{\krestlesspath}{\ensuremath{k \text{\sc-Restless\-Path}}\xspace}

\newcommand{\krestlessmotif}{\ensuremath{k\text{\sc-Restless\-Motif}}\xspace}

\newcommand{\restlessreach}{{\sc Restless\-Reach}\xspace}
\newcommand{\shortrestlessreach}{\ensuremath{k \text{\sc -Restless\-Reach}}\xspace}
\newcommand{\krestlessreach}{\ensuremath{k \text{\sc -Restless\-Reach}}\xspace}

\newcommand{\krestlessmotifreach}{\ensuremath{k\text{\sc-Restless\-Motif\-Reach}}\xspace}
\newcommand{\atmkrestlessreach}{\ensuremath{\textit{atm-} k \text{\sc -Restless\-Reach}}\xspace}

\newcommand{\finegrainedoracle}{{\sc Fine\-Grained\-Oracle}\xspace}
\newcommand{\setcover}{{\sc Set\-Cover}\xspace}

\newcommand{\prepp}[3]{\ensuremath{\mathcal{P}_{{#1},{#2},{#3}}(\vec{x},\vec{y})}\xspace}
\newcommand{\prep}[3]{\ensuremath{\chi_{{#1},{#2},{#3}}(\vec{x},\vec{y})}\xspace}
\newcommand{\qrep}[3]{\ensuremath{\zeta_{{#1},{#2},{#3}}(\vec{z},\vec{y})}\xspace}
\newcommand{\qrepw}[3]{\ensuremath{\zeta_{{#1},{#2},{#3}}(\vec{z},\vec{w},\vec{y})}\xspace}

\newcommand{\fmri}{{f{\sc\large  mri}}\xspace}

\newcommand{\dfs}{{{\sc\large dfs}}\xspace}

\newcommand{\deltaexp}[1]{\ensuremath{{#1^{\uparrow}(\delta)}}\xspace}

\newcommand{\ulg}[1]{\ensuremath{#1^{\downarrow}\xspace}}

\newcommand{\lin}[1]{\ensuremath{L_{\text{in}}(#1)\xspace}}
\newcommand{\lout}[1]{\ensuremath{L_{\text{out}}(#1)\xspace}}

\newcommand{\ie}{{i.e.\xspace}}
\newcommand{\eg}{{e.g.\xspace}}

\newcommand{\yes}{{\sc yes}\xspace}
\newcommand{\no}{{\sc no}\xspace}
\newcommand{\tikzscale}{{0.7}}

\usepackage{tikz,pgfplots,pgfplotstable}
\usetikzlibrary{positioning,fit,shapes}
\usetikzlibrary{decorations.pathreplacing}
\usetikzlibrary{decorations.markings}
\usetikzlibrary{arrows}

\pgfdeclarelayer{background}
\pgfdeclarelayer{foreground}
\pgfsetlayers{background,main,foreground}

\makeatletter
\tikzset{multicircle/.style  args={#1, #2}{%
 alias=tmp@name, %
  postaction={%
    insert path={
     \pgfextra{%
     \pgfpointdiff{\pgfpointanchor{\pgf@node@name}{center}}%
                  {\pgfpointanchor{\pgf@node@name}{east}}%
     \pgfmathsetmacro\insiderad{\pgf@x}%
        \fill[white] (\pgf@node@name.center)  circle (\insiderad-\pgflinewidth);%
        \draw[#2] (\pgf@node@name.center)  circle (\insiderad-\pgflinewidth);%
        \fill[#2] (\pgf@node@name.center)  -- ++(0:\insiderad-\pgflinewidth) arc (0:#1:\insiderad-\pgflinewidth)--cycle;%
        }}}}}
\makeatother

\definecolor{yafaxiscolor}{rgb}{0.3, 0.3, 0.3}
\definecolor{yafcolor1}{rgb}{0.4, 0.165, 0.553}
\definecolor{yafcolor2}{rgb}{0.949, 0.482, 0.216}
\definecolor{yafcolor3}{rgb}{0.47, 0.549, 0.306}
\definecolor{yafcolor4}{rgb}{0.925, 0.165, 0.224}
\definecolor{yafcolor5}{rgb}{0.141, 0.345, 0.643}
\definecolor{yafcolor6}{rgb}{0.965, 0.933, 0.267}
\definecolor{yafcolor7}{rgb}{0.627, 0.118, 0.165}
\definecolor{yafcolor8}{rgb}{0.878, 0.475, 0.686}
\definecolor{yafcolor9}{rgb}{0.965, 0.733, 0.767}

\newlength{\yafaxispad}
\newlength{\yaftlpad}
\newlength{\yaflabelpad}
\newlength{\yafaxiswidth}
\newlength{\yafticklen}
\makeatletter
\def\pgfplots@drawtickgridlines@INSTALLCLIP@onorientedsurf#1{}
\makeatother

\newcommand{\yafdrawxaxis}[2]{
  \pgfplotstransformcoordinatex{#1}\let\xmincoord=\pgfmathresult 
  \pgfplotstransformcoordinatex{#2}\let\xmaxcoord=\pgfmathresult 
  \pgfsetlinewidth{\yafaxiswidth} 
  \pgfsetcolor{yafaxiscolor}
  \pgfpathmoveto{\pgfpointadd{\pgfpointadd{\pgfplotspointrelaxisxy{0}{0}}{\pgfqpointxy{\xmincoord}{0}}}{\pgfqpoint{-0.5\yafaxiswidth}{\yafaxispad}}}
  \pgfpathlineto{\pgfpointadd{\pgfpointadd{\pgfplotspointrelaxisxy{0}{0}}{\pgfqpointxy{\xmaxcoord}{0}}}{\pgfqpoint{0.5\yafaxiswidth}{\yafaxispad}}}
  \pgfusepath{stroke}

}
\newcommand{\yafdrawyaxis}[2]{
  \pgfplotstransformcoordinatey{#1}\let\ymincoord=\pgfmathresult 
  \pgfplotstransformcoordinatey{#2}\let\ymaxcoord=\pgfmathresult 
  \pgfsetlinewidth{\yafaxiswidth} 
  \pgfsetcolor{yafaxiscolor}
  \pgfpathmoveto{\pgfpointadd{\pgfpointadd{\pgfplotspointrelaxisxy{0}{0}}{\pgfqpointxy{0}{\ymincoord}}}{\pgfqpoint{\yafaxispad}{-0.5\yafaxiswidth}}}
  \pgfpathlineto{\pgfpointadd{\pgfpointadd{\pgfplotspointrelaxisxy{0}{0}}{\pgfqpointxy{0}{\ymaxcoord}}}{\pgfqpoint{\yafaxispad}{0.5\yafaxiswidth}}}
  \pgfusepath{stroke}
}

\pgfplotscreateplotcyclelist{yaf}{%
{yafcolor1,mark options={scale=0.75},mark=o}, 
{yafcolor2,mark options={scale=0.75},mark=square},
{yafcolor3,mark options={scale=0.75},mark=triangle},
{yafcolor4,mark options={scale=0.75},mark=o},
{yafcolor5,mark options={scale=0.75},mark=o},
{yafcolor6,mark options={scale=0.75},mark=o},
{yafcolor7,mark options={scale=0.75},mark=o},
{yafcolor8,mark options={scale=0.75},mark=o}} 

\pgfplotsset{axis y line=left, axis x line=bottom,
  tick align=outside,
  compat = 1.3,
  tickwidth=\yafticklen,
  clip = false,
  every axis title shift = 0pt,
    x axis line style= {-, line width = 0pt, opacity = 0},
    y axis line style= {-, line width = 0pt, opacity = 0},
    x tick style= {line width = \yafaxiswidth, color=yafaxiscolor, yshift = \yafaxispad},
    y tick style= {line width = \yafaxiswidth, color=yafaxiscolor, xshift = \yafaxispad},
    x tick label style = {font=\scriptsize, yshift = \yaftlpad},
    y tick label style = {font=\scriptsize, xshift = \yaftlpad},
    every axis y label/.style = {at = {(ticklabel cs:0.5)}, rotate=90, anchor=center, font=\scriptsize, yshift = -\yaflabelpad},
    every axis x label/.style = {at = {(ticklabel cs:0.5)}, anchor=center, font=\scriptsize, yshift = \yaflabelpad},
    x tick label style = {font=\scriptsize, yshift = 1pt},
    grid = major,
    major grid style  = {dash pattern = on 1pt off 3 pt},
  every axis plot post/.append style= {line width=\yafaxiswidth} ,
  legend cell align = left,
  legend style = {inner sep = 1pt, cells = {font=\scriptsize}},
  legend image code/.code={%
    \draw[mark repeat=2,mark phase=2,#1] 
    plot coordinates { (0cm,0cm) (0.15cm,0cm) (0.3cm,0cm) };%
  } 
}

\tikzset{
  on each segment/.style={
    decorate,
    decoration={
      show path construction,
      moveto code={},
      lineto code={
        \path [#1]
        (\tikzinputsegmentfirst) -- (\tikzinputsegmentlast);
      },
      curveto code={
        \path [#1] (\tikzinputsegmentfirst)
        .. controls
        (\tikzinputsegmentsupporta) and (\tikzinputsegmentsupportb)
        ..
        (\tikzinputsegmentlast);
      },
      closepath code={
        \path [#1]
        (\tikzinputsegmentfirst) -- (\tikzinputsegmentlast);
      },
    },
  },
  mid arrow/.style={postaction={decorate,decoration={
        markings,
        mark=at position .6 with {\arrow[#1]{stealth}}
      }}},
}

\title{Restless reachability problems in temporal graphs
}

\author{
  Suhas Thejaswi\thanks{The author contributed to this research during their employment at Aalto University, Finland.}\\
  Max Planck Institute for Software Systems \\
  Kaiserslautern, Germany\\
  \texttt{thejaswi@mpi-sws.org} \\
  \And
  Juho Lauri \\
  Helsinki, Finland \\
  \texttt{juho.lauri@gmail.com} \\
  \And
  Aristides Gionis \\
  Division of Theoretical Computer Science \\
  KTH Royal Institute of Technology \\
  Stockholm, Sweden\\
  \texttt{argioni@kth.se}
}

\begin{document}
\maketitle

\begin{abstract}
We study a family of reachability problems under waiting-time restrictions in temporal and vertex-colored temporal graphs.  Given a temporal graph and a set of source vertices, we find the set of vertices that are reachable from a source via a time-respecting path, where the difference in time\-stamps between consecutive edges is at most a resting time.
Given a vertex-colored temporal graph and a multi\-set query of colors, we find the set of vertices reachable from a source via a time-respecting path such that the vertex colors of the path agree with the multi\-set query and the difference in timestamps between consecutive edges is at most a resting time.
These kind of problems have applications in understanding the spread of a disease in a network, tracing contacts in epidemic outbreaks, finding signaling pathways in the brain network, and recommending tours for tourists, among other.

We present an algebraic algorithmic framework based on constrained multi\-linear sieving for solving the restless reachability problems we propose. In particular, parameterized by the length~$k$ of a path sought, we show that the proposed problems can be solved in $O(2^k k m \Delta)$ time and $O(n \Delta)$ space, where $n$ is the number of vertices, $m$ the number of edges, and $\Delta$ the maximum resting time of an input temporal graph.  
The approach can be extended to extract paths and connected subgraphs in both
static and temporal graphs, thus improving the work of Bj{\"o}klund et al.~[ESA~2015]
and Thejaswi et al.~[Big Data~2020].
In addition, we prove that our algorithms for the restless reachability problems in vertex-colored temporal graphs are optimal under plausible complexity-theoretic assumptions.

Finally, with an open-source implementation, we demonstrate that our algorithm scales to large graphs with up to one billion temporal edges, despite the problems being NP-hard. Specifically, we present extensive experiments to evaluate our scalability claims both on synthetic and real-world graphs.
Our implementation is efficiently engineered and highly optimized. For instance, we can solve the restless reachability problem by restricting the path length to $9$ in a real-world graph dataset with over $36$ million directed edges in less than one hour on a commodity desktop with a 4-core {Haswell} CPU.
\end{abstract}

\keywords{Algebraic fingerprinting \and Multilinear sieving \and Restless paths \and Restless reachability \and Temporal paths \and Temporal reachability}

\newpage
\section*{Statement on ethics and integrity policies}

\para{Data and source code availability.} We use real-world data from several sources: the co-presence dataset from socio-patterns \cite{genois2018can}, Koblenz network collections~\cite{koblenz2013konect}, Copenhagen study network~\cite{sapiezynski2019interaction}, and public-transport networks~\cite{kujala2018collection}. These datasets are publicly available and anonymized ensuring that no individual can be identified. Our implementations are available as open source~\cite{ourcode} and are provided under a modest license agreement. 

\para{Ethics approval.} Our work is theoretical in nature, but we conducted experimental evaluation and case studies to demonstrate practical applicability of our work. Further, our work does not involve any human subject study and/or crowd-sourcing, as such approval from ethics committee and/or an institutional review board is not mandated for this research. 

\para{Funding.} Suhas Thejaswi acknowledges support from the European Research Council (ERC) under the European Union's Horizon $2020$ research and innovation programme (\ $945719$), European Unions's SoBigData++ Transnational Access Scholarship, and Nokia Foundation Scholarship ($20220290$).
Aristides Gionis is supported by the ERC Advanced Grant REBOUND (834862), the EC H2020 RIA project SoBigData++ (871042), and the Wallenberg AI, Autonomous Systems and Software Program (WASP) funded by the Knut and Alice Wallenberg Foundation.

\para{Conflict of interest.} Authors declare no conflict of interest.

\section{Introduction} 
Graphs are used to model many real-world problems such as  information propagation in social networks~\cite{bakshy2012role,jackson2005diffusion},  spreading of epidemics~\cite{anand2017tracking,rozenshtein2016reconstructing,pastor2015epidemic},  protein interactions~\cite{alon2008biomolecular,schwikowski2000network},  activity in brain networks~\cite{bullmore2009complex,de2010temporal,glerean2016reorganization}, and design of nano-structures using DNA~\cite{benson2015dna}. However, real-world problems often entail complex interactions whose semantics  are not captured by usual simple graph models. As such, over the years, researchers have enriched graph models by introducing ($i$) node and edge attributes, giving rise to attributed graphs~\cite{perozzi2014focused,tong2007fast} or ($ii$) edge timestamps, giving rise to temporal graphs~\cite{holme2012temporal,michail2016introduction}. In particular, temporal graphs are used to model complex phenomena and network dynamics in a wide range of applications related to social networks, genealogical trees,  transportation, and telecommunication networks.

As with any graph model, connectivity is a fundamental problem in temporal graphs, \ie, whether two vertices are connected by a time-respecting path or a temporal path---a path in which consecutive edges have non-decreasing timestamps~\cite{holme2012temporal,michail2016introduction}.
An extension to connectivity is reachability, 
where the goal is to find the (temporal) connectivity between all pairs of vertices~\cite{holme2005network}. 
Some variants of connectivity and reachability problems such as finding the path that minimizes the length or arrival time can be solved in polynomial time~\cite{cooke1966shortest,wu2016efficient,holme2005network}.  However, Casteigts et al.~\cite{casteigts2021finding} showed that a variant of the connectivity problem with resting (or waiting) time restrictions---the time difference of consecutive edges is at most a resting time---known as the restless temporal path problem (or more simply a restless path if the context is clear) is \np-hard. In this work, we study a family of connectivity and reachability problems in temporal graphs with resting time restrictions. Specifically, we study the following problems.

\para{Restless path problems.} The restless path problem (\restlesspath) asks if there exists a restless path connecting a source and a destination. Extending this, the short restless path problem (\krestlesspath) asks if there exists a restless path with a specified length, $k-1$, connecting the source and the destination, and the short restless path motif problem (\krestlessmotif) asks if there exists a restless path in a vertex-colored temporal graph, whose vertex colors match with the given multiset of colors of size $k$. 
 
\para{Restless reachability problems.} The restless reachability problem (\restlessreach) asks to find a set of vertices for which there exists a restless path connecting from a given source to the vertex. Extending this, the short restless reachability problem (\krestlessreach) asks to find a set of vertices for which there exists a restless path with a specified length, $k-1$, connecting a given source to the vertex, and the short restless path motif reachability (\krestlessmotifreach) asks to find a set of vertices for which there exists a restless path whose vertex colors match with the given multiset of colors of size $k$, in a vertex-colored temporal graph.

Restless connectivity problems (\ie, restless walks and restless paths) have been extensively studied for estimating infections in an epidemic outbreak~\cite{casteigts2021finding,himmel2019efficient}. Here, each vertex in the graph associates to a person and each temporal edge to a social interaction between two persons at a specific timestamp.
The resting time of a vertex is viewed as the time until which the virus/disease can propagate from that vertex---after exceeding the resting time, the virus/disease becomes inactive and stops propagating---in particular, the resting time of a vertex captures the recovery state of susceptible-infected-recovered (SIR) model of disease propagation~\cite{kermack1927contribution}.
Given the source of an infection, we want to find the set of people who may also be infected with the disease. Additionally, epidemiologists would be interested in a tool that allows them to evaluate different immunization strategies, such as, computing efficiently the spread of the disease when a selected subset of the population has been immunized to the disease, \eg, vaccinated or quarantined.
Beyond epidemiology, restless reachability has applications in finding signaling pathways in brain networks and tour recommendations. See Appendix~\ref{app:motivation} for further motivation.

All problems considered in this work are \np-hard and as such it is likely that they admit no polynomial-time algorithm. In such cases, it is typical to resort to heuristics or approximation algorithms, which still run in reasonable time but compromise the quality of the solution. In contrast, we consider exact algorithms for solving restless reachability problems in both temporal and vertex-colored temporal graphs. To also ensure fast runtime, our algorithms are exponential in the length of the path sought. More precisely, we show that when the path length is small enough our solution scales to massive graphs with up to one billion temporal edges.

It is worth noting that Casteigts et al.~\cite{casteigts2021finding} studied the \restlesspath problem from a complexity-theoretic point of view. For instance, they proved that the problem remains \np-hard on highly structured graphs such as complete graphs with exactly one edge removed. Despite several negative results, the authors pinpoint some parameters $p$ of the problem for which the problem admits an algorithm running in time $f(p)\,n^{\bigO(1)}$, where $f(p)$ is some computable function depending solely on the parameter $p$.
Namely, these parameters are the feedback edge number (FEN), the timed feedback vertex number (TFVN) and the length of the restless path. For FEN, the algorithm they describe runs in time $2^{\bigO(p)}\,n^{\bigO(1)}$ and for TFVN, the algorithm runs in time $6^q\,q!\,n^{\bigO(1)}$, where $p$ is the FEN and $q$ is the TFVN of the input $n$-vertex temporal graph.%
\footnote{In the field of parameterized algorithms, when devising algorithms for \np-hard problems, it is usual to ignore the exact form of any polynomial factors in the runtime and concentrate on the likely unavoidable exponential dependency on the parameter under consideration. That is, when the exact polynomial factor of the algorithm is not worked out (or omitted as non-essential) it is often represented as merely $|x|^{\bigO(1)}$ meaning $|x|^c$ for some $c \geq 1$, where $x$ is the input and $|x|$ is its size.}
In order for these algorithms to be scalable,%
\footnote{As the problem is \np-hard, the function $f(p)$ is necessarily exponential unless \p = \np.}  the corresponding parameter values should be small in practice on relevant instances. Unfortunately, as we show in Appendix~\ref{appendix:dataset-statistics} (see Table~\ref{table:exp:datasets}), there are real-world instances where this is not the case. Therefore, on such instances, it appears that the only known parameterized algorithm for the problem that has hope of being practical is one that pushes the unavoidable exponential dependency into the length of the restless path.%
\footnote{Casteigts et al.~\cite{casteigts2021finding} also presented a $2^k\,n^{\bigO(1)}$ algorithm for \restlesspath with respect to the parameter $k$, which is the length of the path. We compare our approach with theirs in Section~\ref{sec:related}.}
Our key contributions in this paper are as follows:

\squishlisttight
\item We present a generating function for generating restless walks, and a space-efficient algorithm based on constrained multilinear sieving~\cite{bjorklund2016constrained} for solving \krestlessmotif and \krestlessmotifreach in time $\bigO(2^k k m \Delta)$ and $\bigO(n\Delta)$ space,
where $n$ is the number of vertices, $m$ is the number of edges, 
$k-1$ is the path length, and $\Delta$ is the maximum resting time. 
Furthermore, we show that our algorithm solves \krestlesspath and \krestlessreach in $\bigO(2^k k m \Delta)$ time using $\bigO(n\Delta)$ space. Throughout this paper, we call this algorithm a decision oracle as it returns a \yes/\no answer with no explicit solution.

\item Next, we develop the decision oracle into a fine-grained decision oracle capable of reporting the set of vertices that are reachable via a restless path from given source vertices along with the set of timestamps at which the vertices are reachable. Further, we exploit this to extract optimal solutions (\ie, solutions in which the maximum timestamp on the restless path is minimized) for \krestlessmotif and \krestlesspath.
Notably, our solution improves upon the earlier work of Thejaswi et al.~\cite{thejaswi2020pattern,thejaswi2020finding} by reducing the number of queries by a factor of $\log \tau$. Further, for extracting a solution, we reduce the number of queries from the work of  Bj{\"o}rklund et al.~\cite{bjorklund2014fast} from $\bigO(k \log n)$ to $k$.
In total, our extraction algorithm runs in $\bigO(2^k k m \Delta)$ time and $\bigO(n\tau)$ space, where $\tau$ is the maximum timestamp.

\item As a consequence of our fine-grained decision oracle, our algorithm can answer more detailed queries. In particular, it can answer whether a given vertex $u$ is reachable from source $s$ at timestamp $i$ with a restless path of length $\ell$. 
Such a fine-grained oracle can be used to solve multiple variants of the restless path problem, such as finding a restless path that minimizes  the path length, the arrival time, or the total resting time.
Another key strength of our algorithm is its adaptability to multiple variants of the restless path problem. For instance, the algorithm is not limited to a single source, but can be extended to include multiple sources.

\item We prove that our algorithms for \krestlessmotif and \krestlessmotifreach are optimal under plausible complexity-theoretic assumptions. More precisely, we prove that there exists no $\bigO^*((2-\epsilon)^k)$-time algorithm\footnote{The notation $\bigO^*$ hides the factors polynomially bounded in the input size.} for \krestlessmotif or \krestlessmotifreach for any $\epsilon > 0$, assuming the so-called Set Cover Conjecture~\cite{cygan2016problems}, a precise definition is given in Section~\ref{sec:algo:infeasibility}.

\item With an open-source implementation we demonstrate that our algorithms scale to graphs with up to ten million edges on a commodity desktop equipped with an $4$-core Haswell CPU. When scaled to a computing server with $2 \times 12$-core Haswell CPU, the algorithm scales to graphs with up to one billion temporal edges~\cite{ourcode}. As a case study, we model disease spreading in social gatherings as a \shortrestlessreach problem and perform experiments to study the propagation of the disease using the co-presence datasets from socio-patterns~\cite{genois2018can}. We also perform experiments to check the change in the disease spreading pattern  with the presence of people with immunity.
\squishend

\subsection{Related work} \label{sec:related}

Our works builds upon the further related work on multilinear sieving and temporal reachability.

\para{Multilinear sieving.} Algebraic algorithms based on multilinear sieving for solving path problems in static graphs are due to Koutis and Williams~\cite{koutis2008faster,williams2009finding,koutis2009limits}. Bj{\"o}rklund et al.~\cite{bjorklund2017narrow} improved the technique using narrow sieves to get a $\bigO^*(1.66^n)$ time algorithm for the Hamiltonian path problem.  The sieving technique was extended by Bj{\"o}rklund et al.~\cite{bjorklund2016constrained} to find paths and subgraphs that agree with a multiset of colors. A practical implementation of the sieve was provided by Bj{\"o}rklund et al.~\cite{bjorklund2015engineering}. Furthermore, its parallelizability to vector-parallel architectures and scalability to large subgraph sizes was shown by Kaski et al.~\cite{kaski2018engineering}. Note that these algorithms only handle static graphs.
More recently, Thejaswi et al.~\cite{thejaswi2020pattern, thejaswi2020finding} extended the sieving technique to solve a family of pattern-detection problems in temporal graphs.

\para{Reachability in temporal graphs.}
Path problems in temporal graphs are well-studied within different communities, including algorithms, data mining, and complex networks. The problem variants that seek to find a path that minimizes different objectives such as path length.
In recent years, there has been an emphasis on more expressive temporal path problems  due to their applicability in various fields~\cite{anand2017tracking,bullmore2009complex,rozenshtein2016reconstructing,von2009public,holme2005network,sengupta2019arrow,dondi2023colorful,cohen2021tropical}.

The problem of finding a temporal path under waiting-time restrictions was introduced by Dean~\cite{dean2004algorithms},  who studied a polynomial-time solvable variant of the problem. Himmel et al.~\cite{himmel2019efficient} considered the restless walk problem where several visits to a vertex are allowed and presented polynomial-time algorithms.
\citet{thejaswi2020finding}[Section Path motif problem with delays] studied a variant of the temporal path problem that considers both transition time---the time required to transit from a vertex to its neighbor---and delay time---the minimum time that must be spent at each vertex before proceeding---and presented a $\bigO^*(2^k)$ algorithm. In their setting, the time difference of consecutive edges must be \emph{at~least} a vertex-specific delay time, in contrast to restless paths, where the time difference must be \emph{at most} a vertex-specific waiting time.

More recently, Casteigts et al.~\cite{casteigts2021finding} showed that \restlesspath is \np-hard and also that \krestlesspath is \wone-hard parameterized by either the feedback vertex number or the pathwidth of the input graph.
 The authors presented algorithms with running time $\bigO(2^k n^{\bigO(1)})$ and $\bigO(2^n n^{\bigO(1)})$ for \krestlesspath and \restlesspath, respectively, where $n$ is the number of vertices in the input temporal graph and $k$ the length of the path.%
\footnote{To further elaborate, Casteigts et al.~\cite{casteigts2021finding} builds upon the work of Koutis and Williams~\cite{koutis2009limits} using their result as black-box. While this is sufficient to establish fixed-parameter tractability with respect to $k$, the exact polynomial factors are not fully detailed. Based on our calculations the polynomial factor is $n^{\bigO(1)} \approx \bigO(kn + k m \Delta)$ and the space complexity is $\bigO(kn\tau)$.} In comparison to Casteigts et al.~\cite{casteigts2021finding}, our approach can decide the existence of a solution in time $\bigO(2^k k m \Delta)$ and $\bigO(2^n n m \Delta)$, where $n$ is the number of vertices, $m$ is number of temporal edges, and $\Delta$ is the maximum resting time. 
It should be noted that their algorithm only reports the existence of a restless path between a source and a destination with a \yes/\no answer but does not return an explicit solution. In many practical scenarios, such as where further analysis of the path is required, this is insufficient. 
In a recent work building on our research, Zschoche~\cite{zschoche2022restless} presented a randomized $\bigO(4^{k-d} (k-d)^2 m^3 \Delta)$-time algorithm to find a shortest restless path (\ie, one minimizing the length of the path) between a source $s$ and destination $z$, where $d$ is the length of the shortest temporal path from $s$ to $z$.

\restlesspath and its variants studied in this work are self-reducible. By extending the approach of Bj{\"o}rklund et al.~\cite{bjorklund2014fast}, extracting a optimal $k$-restless path (\ie, one minimizing the maximum timestamp) can be done with $\bigO(k \log n \log \tau)$ queries to the decision oracle, followed by $\bigO(k!)$ steps to identify the path.\footnote{See Section~\ref{sec:algo:extraction} for a description of extracting a solution using self-reducibility.} However, using our fine-grained oracle we can extract a solution with $k-1$ queries.
As a result, an optimal restless path can be extracted in $\bigO(2^k k m \Delta)$ and $\bigO(2^n n m \Delta)$ time for \krestlesspath and \restlesspath, respectively.
The \restlessreach and \krestlessreach problems generalize \restlesspath and \krestlesspath problems, respectively. With a decision oracle for \restlesspath (or \krestlesspath), the set of reachable vertices from a source can be identified with $\bigO(n)$ queries, by repeatedly checking the existence of a restless path between source and each vertex. However, our method can answer reachability queries in a single computation, without having to iterate over each vertex. In addition, our method solves a general variant of restless path and restless reachability problems with color constraints on the vertices.

\xhdr{Organization of the paper} The rest of this paper is organized as follows. We introduce the necessary notation in Section~\ref{sec:terminology}, and we formally define the reachability problems for temporal graphs and vertex-colored temporal graphs in Section~\ref{sec:problems}. Afterwards, our algorithmic solution is presented in Section~\ref{sec:algorithm}. Our empirical evaluation on a large collection of synthetic and real-world graphs is discussed in Section~\ref{sec:experiments}. Finally, Section~\ref{sec:conclusion} offers a short conclusion, limitations and directions for future work. 

\section{Terminology} \label{sec:terminology}
In this section we introduce our notation and basic terminology. For a positive integer $k$ we write $[k]=\{1,\dots,k\}$.
A list of symbols used in this paper is available in Appendix~\ref{app:symbols} Table~\ref{table:symbols}.

\para{Static graphs.}
A static undirected graph $G$ is a tuple $(V,E)$ where $V$ is a set of vertices and $E$ is a set of unordered pairs of vertices called edges.
A static directed graph $G$ is a tuple $(V,E)$ where $V$ is a set of vertices and $E$ is a set of ordered pair of vertices called edges.
A static walk between any two vertices is an alternating sequence of vertices and edges $u_1 e_1 u_2 \dots e_{k} u_{k}$ such that there exists an edge $e_i=(u_i, u_{i+1}) \in E$ for each $i \in [k-1]$.  We call the vertices $u_1$ and $u_{k}$ the start and end vertices of the walk, respectively. We refer to walk $W$ as $(s,d)$-walk for $u_1=s$ and $u_{k}=d$. The length of a static walk is the number of edges in the walk. A static path is a static walk with no repetition of vertices.

\para{Undirected temporal graphs.}
An undirected temporal graph $G$ is a tuple $(V,E)$, where $V$ is a set of vertices and $E$ is a set of undirected temporal edges. An undirected temporal edge is a tuple $(u,v,j)$ where $u, v \in V$ and $j \in [\tau]$ is a timestamp, where $\tau$ is the maximum timestamp in $G$. Note that, by definition, two undirected edges $(u,v,j)$ and $(v,u,j)$ are equivalent.
A vertex $u$ is adjacent to vertex $v$ and vice versa at timestamp $i$ in an undirected graph $G$ if there exists an undirected edge $(u,v,i) \in E$.
A temporal walk $W$ is an alternating sequence of vertices and temporal edges 
$u_1 e_1 u_2 e_2 \dots e_{k-1} u_{k}$ such that $e_{i} \in E$ for all $i \in [k-1]$ and for any two edges $e_i=(u_i,u_{i+1},j_i)$, $e_{i+1}=(u_{i+1},u_{i+2},j_{i+1})$ in $W$, it is $j_i \leq j_{i+1}$. 
We say the walk $W$ reaches vertex $u_k$ at time $j_{{k-1}}$.
The vertices $u_1$ and $u_{k}$ are called source and destination vertices of $W$, respectively. The vertices $\{u_2,\dots,u_{k-1}\}$ are called in-vertices (or equivalently, internal vertices) of $W$.
We refer to the temporal walk $W$ as $(s,d)$-temporal walk with source $s=v_1$ and destination $d=v_{k}$. The vertex set and edge set of walk $W$ is denoted as $V(W)$ and $E(W)$, respectively.
%
%
The length of a temporal walk is the number of edges in the temporal walk. A temporal path is a temporal walk with no repetition of vertices.

\para{Directed temporal graphs.}
A directed temporal graph $G$ is a tuple $(V,E)$, where $V$ is a set of vertices and $E \subseteq V \times V \times [\tau]$ is a set of directed temporal edges at discrete time steps, where $\tau$ is the maximum timestamp.
A directed temporal edge is a tuple $(u,v,j)$ where $u, v \in V$ and $j \in [\tau]$ is a timestamp. 
A directed edge $(u,v,j)$ is referred as an outgoing or departing edge for $u$ and an incoming or arriving edge for $v$ at time $j$.
A vertex $u$ is an in-neighbor to $v$ at time $j$ if $(u,v,j) \in E$, similarly a vertex $v$ is an out-neighbor to $u$ at time $j$ if $(u,v,j) \in E$. The set of in-neighbors to $v$ at time $j$ is denoted by $N_j(v)= \{u \mid (u,v,j) \in E\}$ and the set of out-neighbors to $u$ at time $j$ is denoted by $N'_j(u)=\{v \mid (u,v,j) \in E\}$.
%
%

Unless stated otherwise, by a temporal graph we always mean a directed temporal graph. Similarly, by a temporal walk we always mean a directed temporal walk, and by a temporal path we always mean a directed temporal path. To distinguish a temporal graph from a static graph, we sometimes explicitly call a graph non-temporal or static.

\para{Restless walk.}
A restless temporal walk, or simply restless walk,  is a temporal walk such that for any two consecutive edges  $e_{i}=(u_{i},\allowbreak u_{i+1},\allowbreak j_{i})$ and $e_{i+1}=(u_{i+1},u_{i+2},j_{i+1})$,  it holds that $j_{i+1}-j_i \leq \delta(u_{i+1})$,  where the function  $\delta: V \rightarrow [\Delta]$ defines a vertex-dependent waiting time, with $\Delta \in \nnegintegers$ being the {maximum waiting time}.
In other words, in a restless temporal walk, it is not allowed to wait more than a vertex-dependent amount of time in each vertex.
A restless path is a restless walk with no repetition of vertices. 
%

\para{Coloring.}
Let $C$ to be a set of colors. A vertex-colored temporal graph $G=(V,E)$ is a temporal graph with a vertex-coloring function $c: V \rightarrow C$, which maps each vertex $u \in V$ to a subset of colors in $C$.
Let $M$ be a multi\-set of colors and $W$ be a walk. The walk $W$ is properly colored (or is said to agree with $M$) if there exists a bijection $f:V(W) \rightarrow M$ such that $\{f(u)\} \cap \{c(u)\} \neq \emptyset$ for all $u \in V(W)$.

\para{Polynomials.}
We assume that our temporal graph contains $n$ vertices and $m$ temporal edges,  and for simplicity we write $V = \{u_1,\dots,u_n\}$ and $E=\{e_1,\dots,e_m\}$.  We introduce a variable $x_{u_i}$ for each $ u_i \in V$ and a variable $y_{e_j}$ for each $e_j \in E$. Let $\mathcal{P}$ be a multi\-variate polynomial such that each monomial $\pazocal{P} \in \mathcal{P}$ is of the form 
$x_{u_1}^{f_1} \dots x_{u_n}^{f_n} y_{e_1}^{g_1} \dots y_{e_m}^{g_m}$.
A monomial $\pazocal{P}$ is multilinear if $f_i, g_j \in \{0,1\}$ for all $i \in [n]$ and $j \in [m]$. The degree (or size) of $P$ is the sum of the degrees of all the variables in $\pazocal{P}$.
Let $C$ be a set of colors and $c: V \rightarrow C$ be a vertex-coloring function. Let $M$ be a multi\-set of colors. For each $s \in C$, let $\mu(s)$ denote the number of occurrences of color $s$ in $M$, noting that $\mu(s) = 0$ if $s \notin M$.  We say that a monomial $x_{u_1}^{f_1} \dots x_{u_n}^{f_n} y_{e_1}^{g_1} \dots y_{e_m}^{g_m}$ is properly colored if for each $s \in C$ it holds that $\mu(s) = \sum_{i \in c^{-1}(s)} d_i$, in other words, the number of occurrences of color $s$ is equal to the total degree of $x$-variables representing the vertices with color $s$.

\section{Restless path and restless reachability problems} \label{sec:problems}
In this section we define the problems that we consider in this paper.   While we define the problems on temporal directed graphs,  we note that our algorithmic approach can be extended to undirected graphs  by replacing each undirected edge with two directed edges in opposite directions. In such case, the asymptotic complexity of the algorithm remains the same.
%
%
%


\para{Restless path problem (\restlesspath).}
Given a temporal graph $G=(V,E)$, a function  $\delta:V \rightarrow \nnegintegers$, a source $s \in V$,  and a destination $d \in V$, the problem asks if there exists a restless path from $s$ to $d$ in~$G$.

\para{Short restless path problem (\krestlesspath).}
Given a temporal graph $G=(V,E)$, a function $\delta:V \rightarrow \nnegintegers$, a source $s \in V$, a destination $d \in V$, and an integer $k \leq n$, the problem asks if there exists a restless path of length $k-1$ from $s$ to $d$ in~$G$. 

\para{Short restless path motif problem (\krestlessmotif).}
Given a temporal graph $G=(V,E)$ with a coloring function $c:V \rightarrow C$ where $C$ is a set of colors, a function $\delta:V \rightarrow \nnegintegers$, a source $s \in V$, a destination $d \in V$, and a multiset $M$ of colors, $|M|=k$, the problem asks if there exists a restless path from $s$ to $d$ in~$G$ such that the vertex colors of the path agree with $M$.

\begin{figure}
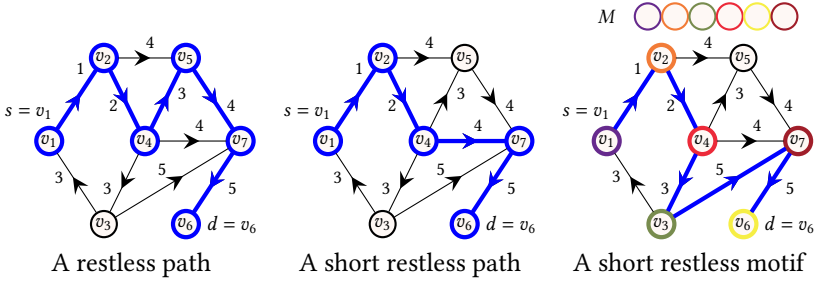

\centering
\setlength{\tabcolsep}{10pt}
\tikzset{every picture/.style={scale=0.7}}
\begin{tabular}{c c c}
\begin{tikzpicture}[scale=\tikzscale,every node/.style={scale=\tikzscale}]]

\input{tikz/tikz-defs}

\node[label] at    ( -0.75,  1) {\large $s=v_1$};
\node[label] at    ( 6.7,  -3) {\large $d=v_6$};

\node[exnode, draw=blue, ultra thick] (v1) at ( 0,    0) {$v_1$};
\node[exnode, draw=blue, ultra thick] (v2) at ( 2,    3) {$v_2$};
\node[exnode, draw=black] (v3) at ( 2,    -3) {$v_3$};
\node[exnode, draw=blue, ultra thick] (v4) at ( 3.5,    0) {$v_4$};
\node[exnode, draw=blue, ultra thick] (v5) at ( 5,    3) {$v_5$};
\node[exnode, draw=blue, ultra thick] (v6) at ( 5,    -3) {$v_6$};
\node[exnode, draw=blue, ultra thick] (v7) at ( 7,    0) {$v_7$};

\path [draw=black, postaction={on each segment={mid arrow=black, ultra thick}}]
(v1) -- node[above, pos = 0.6, inner sep = 10pt] {$1$} (v2)
(v2) -- node[above, pos = 0.6, inner sep = 5pt] {$4$} (v5)
(v2) -- node[left, pos = 0.6, inner sep = 5pt] {$2$} (v4)
(v3) -- node[left, pos = 0.6, inner sep = 5pt] {$3$} (v1)
(v3) -- node[left, pos = 0.7, inner sep = 15pt] {$5$} (v7)
(v4) -- node[left, pos = 0.6, inner sep = 5pt] {$3$} (v3)
(v4) -- node[right, pos = 0.6, inner sep = 5pt] {$3$} (v5)
(v4) -- node[above, pos = 0.6, inner sep = 5pt] {$4$} (v7)
(v5) -- node[right, pos = 0.6, inner sep = 5pt] {$4$} (v7)
(v7) -- node[right, pos = 0.6, inner sep = 10pt] {$5$} (v6)
;

\path [draw=blue, ultra thick, postaction={on each segment={mid arrow=blue, ultra thick}}]
(v1) -- (v2)
(v2) -- (v4)
(v4) -- (v5)
(v5) -- (v7)
(v7) -- (v6)
;
\end{tikzpicture}%
 &
\begin{tikzpicture}[scale=\tikzscale,every node/.style={scale=\tikzscale}]]

\input{tikz/tikz-defs}

\node[label] at    ( -0.75,  1) {\large $s=v_1$};
\node[label] at    ( 6.7,  -3) {\large $d=v_6$};

\node[exnode, draw=blue, ultra thick] (v1) at ( 0,    0) {$v_1$};
\node[exnode, draw=blue, ultra thick] (v2) at ( 2,    3) {$v_2$};
\node[exnode, draw=black] (v3) at ( 2,    -3) {$v_3$};
\node[exnode, draw=blue, ultra thick] (v4) at ( 3.5,    0) {$v_4$};
\node[exnode, draw=black] (v5) at ( 5,    3) {$v_5$};
\node[exnode, draw=blue, ultra thick] (v6) at ( 5,    -3) {$v_6$};
\node[exnode, draw=blue, ultra thick] (v7) at ( 7,    0) {$v_7$};

\path [draw=black, postaction={on each segment={mid arrow=black, ultra thick}}]
(v1) -- node[above, pos = 0.6, inner sep = 10pt] {$1$} (v2)
(v2) -- node[above, pos = 0.6, inner sep = 5pt] {$4$} (v5)
(v2) -- node[left, pos = 0.6, inner sep = 5pt] {$2$} (v4)
(v3) -- node[left, pos = 0.6, inner sep = 5pt] {$3$} (v1)
(v3) -- node[left, pos = 0.7, inner sep = 15pt] {$5$} (v7)
(v4) -- node[left, pos = 0.6, inner sep = 5pt] {$3$} (v3)
(v4) -- node[right, pos = 0.6, inner sep = 5pt] {$3$} (v5)
(v4) -- node[above, pos = 0.6, inner sep = 5pt] {$4$} (v7)
(v5) -- node[right, pos = 0.6, inner sep = 5pt] {$4$} (v7)
(v7) -- node[right, pos = 0.6, inner sep = 10pt] {$5$} (v6)
;

\path [draw=blue, ultra thick, postaction={on each segment={mid arrow=blue, ultra thick}}]
(v1) -- (v2)
(v2) -- (v4)
(v4) -- (v7)
(v7) -- (v6)
;
\end{tikzpicture}%
 & 
\begin{tikzpicture}[scale=\tikzscale,every node/.style={scale=\tikzscale}]]

\input{tikz/tikz-defs}

\node[label] at    ( -0.75,  1) {\large $s=v_1$};
\node[label] at    ( 6.7,  -3) {\large $d=v_6$};

\node[exnode, draw=yafcolor1!100, ultra thick] (v1) at ( 0,    0) {$v_1$};
\node[exnode, draw=yafcolor2!100, ultra thick] (v2) at ( 2,    3) {$v_2$};
\node[exnode, draw=yafcolor3!100, ultra thick] (v3) at ( 2,    -3) {$v_3$};
\node[exnode, draw=yafcolor4!100, ultra thick] (v4) at ( 3.5,    0) {$v_4$};
\node[exnode, draw=black] (v5) at ( 5,    3) {$v_5$};
\node[exnode, draw=yafcolor6!100, ultra thick] (v6) at ( 5,    -3) {$v_6$};
\node[exnode, draw=yafcolor7!100, ultra thick] (v7) at ( 7,    0) {$v_7$};

\node[label] at    ( 0,  4.5) {\large $M$};
\node[exnode, draw=yafcolor1!100] (m1) at ( 1.5,    4.5) {};
\node[exnode, draw=yafcolor2!100] (m2) at ( 2.5,    4.5) {};
\node[exnode, draw=yafcolor3!100] (m3) at ( 3.5,    4.5) {};
\node[exnode, draw=yafcolor4!100] (m4) at ( 4.5,  4.5) {};
\node[exnode, draw=yafcolor6!100] (m6) at ( 5.5,    4.5) {};
\node[exnode, draw=yafcolor7!100] (m7) at ( 6.5,    4.5) {};

\path [draw=black, postaction={on each segment={mid arrow=black, ultra thick}}]
(v1) -- node[above, pos = 0.6, inner sep = 10pt] {$1$} (v2)
(v2) -- node[above, pos = 0.6, inner sep = 5pt] {$4$} (v5)
(v2) -- node[left, pos = 0.6, inner sep = 5pt] {$2$} (v4)
(v3) -- node[left, pos = 0.6, inner sep = 5pt] {$3$} (v1)
(v3) -- node[left, pos = 0.7, inner sep = 15pt] {$5$} (v7)
(v4) -- node[left, pos = 0.6, inner sep = 5pt] {$3$} (v3)
(v4) -- node[right, pos = 0.6, inner sep = 5pt] {$3$} (v5)
(v4) -- node[above, pos = 0.6, inner sep = 5pt] {$4$} (v7)
(v5) -- node[right, pos = 0.6, inner sep = 5pt] {$4$} (v7)
(v7) -- node[right, pos = 0.6, inner sep = 10pt] {$5$} (v6)
;

\path [draw=blue, ultra thick, postaction={on each segment={mid arrow=blue, ultra thick}}]
(v1) -- (v2)
(v2) -- (v4)
(v4) -- (v3)
(v3) -- (v7)
(v7) -- (v6)
;
\end{tikzpicture}%
 \\
A restless path &  A short restless path & A short restless motif
\end{tabular}
\caption{\label{fig:restless-paths} 
Illustration of restless path problems. A temporal graph on vertices $V=\{v_1,\dots,v_7\}$ with source $s=v_1$, destination $d=v_6$ and resting time of vertices $\delta(v_1)=\dots=\delta(v_7)=2$. Arrows represent the direction of edges and the integer value on each edge corresponds to its timestamp.
On the left, an example of a restless path from  $v_1$ to $v_6$ of length $5$. On the center, an example of a restless path from $v_1$ to $v_6$ when the length of the path is restricted to $4$ i.e., $k=5$. On the right, an example of a short restless (path) motif from $v_1$ to $v_6$ such that the vertex colors of the path agree with the multiset of colors in $M$. Restless paths are highlighted in bold (blue).}
\end{figure}

An illustration of restless path problems is available in
Figure~\ref{fig:restless-paths}.
All these three problems are \np-hard (the hardness of the first two can be found in the paper of Casteigts et al.~\cite{casteigts2021finding}). For the last claim, observe that \krestlesspath is a special case of \krestlessmotif where all the vertices in the graph are colored with a single color and the query multiset is $M=\{1^k\}$. Since \krestlesspath is \np-hard, it follows that \krestlessmotif is \np-hard, as well.


\para{Restless reachability problem (\restlessreach).}
Given a temporal graph $G=(V,E)$, a function  $\delta:V \rightarrow \nnegintegers$, and a source vertex $s \in V$, the problem asks to find the set of vertices $D \subseteq V$ such that for each $d \in D$ there exists a restless path from $s$ to $d$ in~$G$. Clearly, the problem generalizes \restlesspath and is thus computationally~hard.
%
%

\para{Short restless reachability problem (\krestlessreach).}
Given a temporal graph $G=(V,E)$,  a function $\delta:V \rightarrow \nnegintegers$, a source vertex $s \in V$,  and an integer $k \leq n$, the problem asks to find the set of vertices $D \subseteq V$ such that for each $d \in D$ there exists a restless path of length $k-1$ from $s$ to $d$ in~$G$. Given that \krestlesspath is hard, \krestlessreach remains hard, as well.

\para{Short restless motif reachability problem (\krestlessmotifreach).}
Given a temporal graph $G=(V,E)$ with coloring function $c: V \rightarrow C$ where $C$ is a set of colors, a function  $\delta:V \rightarrow \nnegintegers$, and a multiset $M$ of colors such that $|M|=k$, the problem asks to find the set of vertices $D \subseteq V$ such that for each $d \in D$ there exists a restless path from $s$ and $d$ in $G$ such that the vertex colors of the path agree with multiset $M$. Again, a routine observation shows that this problem is computationally hard.

\section{Algorithm} \label{sec:algorithm}
Our approach makes use of the polynomial encoding of temporal walks introduced by Thejaswi et al.~\cite{thejaswi2020pattern}, building on the earlier work of Bj\"{o}rklund et al.~\cite{bjorklund2016constrained} for static graphs.
Because algebraic fingerprinting techniques are not well-known, we first provide some intuition behind our methods before diving into details.

\subsection{A birds-eye view of our approach} 
Algebraic fingerprinting methods have been successfully applied to pattern detection problems on graphs. On a high level, the approach works by encoding a problem instance $G$ as a multivariate polynomial where the variables in the polynomial correspond to entities of $G$ such as a vertex or an edge. The key idea is to design a polynomial encoding, which evaluates to a non-zero term if and only if the desired pattern is present in $G$. We can then evaluate the polynomial by substituting random values to its variables. Now, if one of the substitution evaluates to a non-zero term, it implies that the desired pattern is present in $G$. Since the substitutions are random, the approach can give false negatives. However, the resulting algorithms typically have one-sided error meaning false positives are not possible. Moreover, as is typical, the error probability can be made arbitrarily small (\eg, as unlikely as hardware failure) by repeating the substitutions.

In our case we need to decide the existence of a restless path in a temporal graph~$G$.  Following the approach presented above, we encode all the restless walks in $G$ using a multivariate polynomial where each monomial corresponds to a restless walk and variables in a monomial correspond to vertices and/or edges.  Specifically, we design an encoding where a monomial is multilinear (i.e., no variables in the monomial are repeated) if and only the corresponding restless walk it encodes is a restless path. As such, the problem of deciding the existence of a restless path is equivalent to the problem of deciding the existence of a multilinear monomial in the generated polynomial. The existence of a multilinear monomial in a polynomial, in turn, can be determined using polynomial identity testing, in particular, by evaluating random substitutions for the variables; if one of the substitution evaluates to a non-zero term then there exists at least
one multilinear monomial in the polynomial. Most importantly, {\em note that an explicit representation
of the polynomial encoding can be exponentially large. However, in our approach we do not need to represent the polynomial explicitly, but we only need to evaluate random substitutions efficiently.}

For an interested reader we present a detailed discussion on polynomial encoding of walks in static graphs and temporal graphs in Appendix~\ref{app:encoding-walks}.

\subsection{An overview of the algorithm} \label{sec:algo:overview}
To obtain our algorithm, we proceed by taking the following steps. First, we present a dynamic-programming algorithm to encode restless walks as a polynomial and from this polynomial, show that detecting a multilinear monomial of degree $2\ell-1$ is equivalent to detecting a restless path of length~$\ell-1$, and vice versa. Then, we extend the approach to detecting restless paths with additional color constraints on the vertices.  Finally, we use this algorithm to solve the problems described in Section~\ref{sec:problems}.

In this work we use finite fields for evaluating polynomials, in particular we make use of Galois fields (\GF) of characteristic $2$. An element of $\GF(2^b)$ can be represented as a bit vector of length $b$. We can perform field operations such as addition and multiplication on these bit vectors in time  $\bigO(b \log b)$~\cite{lin2016novel}.
The polynomial $\prep{u}{\ell}{i}$ on variables
$\vec{x}=\left\{x_v: v \in V \right\}$, $\vec{y}=\left\{ y_{uv,\ell,j}: (u,v,j) \in E, \ell \in [k] \right\}$ 
in Equation~(\ref{eq:poly-enc:1}) can be evaluated using a random assignment
$\tilde{x}=\{\tilde{x}_v \in \GF(2^b)\} : x_v \in \vec{x}\}$, $\tilde{y}=\left\{ \tilde{y}_{uv,\ell,j} \in \GF(2^b): y_{uv,\ell,j} \in \vec{y} \right\}$. 
Now we can build an arithmetic circuit that represents an algorithm which evaluates the polynomial $\prep{u}{\ell}{i}$ for  input $(\tilde{x}, \tilde{y})$ in time linear in the number of gates in the circuit. Most importantly, observe that the expanded expression of the polynomial $\prep{u}{\ell}{i}$ can be exponentially large, however the arithmetic circuit evaluating $\prep{u}{\ell}{i}$ can be reduced to size polynomial in number of gates by applying the associativity of addition and distributivity of multiplication over addition.

For detecting the existence of a multilinear monomial in the generated polynomial, the algorithm works by randomly substituting values $(\tilde{x},\tilde{y}) \in \GF(2^b)$, for a suitable $b$, to variables in
$\vec{x}=\left\{x_v: v \in V\right\}$ and
$\vec{y}=\left\{y_{uv,\ell,j} :\allowbreak (u,v,j) \in E^\tau, \ell\in [k]\right\}$.
Specifically, the parameter $b$ is the number of bits used to generate the field variables.
Because multiplication between two field variables is defined as an {\tt XOR} operation, multiplication between two variables with the same value results in a zero term. That is, the monomials corresponding to walks that are not paths evaluate to a zero term,  while the monomials corresponding to paths evaluate to a non-zero term. However, it is possible that an evaluation results in a zero term even though there is a multilinear monomial term corresponding to a path.%
\footnote{A polynomial with degree $2k+1$ has at most $2k + 1$ roots, so there is a chance that the chosen values of $\vec{x}$ and $\vec{y}$ are roots resulting in the polynomial to evaluate to zero even when the desired structure is present in the graph and the false-negative probability is bounded by Schwartz-Zippel lemma~\cite{schwartz1980fast,zippel1979probabilistic}.}
To eliminate the effect of such false negatives we repeat the evaluation of the polynomial with $2^k$ random substitutions. The probability of a false negative then becomes ${2^{-b}}{(2k-1)}$,  where $k-1$ is the length of the temporal walk.  In practice, the false negative probability depends on  the quality of the random number generator. It is important to note, however, that the false positive probability is zero. For full technical details, we refer the interested reader to Bj\"{o}rklund et al.~\cite{bjorklund2016constrained} and Cygan et al.~\cite[Chapter~10]{cygan2015parameterized}.

\subsection{Generating restless walks}
In this section we extend the dynamic-programming recursion to generate a polynomial encoding of restless walks of length $\ell-1$  using encoding of restless walks of length $\ell-2$.

An example is illustrated in Figure~\ref{fig:generating:1}, in which we depict a vertex $u$ with incoming neighbors $N_i(u)=\{v_1,v_2\}$. 
From the definition of a restless walk, it is clear that we can continue the walk from vertices $v_1, v_2$ to vertex $u$ at time $i$ only if we had reached $v_1$ or $v_2$ no earlier than time $i-\delta(v_1)$ and $i-\delta(v_2)$, respectively. 
Let $\prep{u}{\ell}{i}$ denote the encoding of all restless walks of length $\ell-1$ ending at vertex $u$ at time $i$. 

\begin{figure}[t]
\centering
\begin{tikzpicture}[scale=\tikzscale,every node/.style={scale=\tikzscale}]]

\input{tikz/tikz-defs}

\node[exnode] (v1) at (  -1,  1) {$v_1$};
\node[exnode] (v2) at (  -1, -1) {$v_2$};
\node[exnode] (u) at ( 1.5,    0) {$u$};

\node[fill=white] at ( -1,    0.4) {\large $\delta(v_1)$};
\node[fill=white] at ( -1,   -1.6) {\large $\delta(v_2)$};

\node[fill=white, text width=10cm] at ( 7.25, -0.7) {\large 
	$\chi_{u,\ell,i}(\vec{x},\vec{y})  =  x_{u} \displaystyle{\sum_{\substack{j \in \{0,\dots,\delta(v_1)\}\\ i-j>0}}} y_{uv_1,\ell-1,i-j} \chi_{v_1,\ell-1,i-j}(\vec{x},\vec{y}) \,+ $\\
    $\,\,\,\,\,\,\,\,\,\,\,\,\,\,\,\,\,\,\,\,\,\,\,\,\,\,\,\,\,\,\,x_{u} \displaystyle{\sum_{\substack{ \in \{0,\dots,\delta(v_2)\}\\ i-j>0}}} y_{uv_2,\ell-1,i-j} \chi_{v_1,\ell-1,i-j}(\mathbf{x,y}) \notag$\\};


\node [fill=white] at ( 0.5,   0.75) {$i$};
\node [fill=white] at ( 0.5,  -0.75) {$i$};

\path [draw=black,postaction={on each segment={mid arrow=black, ultra thick}}]
(v1) -- (u)
(v2) -- (u)
;
\end{tikzpicture}%
\caption{\label{fig:generating:1}
An illustration of generating restless walks.}
\end{figure}

We generate restless walks of length $\ell-1$ that end at vertex~$u$ at time $i$ using restless walks of length~$\ell-2$ by having
%
%
%
\[
\begin{split}
\prep{u}{\ell}{i} =\; & x_u \sum_{\substack{j \in \{0,\dots,\delta(v_1)\}\\i-j>0}}
                        y_{uv_1,\ell-1,i-j}\,\prep{v_1}{\ell-1}{i-j} \;+ \\
                      & x_u \sum_{\substack{j \in \{0,\dots,\delta(v_2)\}\\i-j>0}} 
                        y_{uv_2,\ell-1,i-j}\,\prep{v_2}{\ell-1}{i-j}.
\end{split}
\]
Generalizing from the previous example, the dynamic-programming recursion is written as
\[ \prep{u}{1}{i} = x_u,
\text{ for each } u \in V \text{ and } i \in [\tau], \text{ and}
\]
\begin{equation}
\label{eq:poly-enc:1}
\prep{u}{\ell}{i} = x_u \sum_{v \in N_i(u)} \sum_{\substack{j \in \{0,\dots,\delta(v)\}\\ i-j>0}} 
          y_{uv,\ell-1,i-j}\,\prep{v}{\ell-1}{i-j},
\end{equation}
for each $u \in V$, $\ell \in \{2,\dots,k\}$ and $i \in [\tau]$.

The following result is fundamental to our approach. 
\begin{lemma}
\label{lemma:algo:1}
The polynomial encoding $\prep{u}{\ell}{i}$ presented in Equation~(\ref{eq:poly-enc:1}) contains a multilinear monomial of degree $2\ell-1$ if and only if there exists a restless path of length $\ell-1$ ending at vertex $u$ reaching at time $i$.
\end{lemma}
\begin{proof}
We prove the claim by induction. For $l=1$, $\prep{u}{1}{i}=x_u$ for every $u \in V$ and $i \in [\tau]$, so the base case holds trivially. Assume that for $\ell-1$, $\prep{u}{\ell-1}{i}$ contains a multilinear monomial of degree $2\ell-3$ if and only if there exists a restless path of length $\ell-2$ ending at vertex $u \in V$ at time $i \in [\tau]$. 

For $\prep{u}{\ell}{i}$, from Equation~(\ref{eq:poly-enc:1}), the $y$-variables in any monomial are not repeated, as they have unique $\ell$ and $i$ subscript. By construction, we include the restless walks from neighbors $v \in N_i(u)$ to construct $\prep{u}{\ell}{i}$, if $v$ was reached no earlier than $i-\delta(v)$. Thus, all walks included in the polynomial encoding $\prep{u}{\ell}{i}$ are restless. 
Moreover, by construction, all walks encoded in $\prep{u}{\ell}{i}$ end at vertex $u$ and have length $\ell-1$, since all walks in $\prep{v}{\ell-1}{i}$ have length $\ell-2$. Suppose $\prep{u}{\ell}{i}$ contains a multilinear monomial with degree $2\ell-1$.
Since all $y$-variables are unique, there must be $\ell$ unique $x$-variables, corresponding to restless path of length $\ell-1$. Because, a monomial is multilinear if and only if it corresponds to a path, as established in~\cite[Lemma~2]{thejaswi2020pattern}.

Conversely, suppose there exists a restless path of length $\ell-1$ ending at vertex $u$ at the latest time $i$, which is encoded as $x_1 \dots x_v y_{uv,\ell-1,i} x_u$, but $\prep{u}{\ell}{i}$ does not contain this multilinear monomial of degree $2\ell-1$. From the construction in Equation~\ref{eq:poly-enc:1}, this implies that $\prep{v}{\ell-1}{j}$ for any $j \leq i - \delta(v)$ must not contain the multilinear monomial $x_1 \dots x_v$ with degree $2\ell - 3$, which corresponds to a restless path of length $\ell-2$ ending vertex $v$ and reaching at time $j$. This contradicts the premise that $\prep{v}{\ell-1}{j}$ contains a multilinear monomial of degree $2\ell-3$ if and only if a restless path of length $\ell-2$ exists, completing the proof.
\end{proof}


The next part of the problem is to detect a multilinear monomial in the polynomial in Equation~(\ref{eq:poly-enc:1}) representing the restless walks.  There is already established theory related to this problem \cite{bjorklund2016constrained,koutis2008faster,koutis2009limits}.  In particular, it is known that by substituting $2^\ell$ random values for the variables in $x$ and $y$, and evaluating them,  if one of the evaluation results in a non-zero term,  it implies that there exists at least one multilinear monomial of degree $2\ell-1$.

For each $u \in V$ and $j \in [\ell]$ we introduce a new variable $z_{u, j}$.  The vector of all variables of $z_{u,j}$ is denoted as $\mathbf{z}$ and the vector of all variables of $y$ is denoted as $\mathbf{y}$.  We write $z_{u}^L = \sum_{j \in L} z_{u,j}$, for $u \in V$, $L \subseteq [\ell]$ and $\mathbf{z}^L=\{z^{L}_{u}: u \in V\}$ for $L \subseteq [\ell]$. The values of the variables $z_{u,j}$ are assigned uniformly at random from $\GF(2^b)$. 
For simplicity we write $V = \{u_1, \dots,u_n\}$.

\begin{lemma}[Multilinear sieving \cite{bjorklund2014determinant}]
\label{lemma:multilinear-sieving}
The polynomial
\begin{equation}
\label{eq:multilinear:1}
\qrep{u}{\ell}{i} = \sum_{L \subseteq [\ell]} \chi_{u,\ell,i}(z_{u_1}^L,\dots,z_{u_n}^L, \vec{y})
\end{equation}
is not identically zero if and only if $\prep{u}{\ell}{i}$
contain a multilinear monomial of degree $2\ell-1$.
\end{lemma}

\begin{lemma}
\label{lemma:multilinear}
Evaluating the polynomial in Equation~(\ref{eq:multilinear:1}) can be done in time $\bigO(2^\ell \ell m \Delta)$ and space~$\bigO(n\Delta)$.
\end{lemma}
\begin{proof}
Recall that $\Delta = \max_{v \in V} \delta(v)$,  $m_i$ is the number of edges at $i \in [\tau]$, and $n$ is the number of vertices. Computing $\prep{u}{j}{i}$ for all $u \in V$ requires $(\Delta+1) m_i$ multiplications and additions. We repeat this for all $i \in [\tau]$ and $j \in [\ell]$,  which requires $\bigO(\ell m \Delta)$ multiplications and additions.  Finally we evaluate the polynomial $\qrep{u}{\ell}{i}$ for all $u \in V$ using $2^\ell$ random substitution of variables in $\mathbf{z}^L=\{z_{u_1}^L,\dots,z_{u_n}^L\}$, for each $L \subseteq [\ell]$, which takes time $\bigO(2^\ell \ell m \Delta)$. So the runtime is $\bigO(2^\ell m \Delta)$.
\change{The dynamic programming scheme to compute $\prep{u}{\ell}{i}$ for all $u \in V$, requires the values of $\prep{v}{\ell-1}{i-j}$ for all $v \in V$, $j \in [i-\delta(v),\dots,i]$. Since $\Delta = \max_{v \in V} \delta(v)$, it follows that the space requirement is $\bigO(n \Delta)$.}
\end{proof}

In the next section, we introduce color constraints for the vertices in the restless path. More precisely, given a vertex-colored temporal graph $G=(V,E)$ with a coloring function $c:V \rightarrow C$, where $C$ is a set of colors, and a multiset $M$ of colors, we consider the problem of deciding the existence  of a restless path such that the vertex colors of the path agree with the multiset of colors in $M$. This generalization of the restless path problem with color constraints will be used to solve restless reachability problems in the later sections.


\subsection{Introducing vertex-color constraints}
Given a multiset of colors $M$, we extend the multilinear sieving technique to  detect the existence of a multilinear monomial, such that colors corresponding to the $x$-variables agree with the colors in multiset~$M$.
Recall our earlier definition where $\mu(s)$  denotes the number of occurrences of color $s$ in a multiset $M$. Furthermore, for each $s \in C$,  let $S_s$ denote the set of $\mu(s)$ shades of the color $s$,  with $S_s \cap S_{s'} = \emptyset$ for all $s \neq s'$.  In other words, for each color $s \in C$ we create $\mu(s)$ shades,  so that any two distinct colors have different shades. As an example, for the color multiset $M=\{1,1,2\}$ we have $S_1=\{1_1,1_2\}$ and $S_2=\{2_1\}$, so that $S_1 \cap S_2 = \emptyset$.

For each $u \in V$ and $d \in S_{c(u)}$ we introduce a new variable $\gamma_{u, d}$. For each $d \in \cup_{s \in C} S_s$ and each label $j \in [\ell]$ we introduce a new variable $\omega_{j,d}$. The values of variables $\gamma_{u,d}$ and $\omega_{d,j}$ are drawn uniformly at random from the Galois field $\GF(2^b)$. We write 
\begin{equation}
\label{eq:multilinear:3}
z_{u,j} = \sum_{d \in S_{c(u)}} \gamma_{u,d} \omega_{d,j}, \text{\: and \:} 
z_u^L = \sum_{j \in L} z_{u,j},
\end{equation}
for $L \subseteq [\ell]$, $u \in V$.
%
The following lemma
extends Lemma~\ref{lemma:multilinear-sieving}
in the case of vertex-color constraints.

\begin{lemma}[Constrained multilinear sieving~\cite{bjorklund2016constrained}]
\label{lemma:constraint-multilinear-sieving}
The polynomial  $\prep{u}{\ell}{i}$ contain a multilinear
monomial of degree $2\ell-1$ and it is properly colored if and only if the polynomial
%
\begin{equation}
\label{eq:multilinear:2}
\qrepw{u}{\ell}{i} = 
          \sum_{L \subseteq [l]} \chi_{u,\ell,i}(z_{u_1}^L,\dots,z_{u_n}^L,\vec{y})
\end{equation}
is not identically zero. %
\end{lemma}

From Lemma~\ref{lemma:constraint-multilinear-sieving}, we can determine the existence of a multilinear monomial in  $\qrepw{u}{\ell}{i}$, by making $2^\ell$ random substitutions of the new variables $\mathbf{z}$ in Equation~\ref{eq:multilinear:2}. As detailed in Bj{\"o}rklund et al.~\cite{bjorklund2016constrained} these substitutions can be random for a low-degree polynomial that is not identically zero with only few roots. If one evaluates the polynomial at a random point, one is likely to witness that it is not identically zero~\cite{schwartz1980fast, zippel1979probabilistic}. This gives rise to a randomized algorithm with a false negative probability of $2^{-b}(2\ell-1)$, where the arithmetic is over the Galois field $\GF(2^b)$. Here, again, $b$ is the number of bits used to generate the random values. We make use of this result for designing our algorithms.

\begin{lemma}
\label{lemma:multilinear:2}
Evaluating the polynomial in Equation~(\ref{eq:multilinear:2}) can be done in time $\bigO(2^\ell \ell m \Delta)$ and space~$\bigO(n \Delta)$.
\end{lemma}

The proof of Lemma~\ref{lemma:multilinear:2} is similar to the one of Lemma~\ref{lemma:multilinear}. From Lemma~\ref{lemma:constraint-multilinear-sieving} we obtain an algorithm to detect the existence of a restless path ending at vertex $u \in V$ at time $i$ and the vertex colors of the restless path agree with the colors in multiset $M$.

\subsection{A fine-grained decision oracle} 
In this section we present a \emph{fine-grained} evaluation scheme to evaluate the polynomial in Equation~(\ref{eq:multilinear:2}). 

\para{Intuition.} 
Consider the graph illustrated in Figure~\ref{fig:fine-grained:1}, with vertex set $V=\{v_1,\dots,v_5\}$ with resting time $\delta(v_1) = \dots = \delta(v_5) = 2$, \ie, we can wait at most~$2$ time steps at any vertex.  In order to decide whether there exists a restless path of length $3$ (i.e., $\ell=4$) in the graph, we need to evaluate the polynomial
$\chi(\vec{x},\vec{y})=\sum_{u \in \{v_1,\dots,v_5\}} \sum_{i \in [\tau]} \prep{u}{4}{i}$.
However, to decide if there exists a restless path of length $3$ ending at vertex $v_5$ it is sufficient to evaluate the polynomial $\sum_{i \in [\tau]} \prep{v_5}{4}{i}$.  Furthermore, to decide if there exists a restless path of length $3$ ending at vertex $v_5$ at time $5$, we can restrict the evaluation to the polynomial $\prep{v_5}{4}{5}$. 
Similarly, it suffices to evaluate the polynomial $\prep{u}{\ell}{i}$ to determine whether there exists a path of length $\ell-1$ ending at vertex $u$ at time $i$.

\begin{figure}
\centering
\begin{tikzpicture}[scale=\tikzscale,every node/.style={scale=\tikzscale}]]

\input{tikz/tikz-defs}

\node[exnode, ultra thick, draw=blue] (v1) at (   0, 0) {$v_1$};
\node[exnode] (v2) at (   3, 0) {$v_2$};
\node[exnode, ultra thick, draw=blue] (v3) at (   0, -3) {$v_3$};
\node[exnode, ultra thick, draw=blue] (v4) at (   3, -3) {$v_4$};
\node[exnode, ultra thick, draw=blue] (v5) at (   6, -3) {$v_5$};

\node[fill=white] at (  1.5, -0.5) {\large $1$};
\node[fill=white] at (  3.5, -1.5) {\large $1$};
\node[fill=white] at (  -0.5, -1.5) {\large $2$};
\node[fill=white] at ( 1.5, -2.5) {\large $3$};
\node[fill=white] at ( 4.5, -2.5) {\large $5$};

\node[fill=white] at ( 0.6, 0.7) {\large $\chi_{v_1,4,i}(\vec{x},\vec{y})$};
\node[fill=white] at ( 3.6, 0.7) {\large $\chi_{v_2,4,i}(\vec{x},\vec{y})$};
\node[fill=white] at ( 0.6, -3.6) {\large $\chi_{v_3,4,i}(\vec{x},\vec{y})$};
\node[fill=white] at ( 3.6, -3.6) {\large $\chi_{v_4,4,i}(\vec{x},\vec{y})$};
\node[fill=white] at ( 6.6, -3.6) {\large $\chi_{v_5,4,i}(\vec{x},\vec{y})$};

\path [draw=black, postaction={on each segment={mid arrow=black, ultra thick}}]
(v1) -- (v2)
(v2) -- (v4)
;

\path [draw=black,ultra thick, color=blue, postaction={on each segment={mid arrow=blue, ultra thick}}]
(v1) -- (v3)
(v3) -- (v4)
(v4) -- (v5)
;

\node[fill=white, text width=7cm] at ( 11, -1.2) {\large
$\displaystyle{ \chi(\vec{x},\vec{y}) =  \sum_{u \in \{v_1,\dots,v_5\}} \sum_{i \in
[\tau]} \chi_{u,4,i}}(\vec{x},\vec{y}) $};
\end{tikzpicture}%
\caption{\label{fig:fine-grained:1} An example to illustrate the fine-grained evaluation scheme. The resting time of vertices is $\delta(v_1) = \dots = \delta(v_5) = 2$. A restless path of length $3$ (i.e, $\ell=4$) from vertex $v_1$ to vertex $v_5$ is highlighted in bold.}
\end{figure}

\para{Fine-grained decision oracle.}
Let us generalize the fine-grained evaluation scheme to any temporal graph using the intuition presented. Instead of evaluating a single polynomial
$\chi(\mathbf{x},\mathbf{y}) = \sum_{u \in V} \sum_{\ell \in [k]} \sum_{i \in[\tau]} \prep{u}{\ell}{i}$, 
we work with a set of $n \tau$ polynomials $\left\{\prep{u}{\ell}{i}: u \in V, i \in [\tau]\right\}$ and evaluate each $\prep{u}{\ell}{i}$ independently.
Observe carefully that our generating function in Equation~(\ref{eq:poly-enc:1}) generates a polynomial encoding of all restless walks $\prep{u}{\ell}{i}$ for each $u \in V$ and $i \in [\tau]$ independently for a fixed $\ell$. If the corresponding evaluation polynomial $\qrepw{u}{\ell}{i}$ evaluates to a non-zero term, it implies that there exists a restless path of length $\ell-1$ ending at vertex $u$ at time $i$ and the vertices in the path agree with the multiset of colors in $M$. Using this fine-grained evaluation scheme, we obtain a set of timestamps $R_u = \{i : \qrepw{u}{\ell}{i} \neq 0 \text{ for all } i \in [\tau]\}$ of restless paths ending at vertex $u \in V$ and satisfying the color constraints in $M$. The pseudocode is presented in Algorithm~\ref{algo:1}.

The term \emph{fine-grained oracle} is used to differentiate it from a decision oracle,   which only reports the existence of a restless path with a {\yes}/{\no} answer.  Our fine-grained decision oracle reports more insightful details,  for instance it can answer if there exists a restless path of length $\ell-1$ ending at vertex $u \in V$ at time $i \in [\tau]$ satisfying the color constraints specified in $M$ with a {\yes}/{\no} answer.
A single run of the fine-grained oracle is sufficient to obtain the set of vertices $D \subseteq V$ such that there exists a restless path of length $\ell-1$ ending at each $u \in D$. \change{Additionally, we can obtain all reachable timestamps $R_{u,i}=1$ if $\qrepw{u}{k}{i} \neq 0, i \in [\tau]$ and for each $u \in V$ using a single query to the fine-grained oracle, but this would require $\bigO(n\tau)$ space.}

This re-design of the polynomial evaluation scheme improves the runtime for solving  the temporal path problems described in previous  work~\cite{thejaswi2020pattern,thejaswi2020finding},  where the authors search for a temporal path that contains colors specified in the query.  That is, by replacing their decision oracle with our fine-grained decision oracle, we reduce the number of queries by a factor of $\log \tau$. This, in turn, reduces the total runtime of their solution for detecting an optimal solution from $\bigO(2^\ell \ell (n\tau + m) \log \tau)$ to $\bigO(2^\ell \ell (n\tau + m))$.
Even though the theoretical improvement is modest, it is important to note that for large values of $\ell$ a single run of the decision oracle can take hours to complete, so the practical improvement is significant (for precise results, see Section~\ref{sec:experiments} and Table~\ref{table:exp:realworld:2}, as well as previous work~\cite{kaski2018engineering}).

\begin{algorithm}[t]
{\footnotesize
\DontPrintSemicolon
\caption{\sc FineGrainedOracle($G=(V,E), c, k, M$)}
\label{algo:1}

\KwIn{$G=(V,E)$, input graph \\ 
\Indp \Indp $c$, coloring function\\
$k$, length of path\\
$M$, multiset of colors}

\KwOut{$R$, minimum reachability time\\
\Indp \Indp $D$, set of reachable vertices}

\For{$L \subseteq [k]$}{
  $\mathbf{z}^L \gets \textsc{\small Get-z-assignment}(c, M, L)$ \tcp*{Eq.\,(\ref{eq:multilinear:3})}
  \For{$u \in V, i \in \tau$} {
      $\qrepw{u}{1}{i} \gets z_{u}^{L}$ 
  }
}

\For{$\ell \in \{2,\dots,k\}$}{
  \For{$ u \in V, i \in [\tau]$} {
      $\qrepw{u}{\ell}{i} \gets 0$\\
      \For{$L \subseteq [k]$}{
        $\mathbf{z}^L \gets \textsc{\small Get-z-assignment}(c, M, L)$ \tcp*{Eq.\,(\ref{eq:multilinear:3})}
        $\qrepw{u}{\ell}{i} \gets \qrepw{u}{\ell}{i} \oplus
            \chi_{u,\ell,i}(\mathbf{z}^L,\vec{y})$ \tcp*{Eq.\,(\ref{eq:multilinear:2})}
      }
  }
}

$D \gets \emptyset$,
\change{$R \gets \{R_{u} \gets \infty: u \in V\}$}\\
\For{$u \in V, i \in [\tau]$}{
  \If{$\qrepw{u}{k}{i} \neq 0$}{
\change{
    $R_u \gets \min(R_u, i)$ \tcp*{$R_{u,i} \gets 1$ to obtain all reachable timestamps with space $\bigO(n\tau)$}}
    \If{$u \notin D$}{
      $D \gets D \cup \{u\}$
    }
  }
}
\Return $R, D$
}
\end{algorithm}

Let us then turn to our algorithmic results.
\begin{theorem}
\label{theorem:algo:restlessmotifreach}
There exists a randomized algorithm for solving \krestlessmotifreach problem in time 
$\bigO(2^k k m \Delta)$ and \change{space $\bigO(n \Delta)$}.
\end{theorem}
\begin{proof}
Given an instance $(G,c,k,M,s)$ of \krestlessmotifreach we build a graph $G'=(V',E')$ such that 
$V=V \cup \{s'\}$,
$E'=E \cup \{(s',s,i): (s,u,i) \in E\}$, 
$c'(u) = c(u)$ for all $u \in V$, $c'(s')=k+1$, 
$M'=M \cup \{c'(s')\}$ and query the \finegrainedoracle with instance $(G',c',k+1,M')$.
The construction is depicted in Figure~\ref{fig:algo:krestlessmotifreach}.
In the instance $(G',c',k+1,M')$, the origin of the graph is enforced by introducing an additional vertex $s'$ adjacent to $s$.
Since $s'$ is assigned a unique color, if there is a resting path agreeing with $M'$, then the path must originate from $s'$ and pass through~$s$.  If there exists a restless path originating from $s'$ and ending at $u \in V \setminus \{s, s'\}$ such that the vertex colors of the path agree with $M'$, then we have a restless path originating from $s$ and ending at $u$ such that the vertex colors of the path agree with $M$.
As the graph $G'$ will have at most $2m$ edges and $n+1$ vertices, we have obtained an algorithm for solving \krestlessmotifreach using $\bigO(2^k k m \Delta)$ time and $\bigO(n \Delta)$ space.
\end{proof}


\begin{figure}
\centering
\tikzset{every picture/.style={scale=0.9}}
\setlength{\tabcolsep}{30pt}
\begin{tabular}{c c}
\begin{tikzpicture}[scale=\tikzscale,every node/.style={scale=\tikzscale}]]

\input{tikz/tikz-defs}

\node[exsquare, draw=green] (v1) at (  0,  0) {$v_1$};
\node[exnode, draw=black] (v2) at ( -2, -2) {$v_2$};
\node[exnode, draw=black] (v3) at (  2, -2) {$v_3$};
\node[exsquare, draw=green] (v4) at (  0, -4) {$v_4$};

\node[fill=white] at (  -1.4, -0.8) {\large $1$};
\node[fill=white] at ( 0.3, -1) {\large $3$};
\node[fill=white] at ( 1.7, -0.7) {\large $1,2$};
\node[fill=white] at (  -0.5, -1.6) {\large $2$};
\node[fill=white] at ( -1.7, -3) {\large $1,2$};
\node[fill=white] at ( 0.3, -3) {\large $2$};
\node[fill=white] at ( 1.6, -3.5) {\large $1$};

\node[fill=white] at ( 0, 0.6) {\large $s = v_1$};

\node[fill=white] at ( -2, -5) {\large $M\,:$};
\node[exsquare, draw=green] at (-1, -5) {$g$};
\node[exnode, draw=black] at (0, -5) {$b$};
\node[exnode, draw=black] at (1, -5) {$b$};

\path [draw=black, postaction={on each segment={mid arrow=black, ultra thick}}]
(v2) -- (v1)
(v1) to [bend left=16] (v3)
(v3) to [bend left=16] (v1)
(v2) -- (v4)
(v3) -- (v2)
(v4) to [bend right=16] (v3)
(v3) to [bend right=16] (v4)
;

\end{tikzpicture}%
 &
\begin{tikzpicture}[scale=\tikzscale,every node/.style={scale=\tikzscale}]]

\input{tikz/tikz-defs}

\node[exdiamond, draw=red] (v0) at (  -2,  0) {$s'$};
\node[exsquare, draw=green] (v1) at (  0,  0) {$v_1$};
\node[exnode, draw=black] (v2) at ( -2, -2) {$v_2$};
\node[exnode, draw=black] (v3) at (  2, -2) {$v_3$};
\node[exsquare, draw=green] (v4) at (  0, -4) {$v_4$};

\node[fill=white] at ( -1, 0.4) {\large $1,2$};
\node[fill=white] at (  -1.4, -0.8) {\large $1$};
\node[fill=white] at ( 0.3, -1) {\large $3$};
\node[fill=white] at ( 1.7, -0.7) {\large $1,2$};
\node[fill=white] at (  -0.5, -1.6) {\large $2$};
\node[fill=white] at ( -1.7, -3) {\large $1,2$};
\node[fill=white] at ( 0.3, -3) {\large $2$};
\node[fill=white] at ( 1.6, -3.5) {\large $1$};

\node[fill=white] at ( 0, 0.6) {\large $s = v_1$};

\node[fill=white] at ( -2, -5) {\large $M'\,:$};
\node[exsquare, draw=green] at (-1, -5) {$g$};
\node[exnode, draw=black] at (0, -5) {$b$};
\node[exnode, draw=black] at (1, -5) {$b$};
\node[exdiamond, draw=red] at ( 2, -5) {$r$};

\path [draw=black, postaction={on each segment={mid arrow=black, ultra thick}}]
(v0) -- (v1)
(v2) -- (v1)
(v1) to [bend left=16] (v3)
(v3) to [bend left=16] (v1)
(v2) -- (v4)
(v3) -- (v2)
(v4) to [bend right=16] (v3)
(v3) to [bend right=16] (v4)
;

\end{tikzpicture}%
\end{tabular}
\caption{\label{fig:algo:krestlessmotifreach}
An illustration of graph construction to solve \krestlessmotifreach. An instance of \krestlessmotifreach with a multiset of colors (left) and the transformation of the graph and the multiset (right). For illustrative purpose, we denote the vertices with color \underline{r}ed, \underline{b}lack and \underline{g}reen using diamond, circular and square shapes, respectively.}
\end{figure}

An algorithm for \krestlessreach is obtained by transforming it to a \krestlessmotifreach instance. More precisely, the algorithm works by constructing a vertex-colored instance that encodes the source, whereas a multiset of colors encodes the path length and the solution to \restlessreach following as a special case.

\begin{theorem}
\label{theorem:algo:krestlessreach}
There exists a randomized algorithm for \krestlessreach in time $\bigO(2^k k m \Delta)$ and \change{space $\bigO(n \Delta)$}.
\end{theorem}
\begin{proof}
Given an instance $(G,s,k)$ of \krestlessreach, we introduce a coloring function $c:V \rightarrow \{1,2\}$ such that $c(s) = 2$ and $c(v) = 1$ for all $v \in V \setminus \{s\}$. We obtain a graph $G'=(V, E \setminus \{(u,v,i) \in E : v = s\})$ by removing all incoming edges to $s$ in $G$, and by setting the multiset $M=\{1^{k-1}\} \cup \{2\}$. We query the \finegrainedoracle with instance $(G', c, k , M)$. The transformation is illustrated in Figure~\ref{fig:algo:krestlessreach}.
In the instance $(G', c, k, M)$, the origin of the restless path is enforced by removing all incoming edges to $s$ and coloring $s$ with a unique color. If we have a restless path ending at $u \in V \setminus \{s\}$ and agreeing with multiset $M$, it implies that the temporal path originates from $s$ and ends at $u$.
The graph $G'$ has $n+1$ vertices and $m$ edges, so we have a $\bigO(2^k k m \Delta)$ time and $\bigO(n \Delta)$ space algorithm for solving \krestlessreach.
\end{proof}

\begin{figure}
\centering
\tikzset{every picture/.style={scale=0.9}}
\setlength{\tabcolsep}{30pt}
\begin{tabular}{c c}
\begin{tikzpicture}[scale=\tikzscale,every node/.style={scale=\tikzscale}]]

\input{tikz/tikz-defs}

\node[exnode, draw=black] (v1) at (  0,  0) {$v_1$};
\node[exnode, draw=black] (v2) at ( -2, -2) {$v_2$};
\node[exnode, draw=black] (v3) at (  2, -2) {$v_3$};
\node[exnode, draw=black] (v4) at (  0, -4) {$v_4$};

\node[fill=white] at (  -1.4, -0.8) {\large $1$};
\node[fill=white] at ( 0.3, -1) {\large $3$};
\node[fill=white] at ( 1.7, -0.7) {\large $1,2$};
\node[fill=white] at (  -0.5, -1.6) {\large $2$};
\node[fill=white] at ( -1.7, -3) {\large $1,2$};
\node[fill=white] at ( 0.3, -3) {\large $2$};
\node[fill=white] at ( 1.6, -3.5) {\large $1$};

\node[fill=white] at ( 0, 0.6) {\large $s = v_1$};

\node[fill=white] at ( -2, -5.15) {};

\path [draw=black, postaction={on each segment={mid arrow=black, ultra thick}}]
(v2) -- (v1)
(v1) to [bend left=16] (v3)
(v3) to [bend left=16] (v1)
(v2) -- (v4)
(v3) -- (v2)
(v4) to [bend right=16] (v3)
(v3) to [bend right=16] (v4)
;

\end{tikzpicture}%
 &
\begin{tikzpicture}[scale=\tikzscale,every node/.style={scale=\tikzscale}]]

\input{tikz/tikz-defs}

\node[exdiamond, draw=red] (v1) at (  0,  0) {$v_1$};
\node[exnode, draw=black] (v2) at ( -2, -2) {$v_2$};
\node[exnode, draw=black] (v3) at (  2, -2) {$v_3$};
\node[exnode, draw=black] (v4) at (  0, -4) {$v_4$};

\node[fill=white] at ( 1.7, -0.7) {\large $1,2$};
\node[fill=white] at (  -0.5, -1.6) {\large $2$};
\node[fill=white] at ( -1.7, -3) {\large $1,2$};
\node[fill=white] at ( 0.3, -3) {\large $2$};
\node[fill=white] at ( 1.6, -3.5) {\large $1$};

\node[fill=white] at ( 0, 0.6) {\large $s = v_1$};

\node[fill=white] at ( -2, -5) {\large $M\,:$};
\node[exdiamond, draw=red] at (-1, -5) {$r$};
\node[exnode, draw=black] at (0, -5) {$b$};
\node[exnode, draw=black] at (1, -5) {$b$};

\path [draw=black, postaction={on each segment={mid arrow=black, ultra thick}}]
(v1) to [bend left=16] (v3)
(v2) -- (v4)
(v3) -- (v2)
(v4) to [bend right=16] (v3)
(v3) to [bend right=16] (v4)
;

\end{tikzpicture}%
\end{tabular}
\caption{\label{fig:algo:krestlessreach}
The transformation of \krestlessreach instance (left) to \krestlessmotifreach instance (right). For illustrative purposes we denote vertices with color \underline{r}ed and \underline{b}lack using diamond and circular shapes, respectively.}
\end{figure}

\begin{theorem}
\label{theorem:algo:restlessreach}
There exists a randomized algorithm for \restlessreach in time $\bigO(2^n n m \Delta)$ and \change{space $\bigO(n \Delta)$}.
\end{theorem}
\begin{proof}
For solving \restlessreach, the construction is similar to
Theorem~\ref{theorem:algo:krestlessreach}.
However, we need to make $n-2$ calls to the \finegrainedoracle assuming the maximum length of
the restless path is $n-1$.
Finally, we obtain a set $R' = \{ R'_{u,i}: u \in V, i \in [\tau]\}$
such that $R'_{u,i} = 1$ if there exists a restless path
from $s$ to $u$ ending at time $i$ with length at most $n-1$,
and $R'_{u,i} = 0$, otherwise. The pseudocode is available in
Algorithm~\ref{algo:restlessreach}.

In total, we make $n-2$ \finegrainedoracle calls for each $k \in
\{2,\dots,n-1\}$.
Each run of the oracle takes $\bigO(2^k k m\Delta)$ time.
To summarize, the runtime of the algorithm is
${\sum_{k=2}^{n-1} 2^k k m\Delta}$,
which is  $\bigO(2^n n\allowbreak m\Delta )$.
The space complexity is $\bigO(n\tau)$, completing the proof.
\end{proof}

\begin{algorithm}[t]
\caption{\sc RestlessReach($G=(V,E), s$)}
\label{algo:restlessreach}
{\footnotesize
\KwIn{$G=(V,E)$, input graph\\
\Indp \Indp $s$, source vertex}

\KwOut{$R$, minimum reachability time\\
\Indp \Indp $D$, set of reachable vertices}

$n \gets |V|$\\
\For{$u \in (V \setminus \{s\})$} {
  $c(u) \gets 1$
}
$c(s) \gets 2$\\

$G' \gets (V, E \setminus \{(u,s,i) \in E\})$\\

\For{$k \in \{2,\dots,n-1\}$} {
  $M \gets \{1^{k-1}\} \cup \{2\}$\\
  $R, D \gets \textsc{FineGrainedOracle}(G', c, k, M)$\\
}
\Return{$R, D$}
}
\end{algorithm}

\subsection{Discussion} 
Consider the following variant of the restless reachability problem that we call at-most-$k$-restless reachability problem (\atmkrestlessreach). Given a temporal graph and a source vertex we need to find a set of vertices which are reachable from source via a restless path such that the length of the path is at most $k-1$.
From Theorem~\ref{theorem:algo:restlessreach}, we have a randomized algorithm for solving \atmkrestlessreach in time $\bigO(2^k k m\Delta)$ and space $\bigO(n \Delta)$.
Also note that we can use the algorithms for \krestlessmotifreach, \krestlessreach and \restlessreach as we can solve \krestlessmotif, \krestlesspath and \restlesspath, respectively.

A general variant of \restlessreach with a set of sources $S \subseteq V$   can be reduced to \restlessreach with a single source by introducing an additional vertex $s'$ and connecting all the sources $s \in S$ to $s'$ with a temporal edge. More precisely, given a graph $G=(V,E)$ and set of sources $S \subseteq V$, we construct a graph $G'=(V',E')$ with $V' = V \cup \{s'\}$ and $E'=E \cup \left\{(s',s,i)\mid s\in S \text{~and~} (s,u,i)  \in E\right\}$.  Solving \restlessreach on the graph instance $G'$ with source $s'$ is equivalent to solving \restlessreach with set of sources~$S$.
Additionally, finding the restless path that minimizes the length (shortest path), minimizes the arrival time (fastest path), or minimizes the total waiting time (foremost path) can be computed using the output $\left\{R'_{u,i}: u \in V, i \in [\tau]\right\}$ for each $\ell \in \{2,\dots,k\}$ from the fine-grained oracle. However, enabling such computation requires $\bigO(n \tau k)$ space.

\subsection{Extracting an optimal solution using $k$ queries} \label{sec:algo:extraction}
In this section, we present an algorithm for extracting an optimal solution for \krestlessmotif and \krestlesspath using $k$ queries to the \finegrainedoracle. By optimal we mean that the maximum timestamp in the restless path is minimized.

Our \finegrainedoracle reports the existence of a restless path from a given source to a destination at discrete timestamps with a \yes/\no answer. However, in many cases we also require an explicit solution, \ie, a restless path which actually witnesses the fact. We present two approaches to extracting an optimal solution. Our first approach makes use of self-reducibility of decision oracles and temporal \dfs, based on the previous work of Bj\"{o}rklund et al.~\cite{bjorklund2014fast} and Thejaswi et al.~\cite{thejaswi2020finding}. Our second approach makes use of the fine-grained oracle.
For using self-reducibility, we implement a naive version of the multilinear sieve which only reports a \yes/\no answer without the fine-grained capabilities, that is,  the algorithm does not report the set of vertices and the timestamps at which the vertices are reachable from a  given source via a restless path. We summarize our results in Table~\ref{table:summary:1}.

\para{Self-reducibility and temporal {\small DFS}.}
The approach works in three steps: 
First, we obtain the minimum (optimal) timestamp $t \in [\tau]$ for which there exists a feasible solution. For this, we construct the polynomial encoding of restless walks of length $k-1$ which end at time at most $t' \in [\tau]$ and query the decision oracle for the existence of a solution. Using binary search on the range $[\tau]$, we use at most $\log \tau$ queries to obtain the optimal timestamp.
Next, we extract a $k$-vertex temporal subgraph that contains a restless temporal path. By recursively dividing the graph in to two halves we can obtain the desired subgraph using $\bigO(k \log n)$ queries to the decision oracle in expectation~\cite{bjorklund2014fast}.
Finally, we extract the restless path by performing a temporal {\small DFS} from a given source in the $k$-vertex subgraph. Even though the worst case complexity of the temporal {\small DFS} is $\bigO(k!)$, we demonstrate that the approach is practical.
For technical details of self-reducibility of decision oracles, we refer the interested reader to Bjorklund et al.~\cite{bjorklund2014fast}.

In summary, the overall complexity of extracting an optimal solution using self-reducibility and temporal {\small DFS} is $\bigO((2^k k^2 m \Delta \log n \log \tau) + k!)$.

\para{Using fine-grained oracle.}
As a second approach, extracting a solution can also be done with $k$ queries to the fine-grained oracle. 
Let us present the high-level idea behind fine-grained extraction.  Consider a temporal graph presented in Figure~\ref{fig:finegrained-extraction} with vertex set $V=\{v_1,\dots,v_6\}$ with resting times $\delta(v_1)=\dots=\delta(v_6)=2$. In this example we want to extract a restless path of length $3$ from vertex $v_1$ to vertex $v_6$, if such a restless path exists. For illustrative purposes, we use colors \underline{b}lack and \underline{r}ed and have the multiset of colors $M=\{r,b,b,b\}$. For the first iteration it suffices to verify if $\qrepw{v_6}{4}{i}$ evaluates to a non-zero term for each $i \in [\tau]$. Since there exists a restless path of length $3$ ending at vertex $v_6$ at time $5$ agreeing with the colors in $M$, the corresponding evaluation polynomial $\qrepw{v_6}{4}{5}$ is non-zero.
For the second iteration, delete vertex $v_6$ from the graph and remove a $b$ from $M$ leaving us with $M = \{ r,b,b \}$.  Now check if there exists a restless path of length $2$ ending at any of the neighbors of $v_6$, i.e., $N_5(v_6) = \{v_3, v_5\}$ at any of the timestamps $i \in \{3,4,5\}$. This can be done by verifying if the polynomials $\qrepw{v_3}{2}{i}, \qrepw{v_5}{2}{i}$  evaluate to a non-zero term for timestamps $i \in \{3,4,5\}$. In our case, $\qrepw{v_5}{2}{3}$ evaluates to a non-zero term, which implies that there exists a restless path from $v_1$ to $v_3$ ending at timestamp $3$ agreeing with the colors in $M'$, so we add the edge $(v_5,v_6,5)$ to the solution. We can recursively repeat the second iteration until we reach the vertex $v_1$ to obtain a restless path from $v_1$ to $v_5$.

\begin{figure}
\centering
\tikzset{every picture/.style={scale=0.9}}
\begin{tabular}{c c}
\begin{tikzpicture}[scale=\tikzscale,every node/.style={scale=\tikzscale}]

\input{tikz/tikz-defs}

\node[exdiamond, ultra thick, draw=red] (v1) at (   0, 0) {$v_1$};
\node[exnode, draw=black] (v2) at (   3, 0) {$v_2$};
\node[exnode, draw=black] (v3) at (   6, 0) {$v_3$};
\node[exnode, ultra thick, draw=black] (v4) at (   0, -3) {$v_4$};
\node[exnode, ultra thick, draw=black] (v5) at (   3, -3) {$v_5$};
\node[exnode, ultra thick, draw=black] (v6) at (   6, -3) {$v_6$};

\node[fill=white] at (  1.5, -0.5) {\large $1$};
\node[fill=white] at ( 4.5, -0.5) {\large $2$};
\node[fill=white] at (  3.5, -1.5) {\large $1$};
\node[fill=white] at (  -0.5, -1.5) {\large $2$};
\node[fill=white] at ( 1.5, -2.5) {\large $3$};
\node[fill=white] at ( 4.5, -2.5) {\large $5$};
\node[fill=white] at ( 6.5, -1.5) {\large $5$};

\node[fill=white] at ( 6, -3.6) {\large $\zeta_{v_6,4,i}(\vec{z},\vec{w},\vec{y})$};

\node[fill=white] at ( 1, -4.5) {\large $M:$};
\node[exdiamond, draw=red] at ( 2, -4.5) {$r$};
\node[exnode, draw=black]  at ( 3, -4.5) {$b$};
\node[exnode, draw=black]  at ( 4, -4.5) {$b$};
\node[exnode, draw=black]  at ( 5, -4.5) {$b$};

\path [draw=black, postaction={on each segment={mid arrow=black, ultra thick}}]
(v1) -- (v2)
(v2) -- (v3)
(v3) -- (v6)
(v2) -- (v5)
;

\path [draw=black,ultra thick, color=blue, postaction={on each segment={mid arrow=blue, ultra thick}}]
(v1) -- (v4)
(v4) -- (v5)
(v5) -- (v6)
;

\draw[purple, dashed, ultra thick]  (6.4,-2) arc (100:160:1.9cm);

\end{tikzpicture}%
 &
\begin{tikzpicture}[scale=\tikzscale,every node/.style={scale=\tikzscale}]]

\input{tikz/tikz-defs}

\node[exdiamond, ultra thick, draw=red] (v1) at ( 0, 0) {$v_1$};
\node[exnode, draw=black] (v2) at ( 3, 0) {$v_2$};
\node[exnode, draw=black] (v3) at ( 6, 0) {$v_3$};
\node[exnode, ultra thick, draw=black] (v4) at ( 0, -3) {$v_4$};
\node[exnode, ultra thick, draw=black] (v5) at ( 3, -3) {$v_5$};

\node[fill=white] at (  1.5, -0.5) {\large $1$};
\node[fill=white] at (  4.5, -0.5) {\large $2$};
\node[fill=white] at (  3.5, -1.5) {\large $1$};
\node[fill=white] at (  -0.5, -1.5) {\large $2$};
\node[fill=white] at ( 1.5, -2.5) {\large $3$};

\node[fill=white] at ( 3.6, -3.6) {\large $\zeta_{v_5,3,i}(\vec{z},\vec{w},\vec{y})$};
\node[fill=white] at ( 6, -0.6) {\large $\zeta_{v_3,3,i}(\vec{z},\vec{w},\vec{y})$};

\node[fill=white] at ( 1, -4.5) {\large $M:$};
\node[exdiamond, draw=red] at ( 2, -4.5) {$r$};
\node[exnode, draw=black]  at ( 3, -4.5) {$b$};
\node[exnode, draw=black]  at ( 4, -4.5) {$b$};

\path [draw=black, postaction={on each segment={mid arrow=black, ultra thick}}]
(v1) -- (v2)
(v2) -- (v3)
(v2) -- (v5)
;

\path [draw=black,ultra thick, color=blue, postaction={on each segment={mid arrow=blue, ultra thick}}]
(v1) -- (v4)
(v4) -- (v5)
;

\draw[purple, dashed, ultra thick]  (3.4,-2) arc (100:160:1.9cm);

\end{tikzpicture}%
 \\
iteration-$1$ & iteration-$2$
\end{tabular}
\caption{\label{fig:finegrained-extraction}
Extracting a restless path using the fine-grained decision oracle. On iteration one (left), a \krestlesspath instance with a multiset of colors $M$ and a restless path of length $3$ from vertex $v_1$ to vertex $v_6$ highlighted in bold. On iteration two (right), a \krestlesspath instance with a multiset of colors $M$ and a restless path of length $2$ from vertex $v_1$ to vertex $v_5$ highlighted in bold.}
\end{figure}

A generalization of the approach is described as follows: Let $(G,c,M,k,s,d,\tau)$ be an instance of \krestlessmotif. We build an instance $(G^\ell,c',M^\ell,\ell,s',\tau^\ell)$ of \krestlessmotifreach for each $\ell \in \{k+1,k,\dots,2\}$. %
For the first iteration, where $\ell=k+1$, the graph $G^\ell$ is constructed as described in Theorem~\ref{theorem:algo:restlessmotifreach} to obtain a new source vertex $s'$ and a coloring function $c'$,  $M^\ell=M \cup c'(s')$, and $\tau^\ell = \tau$. The graph construction of \krestlessmotifreach for $\ell=k+1$ is illustrated in Figure~\ref{fig:algo:finegrained-extraction:2}.
We apply the algorithm from Theorem~\ref{theorem:algo:restlessmotifreach} to obtain~$R^\ell$ and~$D^\ell$.

Let $(u_\ell,i_\ell) \in R^\ell$ be a (vertex, minimum timestamp) pair such that  $R^\ell_{u_\ell,i_\ell} = 1$. For the first iteration, where $\ell = k+1$, we remove the vertex $u_{k+1}=d$ and remove the incoming and outgoing edges of $u_{k+1}$ in $G^{k+1}$ to obtain $G^{k}$. 
Let $M^{\ell-1}=M^\ell \setminus c(u_\ell)$ and $\tau^{\ell-1}=i_\ell$. We evaluate the instance $(G^{\ell-1},c',M^{\ell-1},\ell-1,s', \tau^{\ell-1})$ to obtain $R^{\ell-1}$. Let $(u_{\ell-1},i_{\ell-1}) \in R^{\ell-1}$ be a (vertex, timestamp) pair such that
$R^{\ell-1}_{u_{\ell-1}, i_{\ell-1}} = 1$. As $R^\ell_{u_\ell,i_\ell} = 1$ there exists an edge $(u_{\ell-1}, u_{\ell}, i_\ell) \in E(G^{\ell})$ in $G^\ell$. In each iteration, we add the edge $(u_{\ell-1}, u_\ell, i_\ell)$ to the solution and continue the process for each $\ell \in \{k+1,k,\dots,2\}$.
In total we make $k$ queries to the fine-grained oracle. Thus, the runtime of extracting an optimal solution is  $\sum^{k+1}_{\ell=2} 2^\ell \ell m \Delta = \bigO(2^k k m \Delta)$.
Similarly, we have a $\bigO(2^k k m \Delta)$-time and \change{$\bigO(n\tau)$-space} algorithm for extracting an optimal solution for \krestlesspath.

\begin{figure}
\centering
\tikzset{every picture/.style={scale=0.9}}
\setlength{\tabcolsep}{30pt}
\begin{tabular}{c c}
\begin{tikzpicture}[scale=\tikzscale,every node/.style={scale=\tikzscale}]]

\input{tikz/tikz-defs}

\node[exsquare, draw=green] (v1) at (  0,  0) {$v_1$};
\node[exnode, draw=black] (v2) at ( -2, -2) {$v_2$};
\node[exnode, draw=black] (v3) at (  2, -2) {$v_3$};
\node[exsquare, draw=green] (v4) at (  0, -4) {$v_4$};

\node[fill=white] at (  -1.4, -0.8) {\large $1$};
\node[fill=white] at ( 0.3, -1) {\large $3$};
\node[fill=white] at ( 1.7, -0.7) {\large $1,2$};
\node[fill=white] at (  -0.5, -1.6) {\large $2$};
\node[fill=white] at ( -1.7, -3) {\large $1,2$};
\node[fill=white] at ( 0.3, -3) {\large $2$};
\node[fill=white] at ( 1.6, -3.5) {\large $1$};

\node[fill=white] at ( 0, 0.6) {\large $s = v_1$};
\node[fill=white] at ( 0, -4.7) {\large $d = v_4$};

\node[fill=white] at ( -2, -5.5) {\large $M\,:$};
\node[exsquare, draw=green] at (-1, -5.5) {$g$};
\node[exsquare, draw=green] at (0, -5.5) {$g$};
\node[exnode, draw=black] at (1, -5.5) {$b$};
\node[exnode, draw=black] at (2, -5.5) {$b$};

\path [draw=black, postaction={on each segment={mid arrow=black, ultra thick}}]
(v2) -- (v1)
(v1) to [bend left=16] (v3)
(v3) to [bend left=16] (v1)
(v2) -- (v4)
(v3) -- (v2)
(v4) to [bend right=16] (v3)
(v3) to [bend right=16] (v4)
;

\end{tikzpicture}%
 &
\begin{tikzpicture}[scale=\tikzscale,every node/.style={scale=\tikzscale}]]

\input{tikz/tikz-defs}

\node[exdiamond, draw=red] (v0) at (  -2,  0) {$s'$};
\node[exsquare, draw=green] (v1) at (  0,  0) {$v_1$};
\node[exnode, draw=black] (v2) at ( -2, -2) {$v_2$};
\node[exnode, draw=black] (v3) at (  2, -2) {$v_3$};
\node[exsquare, draw=green] (v4) at (  0, -4) {$v_4$};

\node[fill=white] at ( -1, 0.4) {\large $1,2$};
\node[fill=white] at (  -1.4, -0.8) {\large $1$};
\node[fill=white] at ( 0.3, -1) {\large $3$};
\node[fill=white] at ( 1.7, -0.7) {\large $1,2$};
\node[fill=white] at (  -0.5, -1.6) {\large $2$};
\node[fill=white] at ( -1.7, -3) {\large $1,2$};
\node[fill=white] at ( 0.3, -3) {\large $2$};
\node[fill=white] at ( 1.6, -3.5) {\large $1$};


\node[fill=white] at ( -2.5, -5.5) {\large $M'\,:$};
\node[exsquare, draw=green] at (-1.5, -5.5) {$g$};
\node[exsquare, draw=green] at (-0.5, -5.5) {$g$};
\node[exnode, draw=black] at (0.5, -5.5) {$b$};
\node[exnode, draw=black] at (1.5, -5.5) {$b$};
\node[exdiamond, draw=red] at ( 2.5, -5.5) {$r$};

\path [draw=black, postaction={on each segment={mid arrow=black, ultra thick}}]
(v0) -- (v1)
(v2) -- (v1)
(v1) to [bend left=16] (v3)
(v3) to [bend left=16] (v1)
(v2) -- (v4)
(v3) -- (v2)
(v4) to [bend right=16] (v3)
(v3) to [bend right=16] (v4)
;

\end{tikzpicture}%
 \\
\krestlessmotif & \krestlessmotifreach 
\end{tabular}
\caption{\label{fig:algo:finegrained-extraction:2}
The transformation of \krestlessmotif (left) to an instance of \krestlessmotifreach (right) for extracting a solution using the fine-grained oracle. Vertices with color \underline{r}ed, \underline{b}lack and \underline{g}reen are drawn using diamond, circular and square shapes, respectively.
}
\end{figure}

Using a similar construction, we can improve the runtime of extracting an
optimal solution for \pathmotif and \ktemppath introduced
in~\cite{thejaswi2020pattern,thejaswi2020finding} from $\bigO((2^k k (n\tau +
m))(k \log n + \log \tau)) + k!)$ to $\bigO(2^k k (n \tau + m))$. Additionally,
our fine-grained construction can be employed to extract a solution for the
$k$-path problem and the graph motif problem in static graphs by reducing the
number of queries from $\bigO(k \log n)$ to $\bigO(k)$, thus improving the work
of Bj{\"o}rklund et al.~\cite{bjorklund2014fast}.


\subsection{A deterministic $\bigO^*(2^n)$ algorithm for solving \restlesspath}
\label{sec:algo:deterministic}
To obtain a deterministic $\bigO^*(2^n)$-time algorithm for \restlesspath, we proceed as follows. First, we construct a static expansion of the given temporal graph, and then execute a deterministic algorithm for \rainbowpath on the resulting graph. Before proceeding, we introduce this problem formally.

\para{Rainbow path problem in static graphs (\rainbowpath).}
Given a static graph $G'=(V',E')$, a vertex-coloring function $c:V' \rightarrow C'$
and two distinct vertices $s, d \in V'$, the problem asks us to decide if there exists a
rainbow path from $s$ to $d$, that is, a path on which no color repeats. A
deterministic polynomial-space algorithm for \rainbowpath is due to
Kowalik and Lauri~\cite{kowalik2016onfinding}.

\begin{lemma}[\cite{kowalik2016onfinding}, Corollary~5]
\label{lemma:rainbowpath}
There exists a deterministic algorithm for solving \rainbowpath in static graphs
with time $\bigO(2^{|C'|} |C'| (|E'| + |C'|^2)$ and space $\bigO(|V'| + |C'|)$, where
$V'$ is the set of vertices, $E'$ is the set of edges and $C'$ is the set of vertex
colors.
\end{lemma}

\para{$\delta$-expansion.} Let $G=(V,E)$ be a temporal graph with maximum
timestamp $\tau$ with number of vertices $|V|=n$ and number of edges $|E|=m$.
For simplicity we assume that the vertex set is $V=\{v_1,\dots,v_n\}$.
Let $\delta:V \rightarrow \nnegintegers$ be a vertex-dependent waiting time. The
\emph{$\delta$-expansion} of $G$ is a static directed graph $\deltaexp{G}=(V',E')$
where 
%
$V' = \{ v^t_i : v_i \in V, t \in [\tau]~|~\exists e \in E: v_i \in e\}$
and $E' = \{(v^t_i,v^{t+\ell}_j):(v_i,v_j,t) \in E, \ell \in \{0, \dots,
\delta(v_j)\}, t+\ell \leq \tau\}$.\footnote{Note that $V'$ is not a multiset,
that is, $V'$ does not contain duplicate elements.} 
For each vertex $v_i^t \in V'$, we assign the color $c(v_i^t) = i$ and the set
of colors $C=[n]$.
Observe that $|V'| = |E'|=(\Delta + 1)\,|E|$, where
$\Delta = \max_{v_i \in V} \delta(v_i)$. Note that $\deltaexp{G}$ can
be computed in time and space $\bigO(m \Delta)$. 
For an illustration of $\delta$-expansion, see Figure~\ref{fig:delta-expansion}.
Finally, we claim that there exists a \restlesspath in $G$ if and only if there
exists a \rainbowpath in $\deltaexp{G}$.

\begin{figure}
\centering
\tikzset{every picture/.style={scale=0.8}}
\setlength{\tabcolsep}{25pt}
\begin{tabular}{c c}
\begin{tikzpicture}[scale=\tikzscale,every node/.style={scale=\tikzscale}]

\input{tikz/tikz-defs}

\node[exnode, draw=black] (v1) at (   0, 0) {$v_1$};
\node[exnode, draw=black] (v2) at (   3, 0) {$v_2$};
\node[exnode, draw=black] (v3) at ( 4.5, 2) {$v_3$};
\node[exnode, draw=black] (v4) at ( 1.5, 2) {$v_4$};
\node[exnode, draw=black] (v5) at (   6, 0) {$v_5$};

\node[fill=white] at ( 1.5, -0.5) {\large $1$};
\node[fill=white] at ( 4.2,    1) {\large $2$};
\node[fill=white] at ( 3, 2.5) {\large $3$};
\node[fill=white] at ( 1.8, 1) {\large $4$};
\node[fill=white] at ( 4.5, -0.5) {\large $5$};

\path [draw=black, postaction={on each segment={mid arrow=black, ultra thick}}]
(v1) -- (v2)
(v2) -- (v3)
(v3) -- (v4)
(v4) -- (v2)
(v2) -- (v5)
;

\end{tikzpicture}%
 &
\begin{tikzpicture}[scale=\tikzscale,every node/.style={scale=\tikzscale}]]

\input{tikz/tikz-defs}


\node[exnode, draw=yafcolor1!100] (v11) at ( 0,    6) {$v^1_1$};
\node[exnode, draw=yafcolor2!100] (v12) at ( 0,  4.5) {$v^1_2$};

\node[exnode, draw=yafcolor2!100] (v22) at ( 2,  4.5) {$v^2_2$};
\node[exnode, draw=yafcolor3!100] (v23) at ( 2,    3) {$v^2_3$};

\node[exnode, draw=yafcolor2!100] (v32) at ( 4,  4.5) {$v^3_2$};
\node[exnode, draw=yafcolor3!100] (v33) at ( 4,    3) {$v^3_3$};
\node[exnode, draw=yafcolor4!100] (v34) at ( 4,  1.5) {$v^3_4$};

\node[exnode, draw=yafcolor2!100] (v42) at ( 6,  4.5) {$v^4_2$};
\node[exnode, draw=yafcolor3!100] (v43) at ( 6,    3) {$v^4_3$};
\node[exnode, draw=yafcolor4!100] (v44) at ( 6,  1.5) {$v^4_4$};

\node[exnode, draw=yafcolor2!100] (v52) at ( 8,  4.5) {$v^5_2$};
\node[exnode, draw=yafcolor4!100] (v54) at ( 8,  1.5) {$v^5_4$};
\node[exnode, draw=yafcolor5!100] (v55) at ( 8,    0) {$v_5^5$};

\path [draw=black, postaction={on each segment={mid arrow=black, ultra thick}}]
(v11) -- (v12)
(v11) -- (v22)
(v11) -- (v32)
;

\path [draw=black, postaction={on each segment={mid arrow=black, ultra thick}}]
(v22) -- (v23)
(v22) -- (v33)
(v22) -- (v43)
;

\path [draw=black, postaction={on each segment={mid arrow=black, ultra thick}}]
(v33) -- (v34)
(v33) -- (v44)
(v33) -- (v54)
;

\draw (v44) edge[-, draw=black, bend right = 30, postaction={mid arrow=black, ultra thick}]  (v42);
\draw (v44) edge[-, draw=black, bend right = 10, postaction={mid arrow=black, ultra thick}]  (v52);

\draw (v42) edge[-, draw=black, bend left = 8, postaction={mid arrow=black, ultra thick}]  (v55);

\end{tikzpicture}%
 \\
A temporal graph $G$ & $\delta$-expansion of $G$
\end{tabular}
\caption{\label{fig:delta-expansion}
On the left, a temporal graph $G=(V,E,\tau)$ with vertex set $V=\{v_1,\dots,v_5\}$, edge set $E=\{(v_1,v_2,1), (v_2,v_3,2), (v_3,v_4,3), (v_4,v_2,4), (v_2,v_5,5)\}$, maximum timestamp $\tau=5$ and resting time $\delta(v_1)=\dots=\delta(v_5)=2$.
On the right, $\delta$-expansion $\deltaexp{G}$ of $G$. 
For each temporal edge $(v_i,v_j,t) \in E$, we introduce at most $\delta(v_j)+1$ static edges
$\{(v^t_i,v^{t+\ell}_j): \ell \in \{0, \dots, \delta(v_j)\}, t+\ell \leq \tau\}$.
If $(v^{t}_{i},v^{t'}_{j}) \in E'$, then $v^{t}_{i}, v^{t'}_j \in V'$. Finally, all vertices with same subscript i.e., $\{v_i^1,v_i^2,\dots\}$ for each $i \in [n]$ are assigned the same color.
}
\end{figure}

\begin{lemma}
\label{lemma:delta-expansion}
There exists a \restlesspath in the temporal graph $G$ if and only if there
exists a \rainbowpath in the corresponding $\delta$-expansion $\deltaexp{G}$.
\end{lemma}
\begin{proof}
Let $P=v_1e_1v_2,\dots,e_{k-1}v_k$ be a restless path in $G$. We construct a rainbow path $P'=v'_1 e'_1 v'_2 \dots e'_{k-1}v'_k$ as follows. For edges $e_i = (v_i,v_{i+1},t_i), e_{i+1}= (v_{i+1},v_{i+2},t_{i+1}) \in P, i=[k-1]$, we pick a static edge $e'_i = (v_i^{t_i}, v_{i+1}^{t_{i+1}})$ to path $P'$ and vertices $v'_i = v_i^{t_i}, v'_{i+1}= v_{i+1}^{t_{i+1}}$. From construction, we know that $v'_i = v_i{t_i}, i \in [k]$ and $e'_i=(v_i^{t_i}, v_{i+1}^{t_{i+1}}), i \in [k-1]$ exists in $\deltaexp{G}$.  Since $P$ is a path, no vertex repeats in $P$, which implies that no two vertices in $P'$ have the same color. We conclude that $P'$ is a rainbow path.

Conversely, let $P'=v'_1 e'_1 v'_2 \dots e'_{k-1} v'_{k}$ be a rainbow path in $\deltaexp{G}$, we construct a restless path $P=v_1 e_1 v_2 \dots e_{k-1} v_k$ by replacing each edge $e'_i = (v_{i}^{t_i},v_{i+1}^{t_{i+1}})\in P', i=[k-1]$ by $e_i = (v_i,v_{i+1},t_i)$. From construction the waiting time at vertex $v_i$ is at most $t_{i+1} - t_i \leq \delta(v_i)$. Since $P'$ is a rainbow path, it implies that $v_i \neq v_j$ for all $i \neq j$. So $P$ is a restless path, which concludes our proof.
\end{proof}

Combining Lemma~\ref{lemma:rainbowpath} with Lemma~\ref{lemma:delta-expansion}, we obtain the following result.

\begin{theorem}
\label{thm:det-alg}
There exists a deterministic algorithm for solving the \restlesspath problem in
time $\bigO(2^n n(m \Delta + n^2))$ and space $\bigO(m\,\Delta + n)$.
\end{theorem}

Theorem~\ref{thm:det-alg} also implies a deterministic 
$\bigO(2^n n^2 (m \Delta + n^2))$-time
algorithm for solving \restlessreach. However, the deterministic
algorithm presented in this section is of theoretical interest only and is unlikely to scale in practice even for graphs of modest size.
However, note that Lemma~\ref{lemma:rainbowpath} can be replaced with any algorithm for \rainbowpath. Indeed, any improvement in the running time of the algorithm for \rainbowpath will improve the running time of algorithms for \restlesspath and \restlessreach.

\begin{table*}[t]
\caption{A summary of time and space complexity. Here, $n$ is the number of vertices, $m$ is the number of edges, $\tau$ is the maximum timestamp, $k-1$ is the length of path
and $\Delta$ is the maximum resting time. For extraction we use randomized
algorithm as a subroutine.}
\label{table:summary:1}
\centering
\begin{tabular}{l r r}
\toprule
Problem & Time complexity & Space complexity\\
\midrule
{\em {Fine-grained oracle (randomized)}} \\
\cmidrule{1-1}
\restlesspath         & $\bigO(2^n n m \Delta)$ & $\bigO(n\Delta)$ \\
\krestlesspath        & $\bigO(2^k k m \Delta)$ & $\bigO(n\Delta)$ \\
\krestlessmotif       & $\bigO(2^k k m \Delta)$ & $\bigO(n\Delta)$ \\
\restlessreach        & $\bigO(2^n n m \Delta)$ & $\bigO(n\Delta)$ \\
\krestlessreach       & $\bigO(2^k k m \Delta)$ & $\bigO(n\Delta)$ \\
\krestlessmotifreach  & $\bigO(2^k k m \Delta)$ & $\bigO(n\Delta)$ \\
\atmkrestlessreach    & $\bigO(2^k k m \Delta)$ & $\bigO(n\Delta)$ \\
\midrule
{\em Inclusion-exclusion (deterministic)}\\
\cmidrule{1-1}
\restlesspath        & $\bigO(2^n n (m\Delta + n^2))$ & $\bigO(m\Delta + n)$ \\
\restlessreach       & $\bigO(2^n n^2 (m\Delta + n^2))$ & $\bigO(m\Delta + n)$ \\
\midrule
{\em {Extraction} (\krestlesspath)}\\
\cmidrule{1-1}
Self-reducibility + temporal {\small DFS} & $\bigO((2^k k^2 m \Delta \log n \log \tau) + k!)$ & $\bigO(n\Delta)$ \\
Fine-grained extraction          &                           $\bigO(2^k k m \Delta)$ & $\bigO(n\tau)$ \\
\bottomrule
\end{tabular}
\end{table*}

\subsection{Infeasibility of a $\bigO^*((2-\epsilon)^k)$-time algorithm for \krestlessmotif}
\label{sec:algo:infeasibility}

In this section, we prove that under plausible complexity-theoretic assumptions, the algorithms presented for
\krestlessmotif and \krestlessmotifreach are optimal. 

In particular, the assumption that we rely on is the Set Cover
Conjecture~\cite{cygan2016problems} (SCC) which is formulated as follows.  In
the \setcover problem we are given an integer $f$ and a family of sets
$\pazocal{S}$ over the universe $U = \bigcup \pazocal{S}$ with $n = |U|$ and 
$m = |\pazocal{S}|$.  The goal is to decide whether there is a subfamily of at most
$f$ sets $S_1, S_2, \ldots, S_f \in \pazocal{S}$ such that $U =
\bigcup_{i=1}^{f} S_i$, i.e., that the selected sets cover the universe $U$.
The SCC of {Cygan et al.}~\cite{cygan2016problems} states that there is no
algorithm for the \setcover problem that runs in time 
$(2-\epsilon)^n (nm)^{\bigO(1)}$ for any $\epsilon > 0$.  In fact, the fastest known algorithm for
solving \setcover runs in time $\bigO^*(2^n)$, and an algorithm running in time
$\bigO^*((2-\epsilon)^n)$ for any $\epsilon > 0$ has been deemed a major
breakthrough after decades of research on the problem.  Under SCC, exponential
lower bounds for several fundamental problems are known (see
e.g.,~\cite{cygan2016problems,bjorklund2013probably,krauthgamer2019set}).
However, even if SCC turned out to be false, these results (and
Theorem~\ref{theorem:motifalgo:infeasibility}) are still meaningful: instead of
trying to find a faster algorithm for e.g., the arguably richer problem of
\krestlessmotif, one can focus on \setcover which is simpler.

To obtain our result, it is convenient to recall the
\colorfulpath problem in static graphs, and to perform a reduction from that
problem to \krestlessmotif problem.

\mpara{Colorful path problem (\colorfulpath).} Given a static graph $G=(V,E)$
and a coloring function $c:V\rightarrow [k]$, the problem asks if there exists a
path of length $k-1$ in $G$ such that the vertex colors of the path are
different (i.e., such that each color occurs exactly once). The problem is known
to be \np-hard, and known not to admit a $\bigO^*((2 - \epsilon)^k)$ algorithm
for any $\epsilon > 0$ assuming SCC.

\begin{theorem}[Kowalik and Lauri~\cite{kowalik2016onfinding}]
\label{theorem:algo:infeasibility}
Assuming the Set Cover Conjecture, there exists no $\bigO^*((2-\epsilon)^k)$ time algorithm for solving the
\colorfulpath problem for any $\epsilon > 0$.
\end{theorem}

\begin{theorem}
\label{theorem:motifalgo:infeasibility}
If the \krestlessmotif problem has a $\bigO^*((2-\epsilon)^k)$ time algorithm
for any $\epsilon > 0$ then \colorfulpath problem has a
$\bigO^*((2-\epsilon)^k)$ time algorithm.
\end{theorem}
\begin{proof}
Given an instance $I = (G, c, k)$ of \colorfulpath in static graphs, 
we construct an instance $I' = (G', M', c', k')$ of \krestlessmotif in temporal
graph by letting 
$G'=(V',E')$, $V' = V \cup \{s, d\}$, 
$E'= \{(u,v,1): (u,v) \in E\} \cup \{(s,u,1): u \in V\} \cup \{(u,d,1): u \in V\}$,
$c'(u) = c(u)~\text{for all}~u \in V$, $c(s)=k+1$, $c(d) = k+2$, 
$M=\{1,\dots,k+2\}$, and
$\delta: V \rightarrow 1$, $\Delta=1$, $k'=k+2$.
Informally, $G'$ is constructed from $G$ by replacing each edge with a temporal edge with timestamp one, and by making $s$ and $d$ adjacent to a new vertex both receiving a new unique color.
We claim that the instance $I$ of \colorfulpath has a solution if and only if the instance $I'$ of \krestlessmotif has a solution.

Let $P=v_1e_1v_2 \dots e_{k-1}v_k$ be a colorful path
in $G$. 
By construction, we know that for each edge $(u,v) \in E$ we have $(u,v,1) \in E'$, so the path $P'=s e'_0 v_1 e'_1 v_2 \dots e'_{k-1} v_k e'_k d$
exists in $G'$, where $e'_0=(s,v_1,1)$, $e'_k=(v_k,d,1)$, and 
$e'_i=(v_i,v_{i+1}, 1)$ for all $i \in [k-1]$. 
Also, the vertex colors of $\{v_1,\dots,v_k\}$ agree with $\{1,\dots,k\}$ since the $P$ is colorful, so the vertex colors of $\{s,d\} \cup \{v_1,\dots,v_k\}$ agree with $M'$. 
We conclude that $P'$ is a \krestlessmotif in $G'$.
Conversely, let $P'=s e'_0 v_1 e'_1 \dots e'_{k-1} v_k e'_{k+1} d$ be a solution for $I'$ in $G'$. 
We construct a static path $P=v_1 e_1 v_2 \dots e_{k-1}v_k$ by replacing the edges $e'_i=(v_i,v_{i+1},1)$ by
$e_i=(v_i,v_{i+1})$ for all $i \in [k+1]$. 
Since the vertices $\{s,d\} \cup \{v_1,\dots,v_k\}$ agree with colors $M = \{1,\dots,k+2\}$, the vertex colors of $\{v_1,\dots,v_k\}$ agree with colors $\{1,\dots,k\}$ as the colors $k+1$ and $k+2$ only appear once each on $G'$.
Evidently, $P$ is a colorful path in $G$.
\end{proof}

It follows that if we have an algorithm for solving \krestlessmotif with
$\bigO^*((2-\epsilon)^k)$ time for some $\epsilon > 0$, we can use it solve \colorfulpath in
static graphs using the construction described in
Theorem~\ref{theorem:motifalgo:infeasibility} within the same time bound. 
However, from Theorem~\ref{theorem:algo:infeasibility} we know that such an algorithm does not exists unless SCC is false. 
Put differently, assuming SCC, \krestlessmotif does not admit an algorithm running in time  $\bigO^*((2-\epsilon)^k)$ for any $\epsilon > 0$.
Finally, since \krestlessmotifreach problem generalizes the \krestlessmotif problem, the
former problem does not admit an algorithm running in time $\bigO^*((2-\epsilon)^k)$
for any $\epsilon > 0$, assuming SCC.


\section{Experiments} \label{sec:experiments}
In this section, we describe our setup and the experimental results to validate our approach and demonstrate its scalability. Our implementation is available as open source~\cite{ourcode}.

\subsection{Implementation}
A high-level intuition of the implementation is as follows: For variables $\vec{x} = \{x_v : v \in V\}$ and $\vec{y} = \{y_{uv,\ell,i}: (u,v,i) \in E, \ell \in [k]\}$ we assign a value from the Galois field $\GF(2^b)$. Multiplication between any two field variables is defined as an {\tt XOR} operation, likewise, multiplying two variables with same value results in a zero-term. We know that the monomials corresponding to walks that are not paths have at least one repeated variable, so the corresponding monomial evaluates to a zero term, while the monomial corresponding to a path evaluates to a non-zero term since there are no repeated variables. It is possible that a monomial corresponding to a path might evaluate to a zero term resulting in a false negative. To reduce the probability of false negatives, we repeat the evaluation with $2^k$ random assignments for variables $\{x_u: u \in V\}$ and $\left\{y_{uv,\ell,i}: (u,v,i) \in E, \ell \in [k]\right\}$. In theory, the false negative probability of our algorithm is $2^{-b}(2k-1)$.  For our experiments, we choose the field size $b=64$, which makes the false negative probability negligible.

Modern CPUs have very high arithmetic and memory bandwidth, however, the bandwidth comes at the cost of latency. Each arithmetic and memory-access operation is associated with a corresponding latency factor, and often the memory-access latency is orders of magnitude greater than the arithmetic latency. As such, the challenge for efficient implementation engineering is to keep the arithmetic pipeline busy while fetching data from memory for the subsequent arithmetic operations. Memory bandwidth can be improved by using coalesced-memory access, that is, by organizing the memory layout such that the data for consecutive computations are available in consecutive memory addresses. In addition, we can use hardware pre\-fetching to fetch the data required in subsequent computation while performing computation on the data, which is currently in the memory.
Arithmetic bandwidth can be improved using vector extensions, that is, by grouping the data on, which the same arithmetic operations are executed. More precisely, if we are executing the same arithmetic instructions on different operands or data, we can group the operands using vector extensions to execute arithmetic operations in parallel.
For more technical details related to implementation engineering we refer the reader to~\cite{bjorklund2015engineering,kaski2018engineering,thejaswi2020finding}.

Our implementation is written in the C programming language with OpenMP
constructs to achieve thread-level parallelism. Vector parallelism is
achieved by enabling parallel executions of the same arithmetic operations,
which make use of advanced vector extensions (AVX2). Additionally, we use
carry-less multiplication of one quadword ({\sc\large pclmulqdq}) instruction set to enable
fast finite-field arithmetic. 
The finite-field arithmetic implementation we use is from~\cite{bjorklund2015engineering}.

Our engineering effort boils down to implementing the recursions in Equations~(\ref{eq:poly-enc:1}) and~(\ref{eq:multilinear:3}) and evaluating the polynomial  using $2^\ell$ random substitutions for the $x$-variables. Recall that from the construction of the generating function the $y$-variables are unique, so we generate the values of $y$-variables using a pseudo\-random number generator. The values of the $x$-variables are computed using Equation~(\ref{eq:multilinear:3}). Our implementation loops over four variables: the outer most loop is over $[\ell]$, the second loop is over $[\tau]$,  the third loop is over $V$, and the final loop is over  $\{0,\dots,\delta(u)\}$ for $u \in V$.
In Equation~(\ref{eq:multilinear:3}), computing the polynomial $\prep{u}{\ell}{i}$ is independent for each $u \in V$ if we fix~$\ell$ and~$i$, so the algorithm can be  thread-parallelized up to $|V| = n$ threads. We make use of the OpenMP API using the {\tt omp parallel for} construct with {\tt default} scheduling over vertices in $V$ to achieve thread parallelism.  Additionally, performing $2^\ell$ random substitutions of $x$-variables is independent of each other, so each of the $2^\ell$ evaluations can be vector-parallelized. We achieve this by grouping the arithmetic operations on $2^\ell$ random substitutions of variables in $x$ and enabling the vector extensions from AVX2. Recall that our inner-most loop is over $\{0,\dots,\delta(u)\}$, so we arrange the memory layout as $n \times \tau$ to saturate the memory bandwidth. We also employ hardware pre\-fetching by forcing the processor to fetch data for subsequent computations while we are still performing the computation on the data, which is already in the memory.

Our implementation uses $\bigO(n \tau + m)$ memory, which is due to the adjacency list representation of the temporal graphs. 

\para{Preprocessing.}
In the restless reachability problems considered in our work, we compute reachability from a given source vertex to all other vertices without an explicit restriction on the time window. As such, we do not see a straightforward approach to preprocessing the temporal graph $G$ to reduce its size using  heuristic preprocessing techniques such as slicing $G$ within a time window, i.e., considering the edges between a minimum and maximum timestamp window.  Alternatively, we can merge $G$ to obtain a static graph, compute reachability on the static graph, and reconstruct a temporal graph by only using the vertices, which are reachable in the static graph. Such a preprocessing technique is correct since there exists a restless path between any two vertices in a temporal graph only if there exists a path in the corresponding static graph, while the other direction is not always true. However, most of the datasets considered in our experiments have a connected static underlying graph, and therefore we do not see a significant reduction in graph size.

For the restless reachability problems with additional color constraints \ie, for the \krestlessmotif and \krestlessmotifreach problems, we can take advantage of two preprocessing techniques to reduce the graph size: ($i$) by removing all the vertices whose vertex colors do not match with the multiset colors; ($ii$) by merging the temporal graph to a static graph instance, build a vertex-localized sieve on the static graph, and reconstruct the graph using the set of vertices reachable in the static graph instance. Note that these are heuristic approaches to reduce the graph size and we do not claim any theoretical bounds for the reduction in the graph size. For a detailed discussion of preprocessing using vertex-localized sieving we refer the reader to an earlier work~\cite[\S~Preprocessing]{thejaswi2020finding}.

\subsection{Experimental setup} \label{sec:exp:setup}
Here we describe the hardware details and the input graphs used for our experiments.

\para{Hardware.} We make use of two hardware configurations for our experiments. 
\squishlisttight
\item A {\em workstation} with $1 \times 3.2$~GHz Intel Core\,i5-4570\,CPU, Haswell
microarchitecture, $4$ cores, $32$~Gb memory, Ubuntu, and \texttt{gcc}\,v9.1.0.
\item A {\em compute\-node} with $2\times${2.5}~GHz Intel Xeon\,2680\,V3 CPU, 24 cores, 12 cores/CPU, 256~Gb memory, Red Hat, and $\texttt{gcc}$\,v9.2.0.
\squishend

The experiments make use of all the cores. Additionally, we make use of AVX2 and hardware pre\-fetching to saturate the arithmetic and memory bandwidth, respectively.

\para{Input graphs.}
We use both synthetic and real-world graphs in our experiments.  For \emph{synthetic} graphs, we use the temporal-graph generator from Thejaswi et al.~\cite[\S\,9.3]{thejaswi2020finding}, in particular we make use of $d$-regular and power-law graphs.  The regular graphs are generated using the \emph{configuration model}~\cite[\S\,2.4]{bollobas2001random}. The configuration model for power-law graphs is as follows: given non-negative integers $D$, $n$, $w$, and $\alpha < 0$, we generate an $n$-vertex graph such the following properties roughly hold: ($i$) the sum of vertex degrees is~$Dn$; ($ii$) the distribution of degrees is supported at $w$ distinct values with geometric spacing; and ($iii$) the frequency of vertices with degree $d$ is proportional to~$d^\alpha$. The edge timestamps are assigned uniformly at random in the range~$[\tau]$.
Both directed and undirected graphs are generated using the same configuration model, however, for directed graphs the orientation is preserved.
We ensure that the graph generator produces identical graph instances in all the hardware configurations.

For \emph{real-world} graphs, we use the co-presence dataset from socio-patterns \cite{genois2018can}, Koblenz network collections~\cite{koblenz2013konect}, Copenhagen study network~\cite{sapiezynski2019interaction}, and public-transport networks~\cite{kujala2018collection}. For a description of the datasets, see the respective references. The preprocessing details of each dataset is described below.
We pre\-process the datasets to generate a graph by renaming the location identifiers (or vertices) in the range from $1$ to the maximum number of locations (or vertices) available in the dataset. 
If the time values in dataset are Unix timestamps, we approximate the value to the closest second rounded down before assigning an unique timestamp identifier.
For datasets from socio-patterns, we reduce the maximum timestamp $\tau$ by dividing each timestamp by 20 and rename the timestamps in the range from $1$ to the difference between the maximum and minimum timestamps. Since socio-patterns datasets are undirected contact-networks we replace each undirected edge with two directed edges in both directions. 
For the Koblenz datasets we round the timestamp to the closest day. Finally, we present statistics of all datasets used for experiments in Appendix~\ref{appendix:dataset-statistics}.

\subsection{Baselines}
In this section we discuss the baseline approach considered for comparison. 
Additionally, we present careful justification on why certain approaches used to solve temporal reachability problems fail to solve restless reachability problems in Appendix~\ref{appendix:baselines}.

\para{Exhaustive search (baseline).}
We consider an exhaustive-search algorithm based on temporal depth-first-search ({\dfs}), which is a parameterized algorithm with respect to the maximum degree of the graph $d_{\max}$. We perform temporal {\dfs} starting from a source~$s$ by respecting waiting-time constraints and restrict the depth of the search to $k$. We report the minimum reachability time for the vertices that are reachable from~$s$ by at most $k-1$ hops. The time complexity of the exhaustive search algorithm is $\bigO(d_{{\max}}^k)$.
As demonstrated in previous work~\cite{thejaswi2020pattern, thejaswi2020finding}, and also shown in our experiments (see Section~\ref{sec:experiments:results}), the exhaustive search does not scale for large scale-free graphs.  In particular, many real-world graphs exhibit a power-law degree distribution, or more generally, scale-free structure. However, the exhaustive search algorithm is highly practical for some structured graphs such as graphs that are close to being $d$-regular with small maximum degree. In our experiments, we refer to the exhaustive search algorithm as the {\em baseline}.
Furthermore, we implement the $\textsc{FPT}(k)$-algorithm by Casteigts et al.~\cite{casteigts2021finding}, which uses a different polynomial construction, with running time $\bigO(2^k (kn + km \Delta))$ and space complexity $\bigO(kn\tau)$.\footnote{The exact polynomial factors and space complexity are not detailed in their paper; the reported complexities are based on our implementation.}

\para{Discussion.}
We stress that our focus is on exact computation of restless reachability.  As such, we do not compare our proposed algorithms to probabilistic methods, heuristics or approximation algorithms. Further, to the best of our knowledge, there is no publicly available implementation for solving the restless reachability problem that scales to large graphs such as those considered in our experiments. For instance, we argue that the algorithms presented by~\cite{casteigts2021finding} are unlikely to perform adequately in practice given their exponential dependency on a structural parameter that, in many real-world networks, has a high value (see Appendix~\ref{appendix:dataset-statistics} Table~\ref{table:exp:datasets}). 
In addition, it is not immediate how index computation techniques such as 2-hop or 3-hop covers or location based indexing would extend to solve restless reachability. 

Finally, a careful reader might question our focus on \emph{exact computation} by recalling that our algorithm has a false negative probability of $(2k-1)/2^b$. However, by fixing a suitable value for $b$ and potentially running our method multiple times, we can make this probability arbitrarily close to zero. For concreteness, we choose $b=64$ for our experiments, which means that when say $k=10$, the per-vertex false negative probability is less than $2^{59} \approx 5.76 \cdot 10^{-17}$. In comparison, a modern consumer CPU running for at least five days has a 1 in 330 chance of a hardware failure due to a machine-check exception, a 1 in 470 chance of a disk subsystem failure, and 1 in 2700 ($\approx 3.7 \cdot 10^{-4}$) chance of a DRAM memory failure~\cite[Figure~2]{nightingale2011cycles}, all significantly more likely than our algorithm making an error.

\subsection{Experimental results}
\label{sec:experiments:results}

In this section we report our experimental results. The experiments are designed to study the following aspects:
($i$)~scalability of the algorithm to graphs with up to 10 million edges on the workstation configuration;
($ii$)~scalability of the algorithm to large graphs with up to one billion edges on the compute\-node configuration;
($iii$)~computing restless reachability in real-world datasets; and
($iv$)~a case study investigating the effectiveness of  different vertex-selection strategies to act as barriers  and minimize the spread of diffusion processes (\eg, infectious diseases)  through the temporal network. An overview of our experiments is available in Table~\ref{table:exp:results:0}.

In our experimental results, run\-time refers to the empirical running time and memory refers to the peak-memory usage of our implementation.

\begin{table}[t]
\caption{Overview of the experiments. Here $n$ is the number of vertices, $m$ is the number of edges, $d=\frac{m}{n}$ is the average degree, $k$ is the length of path, $\Delta$ is maximum resting time, $\delta$ is the function mapping vertices to resting time, maximum timestamp $\tau = 100$ (fixed) and percentage of vertices considered as separators `Sep (\%)'. For power-law graphs we use $\alpha=-1.0$ (fixed)  and $w = 100$ (fixed).
Recall that for a positive integer $k$ we write $[k]=\{1,\dots,k\}$.
}
\label{table:exp:results:0}
\centering
\footnotesize
\begin{tabular}{l r r r r r c r}
\toprule
Experiment & \multicolumn{1}{c}{$n$} 
& \multicolumn{1}{c}{$m$} 
& \multicolumn{1}{c}{$d$} 
& \multicolumn{1}{c}{$k$} 
& \multicolumn{1}{c}{$\Delta$} 
& \multicolumn{1}{c}{$\delta$} 
& \multicolumn{1}{c}{Sep (\%)}
\\
\midrule
\multicolumn{2}{l}{\em {Figure~\ref{fig:exp:scalability}}}\\
\cmidrule(r{1em}){1-1}
{\sf left} &  $10^2,\dots,10^5$ 
&  $10^4,\dots,10^7$ 
& $10^2$ 
& $10$ 
& $10$ 
& $V \rightarrow \{10\}$ 
& -
\\

{\sf center-left} & $10^3$ 
& $10^5$ 
& $10^2$ 
& $5,\dots,15$ 
& $10$ 
& $V \rightarrow \{10\}$ 
& -
\\

{\sf center-right} & $10^5$ 
& $10^7$ 
& $10^2$ 
& $10$ 
& $10,20,\dots,100$ 
& $V \rightarrow \{\Delta\}$ 
& -
\\

{\sf right} & $10^5,\dots,10^2$ 
& $10^6$ 
& $10,\dots,10^4$ 
& $10$
& $10$ 
& $V \rightarrow \{10\}$ 
& -
\\

\midrule
\multicolumn{2}{l}{\em {Figure~\ref{fig:exp:extraction}}}\\
\cmidrule(r{1em}){1-1}
{\sf left, center} & $10^3,\dots,10^7$ 
& $10^5,\dots,10^9$ 
& $10^2$ 
& $10$
& $10$ 
& $V \rightarrow \{10\}$ 
& -
\\
{\sf right} & $10^2,\dots,10^5$ 
& $10^4,\dots,10^7$ 
& $10^2$ 
& $10$
& $10$ 
& $V \rightarrow \{10\}$ 
& -
\\

\midrule
\multicolumn{2}{l}{\em {Figure~\ref{fig:exp:casteigts-runtime}}}\\
\cmidrule(r{1em}){1-1}
{\sf left, right} & $10^2,\dots,10^5$ 
& $10^4,\dots,10^7$ 
& $10^2$ 
& $10$
& $10$ 
& $V \rightarrow \{10\}$ 
& -
\\

\midrule
\multicolumn{2}{l}{\em {Real-world graphs}}\\
\cmidrule(r{1em}){1-2}
{\sf {Table~\ref{table:exp:realworld:1}}} & -
& -
& -
& $5$
& $10$ 
& $V \rightarrow \{10\}$
& -
\\

{\sf {Table~\ref{table:exp:realworld:2}}} & -
& -
& -
& $10$
& $10$ 
& $V \rightarrow \{10\}$
& -
\\

{\sf {Table~\ref{table:exp:realworld:3}}} & -
& -
& -
& $10$
& $5,10$ 
& $V \rightarrow \{\Delta\}$
& -
\\

{\sf {Figure~\ref{fig:exp:reachability:1}}} & -
& -
& -
& $10$
& $5,20$ 
& $V \rightarrow [\Delta]$
& -
\\

\midrule
\multicolumn{2}{l}{\em {Case study} (Socio-patterns)}\\
\cmidrule(r{1em}){1-2}
{\sf {Figure~\ref{fig:exp:immunization:1}}} & -
& -
& -
& $10$
& $10$ 
& $V \rightarrow [10]$
& $5$
\\

{\sf {Figure~\ref{fig:exp:immunization:2}}} & -
& -
& -
& $10$
& $10$ 
& $V \rightarrow [10]$
& $25$
\\

{\sf {Figure~\ref{fig:exp:immunization:3}}} & -
& -
& -
& $10$
& $10$ 
& $V \rightarrow [10]$
& $5,10,25$
\\

\bottomrule
\end{tabular}
\end{table}

\para{Scalability.}
To demonstrate the scalability of the algorithm we experiment with synthetic graphs. The experiments are performed on five independent random power-law graphs for each configuration specified in Table~\ref{table:exp:results:0}. The source vertex is chosen uniformly at random. All experiments are executed on the workstation configuration using all cores with directed graphs. Note that we demonstrated the scalability of the algorithm for \krestlessreach instances, however for other problem instances the run\-times are similar.

\begin{figure}[t]
\centering
\setlength{\tabcolsep}{0.01cm}
\renewcommand{\arraystretch}{0.01}
\begin{tabular}{c c c c}
  \includegraphics[width=0.25\linewidth]{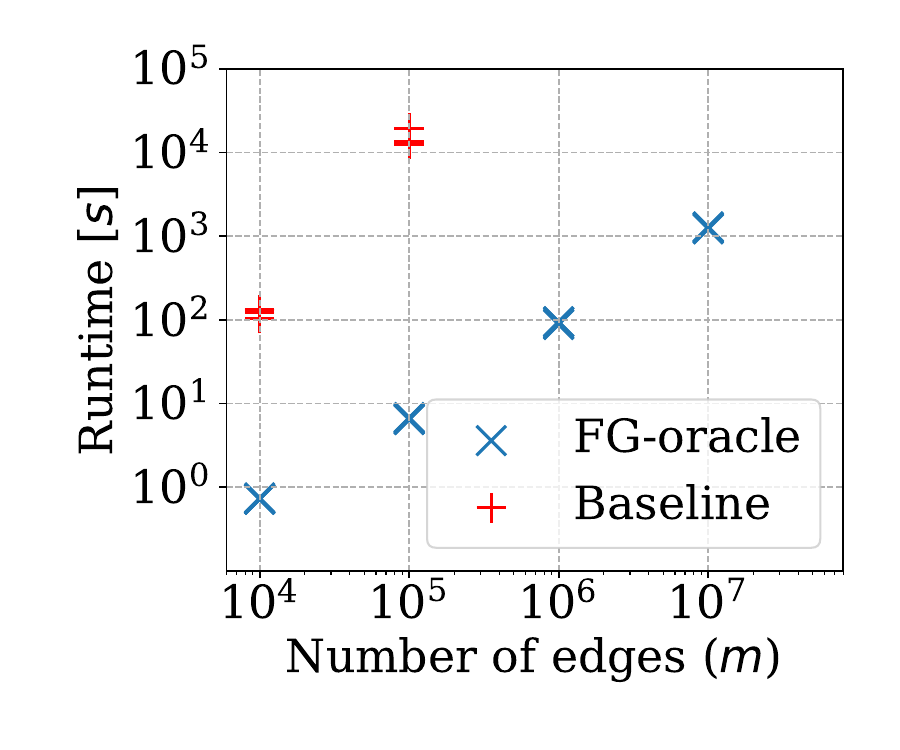} &
  \includegraphics[width=0.25\linewidth]{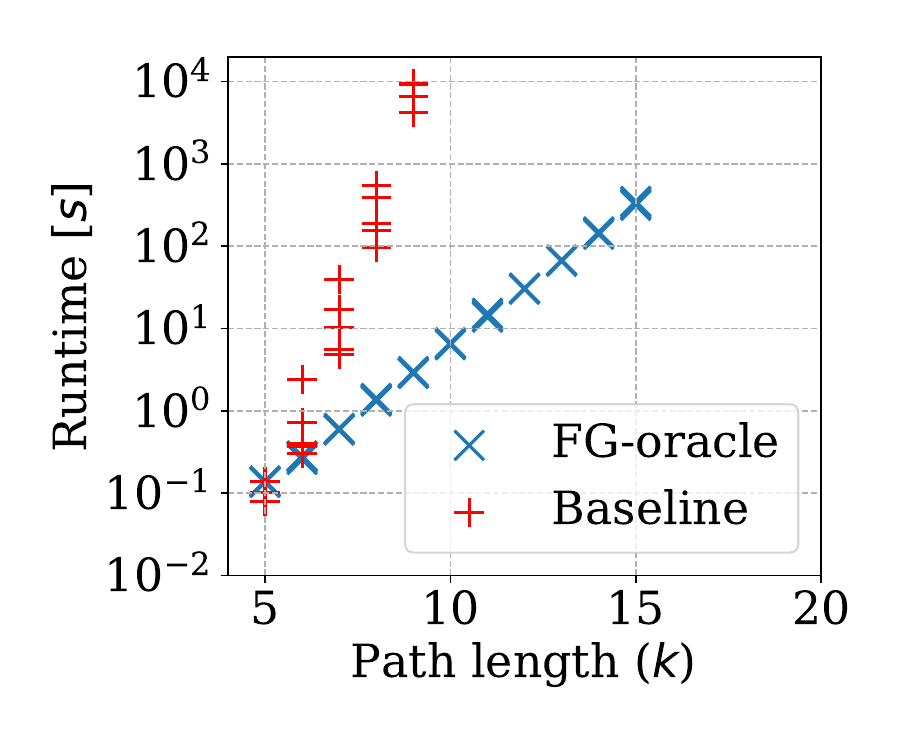}&
  \includegraphics[width=0.25\linewidth]{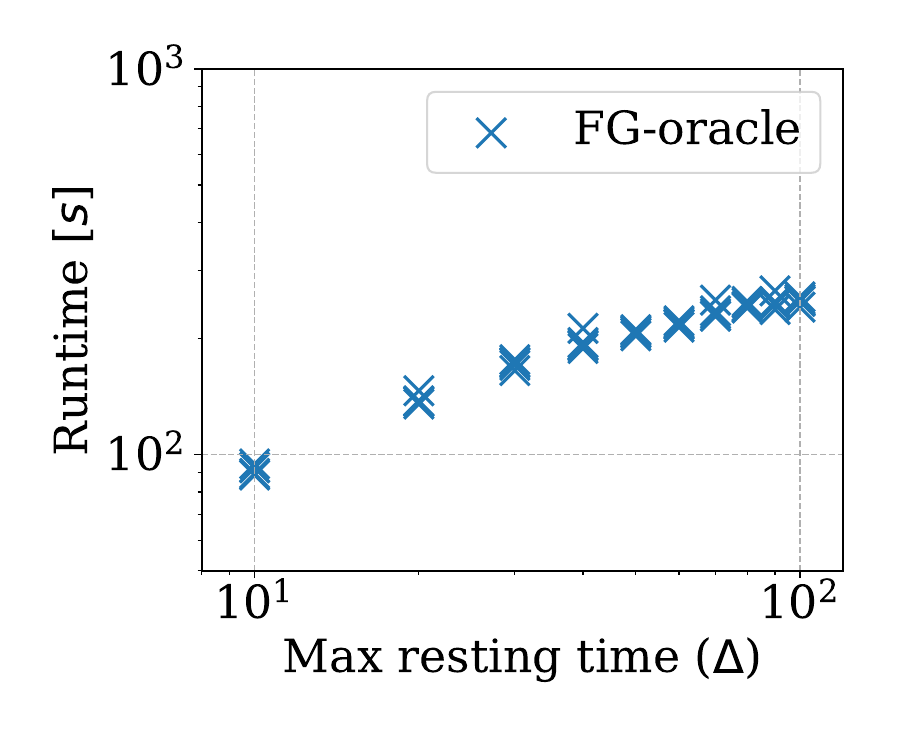} &
  \includegraphics[width=0.25\linewidth]{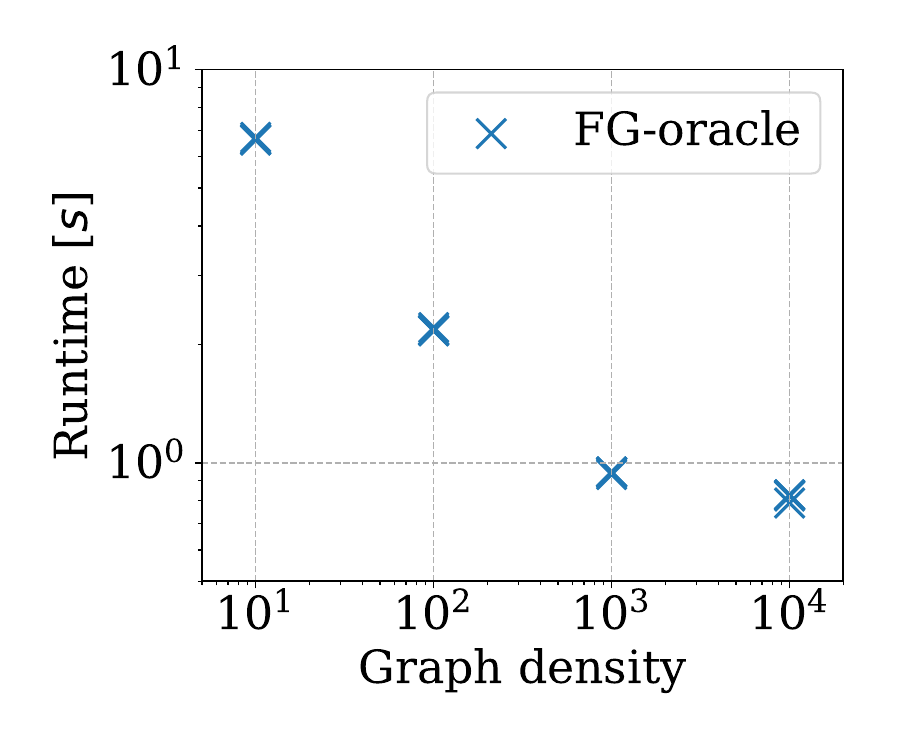}
\end{tabular}
\caption{\label{fig:exp:scalability}
Scalability in synthetic graphs. FG-oracle is fine-grained oracle.
}
\end{figure}

In Figure~\ref{fig:exp:scalability} (left), we compare the run\-time as a function of the number of edges for the fine-grained oracle and the baseline. We observe a linear increase in the run\-time with the the number of edges, as predicted by the theoretical analysis. 
In Figure~\ref{fig:exp:scalability} (center-left), we compare the run\-time as a function of the length of the restless path for the fine-grained oracle and the baseline.  We observe an exponential increase in the run\-time with the increase in length of the restless path, as predicted by the theoretical analysis. The variance in run\-time between the independent graph inputs is very small for the algebraic algorithm as compared to the baseline, which exhibit high variance in run\-time. 
We observe that our baseline fails to report the solution for power-law graphs with $n=10^3$, $m=10^6$, and $k=10$. Note that we terminate experiments that take more than ten hours. The run\-time of the exhaustive search algorithm depends on the degree distribution of the graph: if the temporal {\dfs} visits a high-degree vertex then the baseline algorithm takes long time to complete the execution. However, the baseline is efficient for sparse $d$-regular graphs, where the maximum degree is small.
Again, for scaling with respect to the length of the restless path our baseline failed to report a solution for $k \ge 10$. We terminate the experiments, which take more than ten hours of run\-time.
Our implementation can handle the \krestlessreach problem on a graph instance with one million nodes, ten million edges, and path length $k=10$, in less than 20 minutes using less than 2~gigabytes of working memory on the workstation configuration.

In Figure~\ref{fig:exp:scalability} (center-right), we report the run\-time of the algorithm as a function of the maximum resting time $\Delta$. We do not observe a linear-scaling of the run\-time as the theoretical analysis tells us. A possible explanation is that the graphs used for the experiments are sparse, and there is not enough workload to saturate the empirical arithmetic and memory bandwidth of the hardware, simultaneously.  However, with the increase in the resting time $\Delta$ we need to perform more arithmetic operations there by improving the arithmetic and memory bandwidth.  This is due to the fact that, the implementation enables more streamlining of the memory and arithmetic pipeline of the computer hardware when there is enough workload to parallelize.

Our final scalability experiments report the effect of the graph density on scalability. In Figure~\ref{fig:exp:scalability} (right), we report the run\-time of the algorithm as a function of the graph density. Density of the graph is the ratio of number of edges and the number of vertices. We observe that our implementation performs better for dense graphs. 
Again a possible explanation is that for sparse graphs there is not enough work to keep both the arithmetic and memory pipeline busy, simultaneously. This also presents us a challenge to design an efficient implementation for handling sparse graphs.

\begin{figure}[t]
\centering
\setlength{\tabcolsep}{0.01cm}
\renewcommand{\arraystretch}{0.01}
\begin{tabular}{c c c}
{\bf ~~~~~Extracting a solution} & {\bf ~~~~Memory usage} & {\bf ~~~~Graph topology}\\
\includegraphics[width=0.32\linewidth]{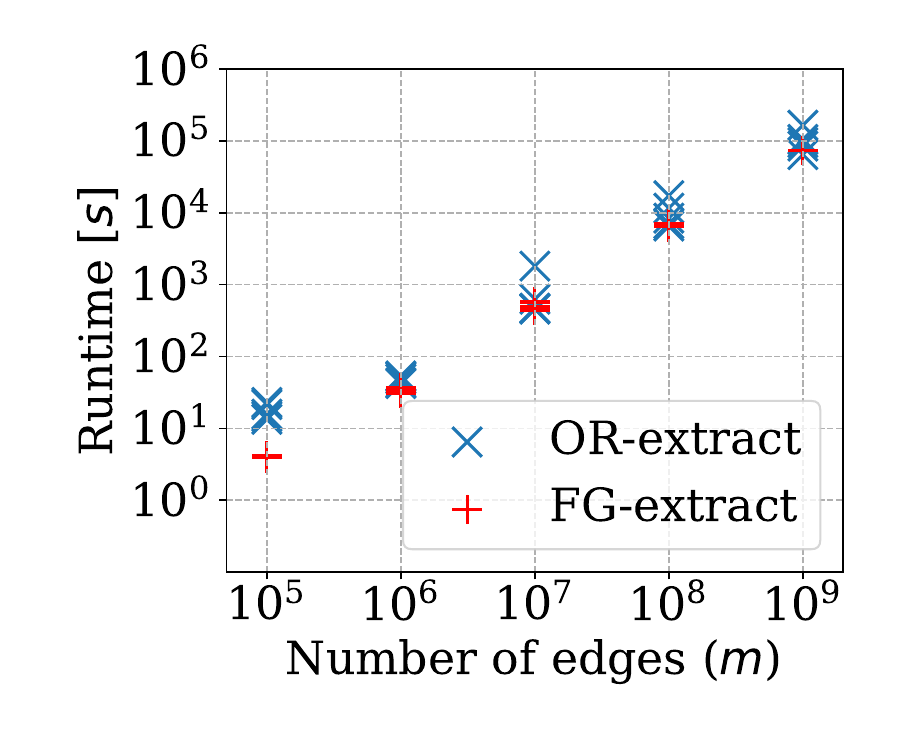} &
\includegraphics[width=0.32\linewidth]{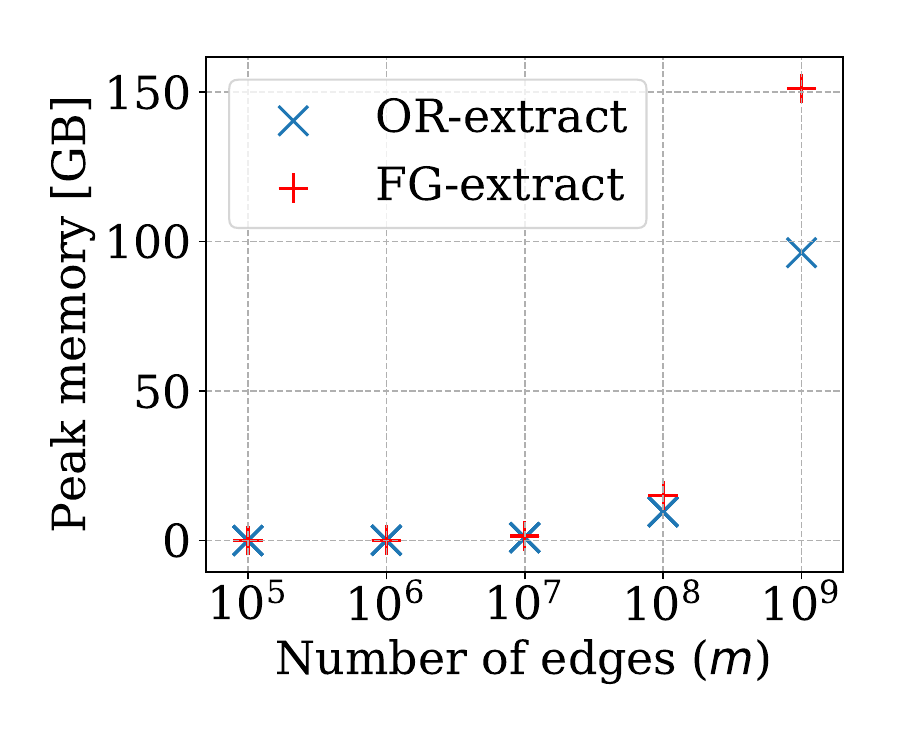} &
\includegraphics[width=0.32\linewidth]{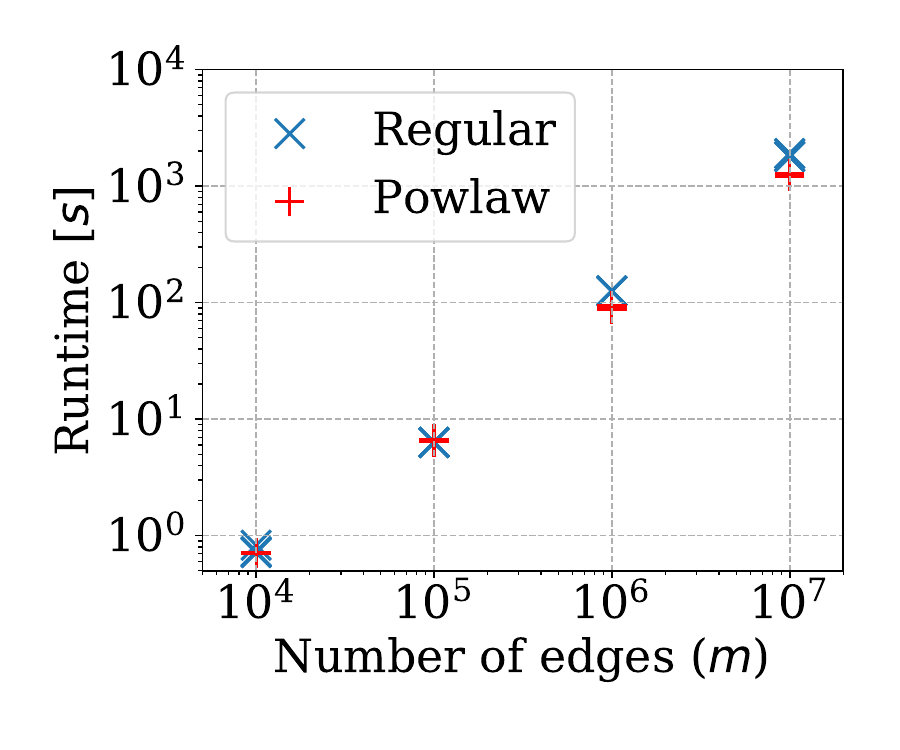}
\end{tabular}
\caption{\label{fig:exp:extraction}
Extracting a solution and topology invariance. OR-extract is solution extraction using decision oracle and FG-extract is solution extraction using fine-grained oracle.
}
\end{figure}

\para{Extracting an optimal solution.}
Our second set of experiments compares the run\-time of the algorithm for extracting an optimal restless path between a source and a destination using the decision oracle  and the fine-grained oracle. Recall that the algorithm only returns a \yes/\no answer for the existence of a solution and we need multiple queries to the oracle to obtain a solution. Using a decision oracle we require $\bigO(k \log n \log \tau)$ queries in expectation to obtain an optimum solution as compared to $k$ queries using a fine-grained oracle (see Section \ref{sec:algo:extraction}).
In Figure~\ref{fig:exp:extraction} we compare the run\-time (left) and the peak-memory usage (center) of the decision oracle and the fine-grained oracle for extracting an optimum solution for five independent random power-law graphs for each configuration of $n=10^3,\dots,10^7$ with fixed values of $d=100$, $\tau=100$, $\Delta=10$, $\delta: V \rightarrow \{\Delta\}$ and $k=10$. The source and the destination are chosen at random. We observe little variance in run\-time for the fine-grained extraction as compared to oracle extraction. For large graphs with hundred million edges the fine-grained extraction is up to four times faster than the oracle extraction. Even though in theory we reduce the number of queries by a factor of $\log n \log \tau$, in practice we do not obtain a significant improvement in the empirical run\-time.  When extracting a solution using the decision oracle, recall that for each query of the oracle we recursively divide the graph into smaller subgraphs. For each smaller subgraph, we build the multilinear sieve and thereby reduce the empirical runtime of each query. In addition, while the the expected number of queries is bounded by $\bigO(k \log n \log \tau)$ in worst case, this bound is not always met in practice. For a given instance, the number of queries required to extract a $k$-vertex subgraph varies depending on the source and the destination, resulting in high variance in the extraction time for extraction via self-reducibility. However, there is considerably less variance in the runtime of the fine-grained extraction approach.

The experiments are executed on the compute\-node configuration using all cores. We report the run\-time of extraction for \krestlessmotif instances, but the run\-times are similar for \krestlesspath instances.

\para{Graph topology.}
Here we study the effect of graph topology on run\-time of the algorithm.  In Figure~\ref{fig:exp:extraction} (right), we report the run\-time of the algorithm for five independent $d$-regular random graph instances for each configuration of $n=10^2,\dots,10^5$, $d=100$, $\tau=100$, $\Delta=10$, $\delta: V \rightarrow \{\Delta\}$, and $k=10$. Power-law graphs with $n=10^2,\dots,10^6$ with fixed values of $d=100$, $k=10$, $\tau=100$, $\Delta=10$, $\delta:V \rightarrow \{\Delta\}$, $\alpha=-1.0$, and $w=100$. We observe no significant change in the run\-time with the change in the graph topology. The experiments are executed on the workstation using all cores. We report the run\-time for \krestlessmotifreach instances, but the run\-times are similar for other problem instances.

\begin{figure}
\centering
\setlength{\tabcolsep}{0.01cm}
\renewcommand{\arraystretch}{0.01}
\begin{tabular}{c c}
{\bf \krestlesspath} & {\bf \krestlessreach} \\
\includegraphics[width=0.35\linewidth]{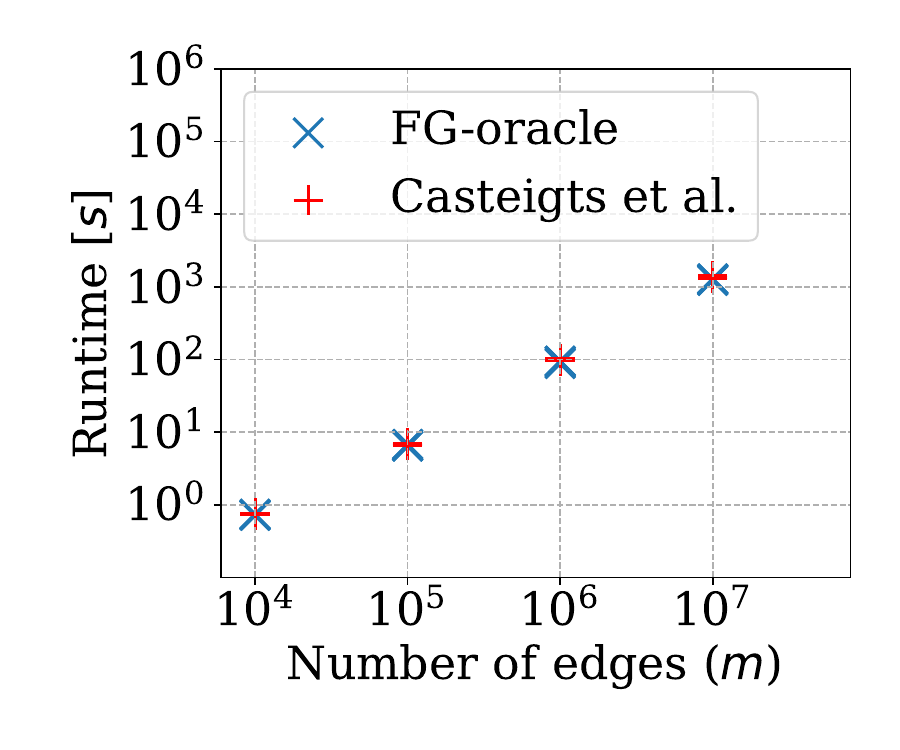} &
\includegraphics[width=0.35\linewidth]{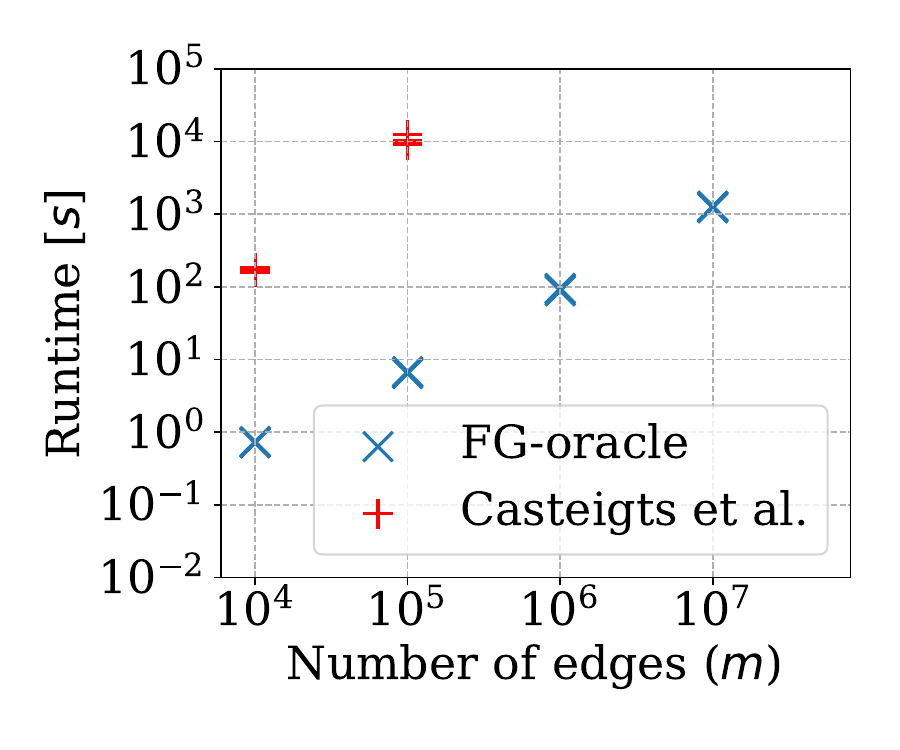}
\end{tabular}
\caption{\label{fig:exp:casteigts-runtime}
Running time comparison of our fine-grained oracle and FPT($k$)-algorithm of \citet{casteigts2021finding} for \krestlesspath and \krestlessreach problems. FG-oracle is fine-grained oracle.}
\end{figure}

\para{Scalability of the FPT($k$)-algorithm by \citet{casteigts2021finding}.}
We compare the run\-time of our fine-grained oracle with the FPT$(k)$-algorithm by \citet{casteigts2021finding} for \krestlesspath and \krestlessmotif problems. Figure~\ref{fig:exp:casteigts-runtime} reports the run\-time for both methods for five independent power-law graph instances for each configuration of $n=10^2,\dots,10^5$ with fixed values of $d=100$, $k=10$, $\tau=100$, $\Delta=10$, $\delta:V \rightarrow \{\Delta\}$, $\alpha=-1.0$, and $w=100$. For \restlesspath, we observe no significant difference in runtime, however, for \krestlessreach the running time of \citet{casteigts2021finding} increased by a factor of $n$, consistent with theoretical expectations. The experiments are executed in the workstation configuration.

\para{Experiments with real-world graphs.}
Our next set of experiments reports the runtime of the algorithm for finding restless reachability and extracting a restless path in real-world datasets. The description of the datasets used for our experiments is available in Section~\ref{sec:exp:setup}.

In Table~\ref{table:exp:realworld:1} and \ref{table:exp:realworld:2} we report the run\-time for finding restless reachability in real-worlds graphs with $k=5$ and $k=10$, respectively. For each dataset we report the maximum run\-time of five independent runs by choosing the source vertex $s \in V$ uniformly at random for each $k \in \{5,10\}$, for fixed value of the maximum resting time $\Delta=10$, and $\delta: V \rightarrow \{\Delta\}$, \ie, the resting time is constant for all the vertices except the source $s$, which has the maximum resting time $\delta(s) = \tau$.
In Column~2 we report the run\-time for solving \krestlessreach, while Column~3 reports the run\-time for solving \restlessreach by restricting the path length to $k-1$. Note that here we solve \atmkrestlessreach where we find the set of vertices that are reachable from a given source via a restless path with length at most $k-1$.  Column~4 is the ratio of Column~3 and Column~2.  Next, in Columns~5 and 6 we report the run\-time of extracting a solution using a decision oracle and a fine-grained oracle, respectively. Column~7 is the ratio of Column~6 and Column~5.  Finally, in Column~8, we report the peak memory usage of the fine-grained extraction. All run\-times are in seconds. The experiments are executed on compute\-node configuration using all cores.

We can solve restless reachability in each real-world graph dataset in Table~\ref{table:exp:datasets} in less than one minute by restricting the length of the restless path $k$ to~5, and in less than two hours using at most 14~Gb of memory for $k=10$. For instance, we can solve restless reachability by limiting $k$ to $10$ in a real-world graph dataset with more than 37 million directed edges and more than 19 thousand timestamps in less than one hour on a Haswell desktop using less than 6~Gb of memory. The reported run\-times are in seconds.

From the results (see Column~3), we see that solving \atmkrestlessreach takes less than twice the empirical running time than that of \krestlessreach, in most of the input graph instance. 
Extracting a solution using fine-grained extraction is effective for large graphs, given that $k$ is not too large. For instance, we obtain an 8-time speedup in computation using fine-grained extraction as compared to oracle extraction in a graph with more than hundred thousand vertices and more than eight hundred thousand edges with $k=10$.

\begin{table}[t]
\caption{Experiments with real-world datasets ($k=5$). For a description of the columns, see the main text.
}
\label{table:exp:realworld:1}
\centering
\footnotesize
\setlength{\tabcolsep}{0.1cm}
\begin{tabular}{l r r r r r r r}
\toprule
& \multicolumn{3}{c}{{\em Reachability}} & 
\multicolumn{3}{c}{{\em Extraction} ($k=5$)} & \\
\cmidrule(lr{1em}){2-4}
\cmidrule(lr{1em}){5-7}
        & \multicolumn{1}{c}{$k=5$}    
        & \multicolumn{1}{c}{$k \le 5$} 
        & \multicolumn{1}{c}{Ratio} 
        & \multicolumn{1}{c}{Fine-grained} 
        & \multicolumn{1}{c}{Decision}  
        & \multicolumn{1}{c}{Speedup} 
        & \multicolumn{1}{c}{Memory} \\ 
Dataset & \multicolumn{1}{c}{(seconds)} 
		& \multicolumn{1}{c}{(seconds)}  
		&       
		& \multicolumn{1}{c}{(seconds)}    
		& \multicolumn{1}{c}{(seconds)} 
		&         
		& \multicolumn{1}{c}{(Gb)} \\ 
\midrule
{\em {Copenhagen}}\\
\cmidrule(r{1em}){1-1}
          {\sf Calls} &  1.72 &  2.66 & 1.55 &  1.71 &  1.80 & 1.05 &  0.48 \\
            {\sf SMS} &  1.91 &  2.59 & 1.36 &  1.84 &  1.88 & 1.02 &  0.48 \\
\midrule
{\em {Socio-patterns}}\\
\cmidrule{1-1}
           {\sf LH10} &  8.18 & 13.06 & 1.60 &  8.24 &  8.27 & 1.00 &  3.28 \\
         {\sf InVS13} & 25.78 & 39.93 & 1.55 & 25.42 & 25.45 & 1.00 &  6.95 \\
         {\sf InVS15} & 39.21 & 59.38 & 1.51 & 42.30 & 38.36 & 0.91 & 13.18 \\
           {\sf SFHH} &  6.91 & 10.36 & 1.50 &  7.18 &  8.45 & 1.18 &  1.47 \\
     {\sf LyonSchool} & 22.39 & 32.56 & 1.45 & 22.58 & 27.33 & 1.21 &  1.70 \\
       {\sf Thiers13} & 51.76 & 73.39 & 1.42 & 51.46 & 55.27 & 1.07 &  5.38 \\
\midrule
{\em {Koblenz}}\\
\cmidrule(r{1em}){1-1}
    {\sf sqwikibooks} &  1.63 &  3.60 & 2.21 &  1.69 &  2.14 & 1.27 &  0.85 \\
   {\sf pswiktionary} &  6.42 & 10.67 & 1.66 &  6.84 &  7.28 & 1.06 &  3.70 \\
   {\sf sawikisource} &  9.96 & 17.21 & 1.73 & 13.60 & 16.63 & 1.22 &  6.58 \\
         {\sf knwiki} & 21.42 & 31.50 & 1.47 & 23.68 & 25.22 & 1.06 & 12.27 \\
       {\sf epinions} &  7.76 & 12.41 & 1.60 &  7.44 & 24.36 & 3.27 &  3.57 \\
\midrule
{\em {Transport}}\\
\cmidrule(r{1em}){1-1}
         {\sf Kuopio} &  0.49 &  0.78 & 1.57 &  0.51 &  0.86 & 1.70 &  0.10 \\
         {\sf Rennes} &  0.76 &  1.45 & 1.92 &  0.74 &  1.21 & 1.63 &  0.25 \\
       {\sf Grenoble} &  0.88 &  1.41 & 1.60 &  0.86 &  1.25 & 1.45 &  0.29 \\
         {\sf Venice} &  1.03 &  1.63 & 1.58 &  1.33 &  1.39 & 1.05 &  0.39 \\
        {\sf Belfast} &  0.82 &  2.31 & 2.81 &  0.89 &  1.31 & 1.47 &  0.31 \\
       {\sf Canberra} &  1.08 &  1.73 & 1.59 &  1.07 &  1.57 & 1.47 &  0.43 \\
          {\sf Turku} &  0.86 &  1.51 & 1.75 &  0.95 &  1.31 & 1.38 &  0.33 \\
     {\sf Luxembourg} &  0.86 &  1.19 & 1.38 &  0.97 &  1.20 & 1.24 &  0.24 \\
         {\sf Nantes} &  1.27 &  1.96 & 1.54 &  1.28 &  1.42 & 1.11 &  0.43 \\
       {\sf Toulouse} &  1.37 &  2.32 & 1.69 &  1.52 &  1.94 & 1.28 &  0.59 \\
        {\sf Palermo} &  1.25 &  1.86 & 1.49 &  1.30 &  1.57 & 1.21 &  0.40 \\
       {\sf Bordeaux} &  1.71 &  2.64 & 1.54 &  1.70 &  1.94 & 1.14 &  0.64 \\
    {\sf Antofagasta} &  0.70 &  0.87 & 1.23 &  0.68 &  1.58 & 2.34 &  0.11 \\
        {\sf Detroit} &  2.63 &  4.13 & 1.57 &  2.50 &  2.93 & 1.17 &  1.22 \\
       {\sf Winnipeg} &  2.33 &  3.33 & 1.43 &  2.31 &  2.69 & 1.17 &  0.94 \\
       {\sf Brisbane} &  3.92 &  5.62 & 1.43 &  3.77 &  4.12 & 1.09 &  1.76 \\
       {\sf Adelaide} &  3.22 &  4.72 & 1.47 &  3.04 &  3.58 & 1.18 &  1.36 \\
         {\sf Dublin} &  2.10 &  3.04 & 1.45 &  2.04 &  2.50 & 1.23 &  0.82 \\
         {\sf Lisbon} &  3.36 &  5.27 & 1.57 &  3.20 &  3.69 & 1.15 &  1.47 \\
         {\sf Prague} &  3.11 &  4.52 & 1.45 &  2.95 &  3.21 & 1.09 &  1.11 \\
       {\sf Helsinki} &  3.69 &  5.06 & 1.37 &  3.49 &  3.97 & 1.14 &  1.46 \\
         {\sf Athens} &  3.54 &  6.29 & 1.78 &  3.40 &  3.91 & 1.15 &  1.46 \\
         {\sf Berlin} &  2.95 &  3.93 & 1.33 &  3.31 &  3.69 & 1.11 &  1.01 \\
           {\sf Rome} &  4.59 &  6.01 & 1.31 &  4.44 &  4.99 & 1.13 &  1.70 \\
      {\sf Melbourne} &  8.24 & 12.61 & 1.53 &  8.56 &  8.18 & 0.96 &  4.00 \\
         {\sf Sydney} & 12.15 & 17.98 & 1.48 & 11.10 & 11.39 & 1.03 &  5.20 \\
          {\sf Paris} &  7.98 & 10.72 & 1.34 &  7.54 &  7.44 & 0.99 &  2.33 \\
\bottomrule
\end{tabular}
\end{table}

\begin{table}[t]
\caption{Experiments with real-world datasets ($k=10$).
For description of the columns, please see the main text.
}
\label{table:exp:realworld:2}
\centering
\footnotesize
\setlength{\tabcolsep}{0.1cm}
\begin{tabular}{l r r r r r r r}
\toprule
& \multicolumn{3}{c}{{\em Reachability}} & 
\multicolumn{3}{c}{{\em Extraction} ($k=10$)} & \\
\cmidrule(lr{1em}){2-4}
\cmidrule(lr{1em}){5-7}
        & \multicolumn{1}{c}{$k=10$}    
        & \multicolumn{1}{c}{$k \le 10$} 
        & \multicolumn{1}{c}{Ratio} 
        & \multicolumn{1}{c}{Fine-grained} 
        & \multicolumn{1}{c}{Decision}  
        & \multicolumn{1}{c}{Speedup} 
        & \multicolumn{1}{c}{Memory} \\ 
Dataset & \multicolumn{1}{c}{(seconds)} 
		& \multicolumn{1}{c}{(seconds)}  
		&       
		& \multicolumn{1}{c}{(seconds)}    
		& \multicolumn{1}{c}{(seconds)} 
		&         
		& \multicolumn{1}{c}{(Gb)} \\ 
\midrule
{\em {Copenhagen}}\\
\cmidrule(r{1em}){1-1}
          {\sf Calls} &    88.98 &   152.91 & 1.72 &    90.60 &   194.03 & 2.14 &  0.48 \\
            {\sf SMS} &    89.32 &   153.67 & 1.72 &    88.36 &   195.87 & 2.22 &  0.48 \\
\midrule
{\em {Socio-patterns}}\\
\cmidrule{1-1}
           {\sf LH10} &   411.13 &   719.49 & 1.75 &   417.43 &   532.08 & 1.27 &  3.28 \\
         {\sf InVS13} &1\,292.91 &2\,342.52 & 1.81 &1\,309.36 &1\,613.37 & 1.23 &  6.95 \\
         {\sf InVS15} &1\,871.23 &3\,414.07 & 1.82 &1\,966.12 &2\,155.92 & 1.10 & 13.18 \\
           {\sf SFHH} &   314.48 &   569.10 & 1.81 &   318.29 &   581.37 & 1.83 &  1.47 \\
     {\sf LyonSchool} &1\,284.53 &2\,294.68 & 1.79 &1\,298.16 &1\,849.85 & 1.42 &  1.70 \\
       {\sf Thiers13} &2\,666.22 &4\,687.66 & 1.76 &2\,663.80 &3\,233.05 & 1.21 &  5.38 \\
\midrule
{\em {Koblenz}}\\
\cmidrule(r{1em}){1-1}
    {\sf sqwikibooks} &    70.78 &   131.26 & 1.85 &    72.20 &   257.57 & 3.57 &  0.85 \\
   {\sf pswiktionary} &   267.32 &   478.48 & 1.79 &   283.14 &   909.79 & 3.21 &  3.71 \\
   {\sf sawikisource} &   464.38 &   856.73 & 1.84 &   495.80 &1\,450.61 & 2.93 &  6.59 \\
         {\sf knwiki} &   909.25 &1\,659.58 & 1.83 &   969.38 &2\,995.04 & 3.09 & 12.28 \\
       {\sf epinions} &   286.11 &   522.78 & 1.83 &   375.55 &3\,051.37 & 8.13 &  3.58 \\
\midrule
{\em {Transport}}\\
\cmidrule(r{1em}){1-1}
         {\sf Kuopio} &    25.08 &    43.45 & 1.73 &    25.42 &   116.30 & 4.58 &  0.10 \\
         {\sf Rennes} &    32.15 &    59.12 & 1.84 &    34.24 &   161.09 & 4.71 &  0.25 \\
       {\sf Grenoble} &    38.38 &    68.99 & 1.80 &    39.15 &   155.41 & 3.97 &  0.29 \\
         {\sf Venice} &    50.07 &    84.05 & 1.68 &    49.62 &   176.52 & 3.56 &  0.39 \\
        {\sf Belfast} &    36.29 &    69.19 & 1.91 &    39.32 &   172.56 & 4.39 &  0.31 \\
       {\sf Canberra} &    48.00 &    86.93 & 1.81 &    48.93 &   182.02 & 3.72 &  0.43 \\
          {\sf Turku} &    39.46 &    75.04 & 1.90 &    43.81 &   158.19 & 3.61 &  0.33 \\
     {\sf Luxembourg} &    37.15 &    65.73 & 1.77 &    39.68 &   140.88 & 3.55 &  0.24 \\
         {\sf Nantes} &    51.82 &    87.37 & 1.69 &    52.35 &   150.44 & 2.87 &  0.43 \\
       {\sf Toulouse} &    62.96 &   113.45 & 1.80 &    62.54 &   203.01 & 3.25 &  0.59 \\
        {\sf Palermo} &    47.20 &    93.09 & 1.97 &    52.74 &   198.76 & 3.77 &  0.40 \\
       {\sf Bordeaux} &    67.52 &   124.61 & 1.85 &    69.16 &   208.33 & 3.01 &  0.64 \\
    {\sf Antofagasta} &    23.72 &    41.53 & 1.75 &    25.26 &   112.66 & 4.46 &  0.11 \\
        {\sf Detroit} &   111.74 &   201.14 & 1.80 &   107.84 &   288.48 & 2.68 &  1.22 \\
       {\sf Winnipeg} &    88.13 &   162.17 & 1.84 &    93.29 &   231.43 & 2.48 &  0.94 \\
       {\sf Brisbane} &   147.66 &   268.12 & 1.82 &   153.42 &   320.82 & 2.09 &  1.76 \\
       {\sf Adelaide} &   122.11 &   221.79 & 1.82 &   124.29 &   312.47 & 2.51 &  1.37 \\
         {\sf Dublin} &    78.42 &   142.13 & 1.81 &    82.51 &   205.09 & 2.49 &  0.82 \\
         {\sf Lisbon} &   134.25 &   246.37 & 1.84 &   137.01 &   332.89 & 2.43 &  1.47 \\
         {\sf Prague} &   108.59 &   199.35 & 1.84 &   107.47 &   245.90 & 2.29 &  1.11 \\
       {\sf Helsinki} &   130.68 &   231.65 & 1.77 &   136.54 &   319.97 & 2.34 &  1.46 \\
         {\sf Athens} &   129.10 &   233.02 & 1.80 &   130.00 &   324.98 & 2.50 &  1.46 \\
         {\sf Berlin} &   102.28 &   179.79 & 1.76 &   105.72 &   233.19 & 2.21 &  1.01 \\
           {\sf Rome} &   159.08 &   287.87 & 1.81 &   164.57 &   368.35 & 2.24 &  1.70 \\
      {\sf Melbourne} &   323.32 &   590.11 & 1.83 &   337.50 &   631.20 & 1.87 &  4.00 \\
         {\sf Sydney} &   411.24 &   732.24 & 1.78 &   427.51 &   866.15 & 2.03 &  5.20 \\
          {\sf Paris} &   223.02 &   403.70 & 1.81 &   224.21 &   477.10 & 2.13 &  2.33 \\
\bottomrule
\end{tabular}
\end{table}

\begin{figure*}[t]
\centering
\includegraphics[width=1.0\linewidth]{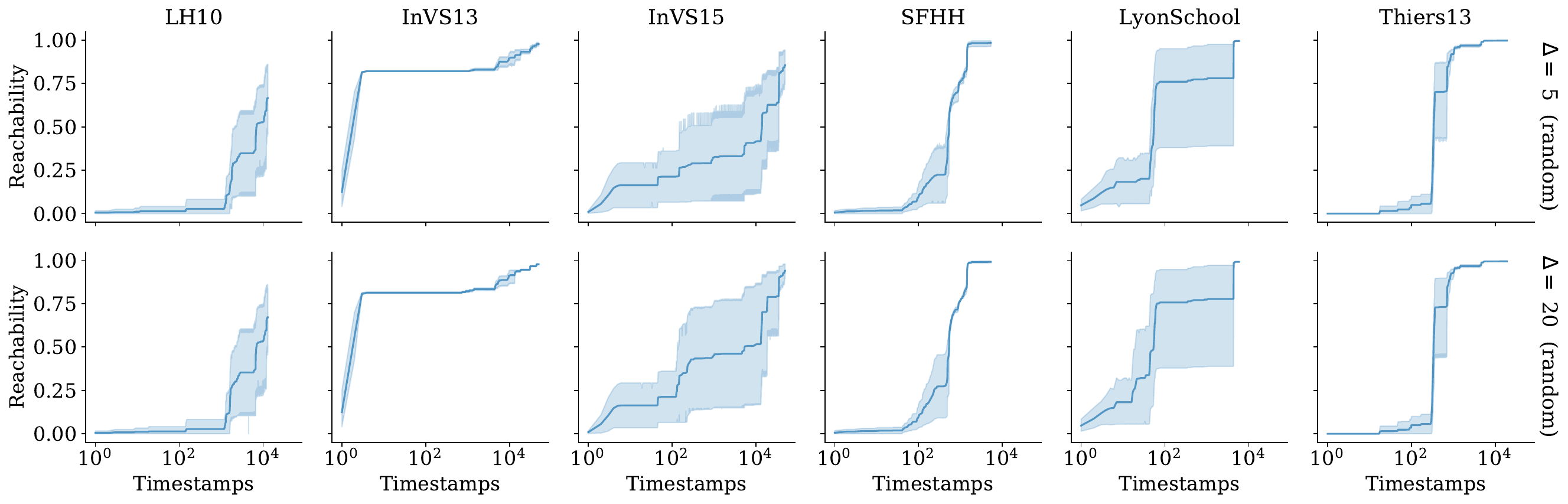}
\caption{\label{fig:exp:reachability:1}
Restless reachability in real-world datasets with $\Delta=5$ (top-row) and $\Delta=20$ (bottom-row) with $\delta:V \rightarrow [\Delta]$ assigned uniformly at random, and $k = 10$ (fixed).
}
\end{figure*}

\para{Reachability.} 
Our next set of experiments studies the restless reachability in socio-patterns dataset. In particular we solve \atmkrestlessreach where we find the set of vertices  reachable from a given source via a restless path of length at most $k-1$. \emph{Reachability} is the ratio of the number of vertices that are reachable with a restless path to the total number of vertices. In Figure~\ref{fig:exp:reachability:1} we report the variance and mean of reachability as a time-series for five independent graph instances for each dataset with $\Delta=5$ (top-row) and $\Delta=20$ (bottom-row).  More precisely, for a given dataset we generate five graph instances by choosing a source vertex uniformly at random and assign the resting times uniformly at random in the range $[\Delta]$. We limit the length of the restless path to $9$, i.e., $k = 10$ and solve \atmkrestlessreach.

We observe high variance in reachability among the independent source vertices for {\sf LH10}, {\sf InVS15}, {\sf LyonSchool}, and {\sf Thiers13} datasets, and smaller variance for {\sf InVS13} and {\sf SFHH}.  We also see that in all datasets, except {\sf LH10}, reachability approaches its maximum value~1, within the total number of timestamps available in each dataset, however, this happens at different times for each dataset, and with a different pace.

In Table~\ref{table:exp:realworld:3}, we report the runtime for solving \atmkrestlessreach in socio-patterns datasets. The reported runtime is  the maximum of five independent runs by choosing the source vertex $s \in V$ uniformly at random with $k=10$ (fixed) and $\Delta=5, 10$, $\delta: V \rightarrow \{\Delta\}$ (fixed). The resting time is constant for all the vertices except the source $s$, which has the maximum resting time $\delta(s) = \tau$. For instance, we can solve \shortrestlessreach by limiting $k=10$ in a real-world graph dataset with more than 37 million directed edges and more than 19 thousand timestamps in less than one hour on workstation configuration using less than 6~Gb of memory. The reported runtimes are in seconds.

\begin{table}[t]
\caption{Computing restless reachability in socio-patterns dataset using the workstation configuration. We report runtime for solving \atmkrestlessreach problem with $k=10$ (fixed).}
\label{table:exp:realworld:3}
\centering
\footnotesize
\setlength{\tabcolsep}{0.1cm}
\begin{tabular}{l r r r r r}
\toprule
Dataset & $n$ & $m$ & $\tau$ & \multicolumn{1}{c}{$\Delta=5$} & \multicolumn{1}{c}{$\Delta=10$}\\
        &     &     &        & (seconds) & (seconds)\\
\midrule
      {\sf LH10} &    73 &     300\,252 & 12\,960 &    156.8\,s &    158.7\\
    {\sf InVS13} &    95 &     788\,494 & 49\,679 &    357.1\,s &    386.5\\
    {\sf InVS15} &   219 &  2\,566\,388 & 49\,679 &    657.2\,s &    733.2\\
      {\sf SFHH} &   403 &  2\,834\,970 &  5\,328 &    245.0\,s &    346.3\\
{\sf LyonSchool} &   242 & 13\,188\,984 &  5\,887 &    853.6\,s & 1\,341.2\\
  {\sf Thiers13} &   328 & 37\,226\,078 & 19\,022 & 2\,083.1\,s & 3\,399.5\\
\bottomrule
\end{tabular}
\end{table}

\subsection{Case study: comparing immunization strategies}
Our final set of experiments studies the change in restless reachability in the presence of a set of barrier vertices $S' \subseteq V$ called \emph{separators}, which must not be included in the restless path. In an epidemic model, the separators can be viewed as a subset of the population, all immune and/or vaccinated. It is known that finding a set of temporal separators with minimum size, which destroy all the restless paths between any two vertices, is $\sum_2^P$-hard~\cite{molter2022thecomplexity}[Theorem~7].

The experiments performed in this case study are to evaluate the effectiveness of our algorithm to answer queries for finding an effective immunization strategy in an epidemic model where the disease propagation is via a restless path. 
To demonstrate this we use two heuristics for finding separators: ($i$) choose vertices at random and ($ii$) choose vertices with maximum temporal degree. Note that these immunization strategies are simple and without theoretical guarantees. Towards this end, we would like to investigate effective approximation or heuristic schemes to find temporal separators, which reduce the fraction of vertices reachable from a given source vertex via a restless path in future work.

Given an instance $(G=(V,E), s, k)$ of \shortrestlessreach and a set $S' \subseteq V$ of separators, we introduce a coloring function $c:V \setminus \{S' \cup s\} \rightarrow \{1\}$, $c(s) = 2$ and $c:S' \rightarrow \{3\}$ and $M=\{1^{\ell-1}, 2\}$ for $\ell \in [k]$. We query the  \finegrainedoracle with instance  $(G'=(V, E\setminus \{u,s,i\}\in E), c,\ell,M)$ for each  $\ell \in \{2,\dots,k\}$. By assigning color  $3 \notin M$ to the separators in $S'$,  we make sure that none of the separators are part of the restless path agreeing the multiset of colors in $M$. Note that here we solve \atmkrestlessreach, in other words we find the set of vertices which are reachable from the source via a restless path of length at most $k-1$.

\begin{figure*}[t]
\centering
\includegraphics[width=1.0\linewidth]{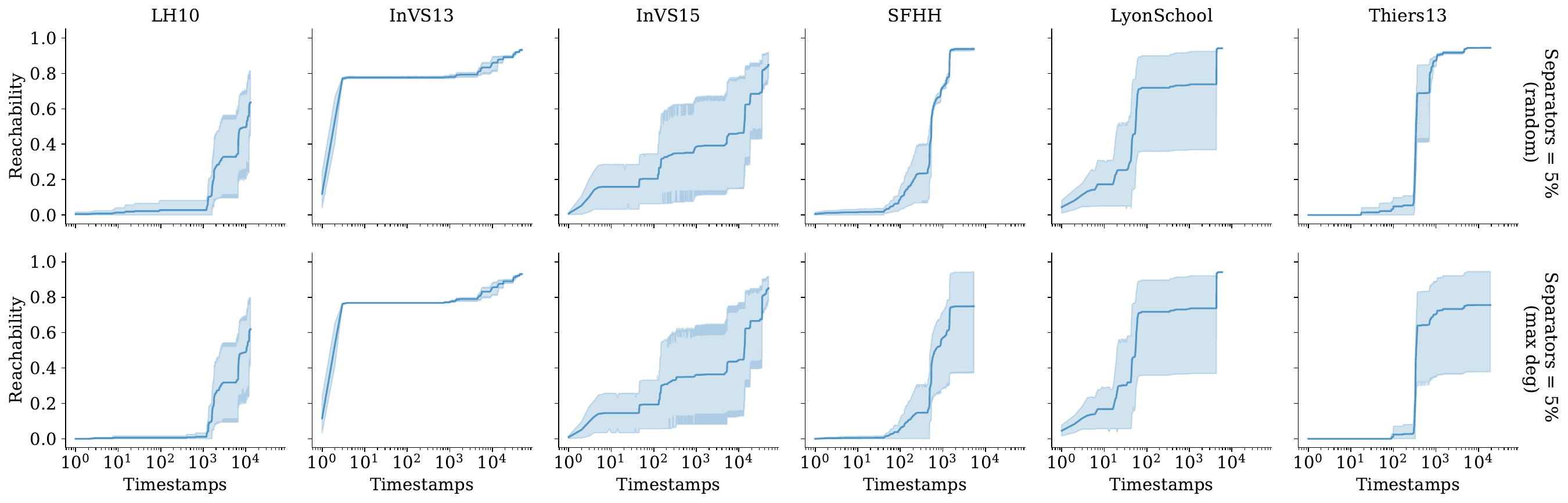}
\caption{\label{fig:exp:immunization:1}
Restless reachability in real-world datasets in the presence of $5\%$ separators. $\Delta=10$, $\delta: V \rightarrow [\Delta]$ assigned uniformly at random and $k=10$ (fixed).
}
\end{figure*}

\begin{figure*}[t]
\centering
\includegraphics[width=1.0\linewidth]{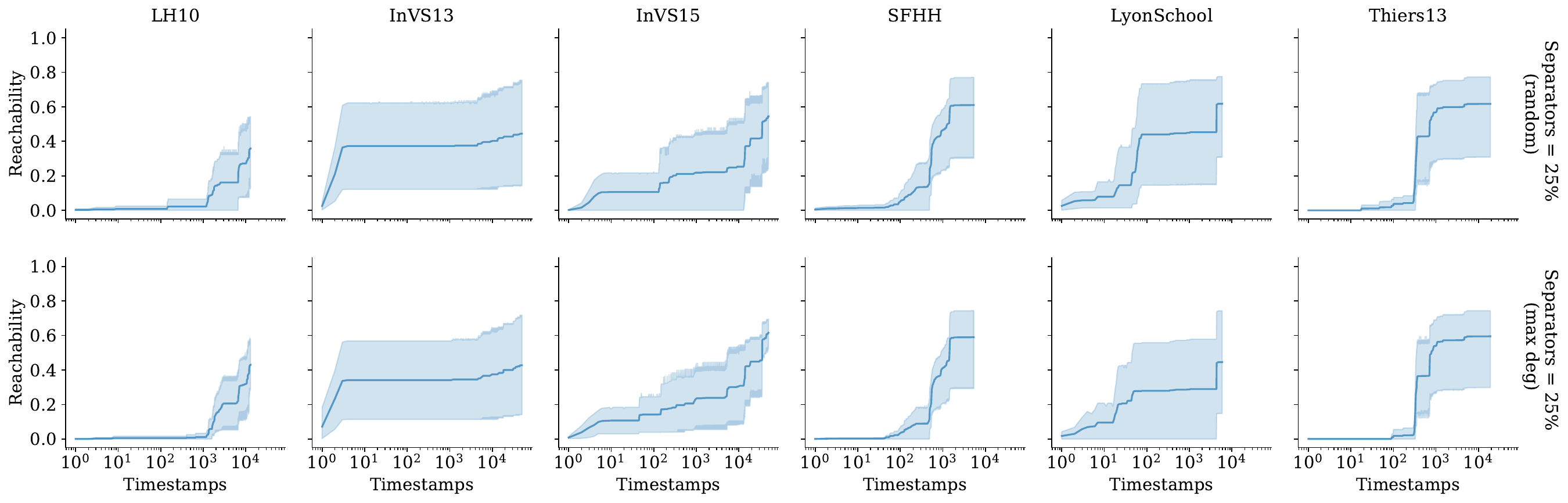}
\caption{\label{fig:exp:immunization:2}
Restless reachability in real-world datasets in the presence of $25\%$ separators. $k=10$, $\Delta=10$, $\delta : V \rightarrow [\Delta]$ assigned uniformly at random.}
\end{figure*}

In Figure~\ref{fig:exp:immunization:1}, we report the variance of reachability in real-world graphs by choosing $5\%$ of the vertices uniformly a random (top-row) and $5\%$ of the vertices with maximum degree (second-row) as separators. Figure~\ref{fig:exp:immunization:2} reports the same experiments with $5\%$ replaced with $25\%$.
For each dataset we generate five graph instances, choose source vertices at random and assign resting times uniformly at random in the range $[\Delta]$ for $\Delta=10$.  Note that we use the same instances in the reachability experiments described above. Reachability is the ratio of number of reachable vertices  to the total number of vertices excluding separators.
From the experimental results we observe high variance in the reachability among independent source vertices and the rate of increase of the reachability with time varies depending on the source vertex. More importantly, even though we expect (empirically) that by choosing vertices with high degree as separators should decrease the reachability as compared to choosing the vertices at random, this is not true for all the datasets. We only observe this phenomenon in {\sf SFHH} and Thiers13 datasets for choosing $5$\% of the vertices as separators and in {\sf InVS15} and {\sf LyonSchool} datasets while choosing $25$\% of the vertices as separators.

\begin{figure*}[t]
\centering
\includegraphics[width=1.0\linewidth]{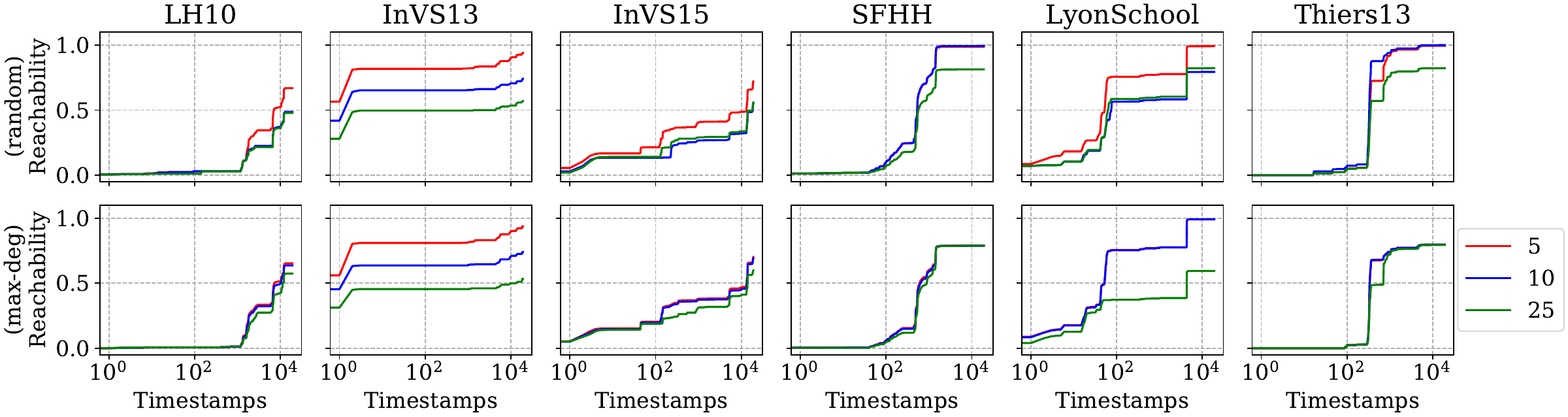}
\caption{\label{fig:exp:immunization:3}
Restless reachability in real-world datasets in the presence of separators. $k=10$ (fixed), $\Delta=10$, $\delta:V \rightarrow [\Delta]$ assigned uniformly at random.
}
\end{figure*}

In Figure~\ref{fig:exp:immunization:3}, we report the variance of reachability in real-world graphs by choosing $5\%$, $10\%$, and $25\%$ of the vertices uniformly a random (top-row) and $5\%$, $10\%$, and $25\%$ of the vertices with maximum degree (second-row) as separators.  Again, for each dataset we generate five graph instances, choose source vertices at random and assign resting time uniformly at random in the range $[\Delta]$ for $\Delta=10$. Note that we use the same instances in the reachability experiments described above.
We observe that reachability reduces with an increase in the number of separators, as expected. Again, even though it is expected (empirically) that choosing vertices with maximum degree to reduce the reachability significantly compared to choosing separators at random, surprisingly enough, this is not true for all datasets. For instance in {\sf LyonSchool} dataset choosing 10\% of the separators at random reduces the average reachability more than choosing vertices with maximum degree as separators. So the heuristic approach of choosing vertices with maximum degrees might not be effective across datasets. This presents us a interesting question of finding a small a set of separators in temporal graphs under resting time restrictions. Also note that the input graph datasets are small world graphs, meaning that the diameter of the underlying graph is small, so the vertices are highly connected (see Table~\ref{table:exp:datasets}).

\section{Conclusions and future work} \label{sec:conclusion}
In this work, we studied a family of reachability problems in temporal and vertex-colored temporal graphs  under waiting-time restrictions.  We presented an algebraic algorithmic framework for solving restless reachability problems we proposed, running in $\bigO(2^k k m \Delta)$ time and $\bigO(n\Delta)$ space. Further, we presented evidence that the algorithms for solving variants of the restless reachability problems involving colors  presented in this work are optimal under certain complexity-theoretic assumptions.
In addition, we engineered an open-source implementation of our algorithm, and demonstrated its viability in experiments on graphs with millions of temporal edges from real-world datasets. Finally, we applied the algorithm we developed in a case study for estimating the change in disease spreading   with the presence of people with immunity. Our finding is that heuristic approaches such as selecting vertices with high degree as separators are not effective in all the graph datasets. Towards this end, we would like to investigate effective ways to choose a small set of separators under waiting time restrictions to contain the spread of the disease in the network.

\xhdr{Future work} 
We demonstrated that our algorithms scale to graphs with one billion edges for path lengths up to $k=10$. Specifically, as our algorithms have running time exponential in $k$, \ie, $\bigO^*(2^k)$, the scalability with respect to $k$ remains limited. Addressing this limitation by extending to larger $k$ could be a direction of future work, potentially adapting the algorithm to vector-parallel architectures such as GPGPUs, as explored in prior work~\cite{kaski2018engineering}.
Further, we believe that our algorithms based on constrained multilinear sieving can be extended to solve other pattern-detection problems in temporal graphs, including finding temporal arborescences, connected temporal subgraphs, and temporal subgraphs with color constraints on the vertices. Additionally, we hypothesize that by extending the narrow-sieve construction proposed by Bj{\"o}rklund et al.~\cite{bjorklund2017narrow} to encode restless \emph{walks} as a polynomial could break the $\bigO^*(2^k)$ barrier for \krestlesspath, potentially reducing the exponent.


\newpage
\bibliographystyle{ACM-Reference-Format}
\bibliography{references}

\newpage
\appendix

{\LARGE \bf Supplementary Material}

\section{Further motivation}\label{app:motivation}
Another application of restless reachability is to find signaling pathways in brain networks. Here we assume that brain regions are represented as vertices, and edges between the regions represent physical proximity, for example, two regions that are physically next to each other or having high signal correlation in an electro-encephalogram (EEG) are connected with an edge. Using functional magnetic resonance imaging (\fmri) we can record the \emph{active} brain regions at different timestamps --- the frequency of the scans is approximately one second \cite{kujala2018brain,thompson2017static}. We introduce a timestamp on an edge if two regions with a static edge are active in consecutive \fmri scans. Finally, we introduce resting time for each region: any incoming signal can be forwarded to another region with in the resting time duration. Given the source of signal origin we would like to find all the regions to which the signal was successfully transmitted. The problem can be abstracted as an instance of  the \restlessreach problem.

Also, consider a tour-recommendation scenario where a traveler is interested in visiting a set of places such as a historic museum, an art gallery, a caf{\'e}, or other. Each location has a maximum time-limit that the traveler is willing to spend. Given a start and an end location we would like to find a travel itinerary satisfying the constraints specified by the traveler. This problem can be modeled as an instance of the \krestlessmotif problem, where each location can be considered as a vertex, the color associated with the vertex represent the type of a location, for example, museum, art gallery, etc., the temporal edges between the vertices represent the transportation links, and the resting time duration associated with each vertex represents the maximum time a traveler is willing to spend at a location.

\section{List of symbols}\label{app:symbols}
\begin{table}[h]
\centering
\footnotesize
\caption{\label{table:symbols} List of symbols.}
\begin{tabular}{l l}
\hline
Symbol & Description \\
\hline
{\em {Graphs}}\\
\cmidrule(r{1em}){1-1}
                      $G=(V,E)$ & graph $G$ with vertex set $V$ and edge set $E$\\
                            $n$ & number of vertices\\
                            $m$ & number of edges\\
                          $m_i$ & number of edges at time stamp $i$\\
                       $\Delta$ & maximum resting time\\
                         $\tau$ & maximum timestamp\\
                            $k$ & length of path or walk \\
$\delta:V \rightarrow [\Delta]$ & restless function mapping vertices to resting time\\
                      $N_i(u)$  & in-neighbors of vertex $u$ at time stamp $i$\\
                     $\ulg{G}$  & underlying static graph of temporal graph $G$\\
                $\deltaexp{G}$  & $\delta$-expansion of temporal graph $G$\\
\hline
{\em {Coloring}}\\
\cmidrule(r{1em}){1-1}
                $C$ & set of colors \\
$c:V \rightarrow C$ & coloring function mapping vertices to colors \\
                $M$ & multiset of colors \\
           $\mu(s)$ & number of occurrences of color $s$ in $M$\\
\hline
{\em {Polynomials}} \\
\cmidrule(r{1em}){1-1}
  $\mathcal{P}$ & polynomial encoding of temporal walks\\
         $\chi$ & polynomial encoding of restless walks\\
        $\zeta$ & evaluation polynomial of restless walks\\
          $x_u$ & $x$-variable encoding vertex $u$\\
$y_{uv,\ell,j}$ & $y$-variable encoding edge $(u,v,j)$ at position $\ell$\\
      $\vec{x}$ & vector of $x$ variables\\
      $\vec{y}$ & vector of $y$ variables\\
      $\vec{z}$ & vector of $z$ variables\\
\hline
\end{tabular}
\end{table}

\newpage
\section{Polynomial encoding temporal walks and temporal paths} \label{app:encoding-walks}
Here we give brief overview of polynomial encoding of walks and paths in static as well as temporal graphs.

\subsection{Monomial encoding of walks}

\para{Static graphs.}
We introduce a set of variables
$\vec{x} = \{x_{v_1},\dots,x_{v_n}\}$ for vertices in $V=\{v_1,\dots,v_n\}$ and a set of
variables $\vec{y} = \{y_{uv,\ell}: (u,v) \in E, \ell \in [k]\}$ such that $y_{uv,\ell}$
corresponds to an edge $(u,v) \in E$ that appears at position $\ell$ in a walk.
Using these variables a walk $W = v_1 e_1 v_2 \dots e_{k-1}v_k$ can be encoded
by the monomial
\[
x_{v_1}\,\,y_{v_1v_2,1}\,\,x_{v_2}\, \dots\, y_{v_{k-1}v_k,k-1}\,\,x_{v_k}.
\]
The monomial encoding of walks in static graphs is illustrated in
Figure~\ref{fig:encoding:walks}. It is easy to see that the encoding is
multilinear, i.e., no
variable in a monomial is repeated, if and only if the corresponding walk is a
path. To encode a walk of length $k-1$, we need $k$ variables of
$\vec{x}$ and $k-1$ variables of $\vec{y}$, for a total of $2k-1$ variables.

\para{Temporal graphs.}
We introduce a set of variables $\vec{x} =\{x_{v_1},\dots,x_{v_n}\}$ for vertices
$V=\{v_1,\dots,v_n\}$ and a set of variables
$\vec{y} = \{y_{uv,\ell,j}: (u,v,j) \in E, \ell \in [k]\}$
such that $y_{uv,\ell,j}$
corresponds to an edge $(u,v,j) \in E$ that appears at position $\ell$ in a
walk. Using these variables a temporal walk
$W = v_1 e_1 v_2 \dots e_{k-1} v_k$ can be encoded using the monomial
\[
x_{v_1}\,\,y_{v_1v_2,1,j_1}\,\,
x_{v_2}\,\,y_{v_2v_3,2,j_2}\, \dots \,
y_{v_{k-1}v_k,k-1,j_{k-1}}\,\,x_{v_k}.
\]

The polynomial encoding of temporal walks is illustrated in
Figure~\ref{fig:encoding:tempwalks}.
A monomial encoding of a temporal walk of length $k-1$ has $2 k-1$ variables.

Again, observe that the encoding
is multilinear if and only if
the corresponding temporal walk is a temporal path.
In fact, it can be shown that a temporal walk is a temporal path if and only if the
corresponding monomial encoding is
multilinear~\cite[Lemma~2]{thejaswi2020pattern}. Likewise, the problem of
deciding the existence of temporal path of length $k-1$ is equivalent to
deciding the existence of a multilinear monomial of degree
$2k-1$~\cite[Lemma~3]{thejaswi2020pattern}. In other words, if we encode all the
walks of length $k-1$ in a temporal graph using a polynomial where each monomial
encodes a walk, then the problem of detecting the existence of a temporal path
of length $k-1$ is equivalent to the problem of detecting existence of a
multilinear monomial of degree $2k-1$ in the encoded polynomial.

\begin{figure}
\centering
\begin{tabular}{c c}
\begin{tikzpicture}[scale=\tikzscale,every node/.style={scale=\tikzscale}]]

\input{tikz/tikz-defs}

\node[exnode] (v1) at ( -0.5, 0) {$v_1$};
\node[exnode] (v2) at (    2, 0) {$v_2$};
\node[exnode] (v3) at (    3, 1.5) {$v_3$};
\node[exnode] (v4) at (    1, 1.5) {$v_4$};

\node[fill=white] at ( -1, -0.5) {\large $x_{v_1}$};
\node[fill=white] at (  1, -0.5) {\large $y_{v_1v_2,1}$};
\node[fill=white] at (  2, -0.5) {\large $x_{v_2}$};
\node[fill=white] at (  3.25, 0.75) {\large $y_{v_2v_3,2}$};
\node[fill=white] at (  3.5,  2) {\large $x_{v_3}$};
\node[fill=white] at (  2,  2) {\large $y_{v_3v_4,3}$};
\node[fill=white] at (  0.5,  2) {\large $x_{v_4}$};
\node[fill=white] at (  0.75,  0.75) {\large $y_{v_4v_2,4}$};





\path [draw=black,postaction={on each segment={mid arrow=black, ultra thick}}]
(v1) -- (v2)
(v2) -- (v3)
(v3) -- (v4)
(v4) -- (v2)
;
\end{tikzpicture}%
 &
\begin{tikzpicture}[scale=\tikzscale,every node/.style={scale=\tikzscale}]]

\input{tikz/tikz-defs}

\node[exnode] (v1) at ( -0.5, 0) {$v_1$};
\node[exnode] (v2) at (    2, 0) {$v_2$};
\node[exnode] (v3) at (  4.5, 0) {$v_3$};

\node[fill=white] at ( -1, -0.5) {\large $x_{v_1}$};
\node[fill=white] at (  1, -0.5) {\large $y_{v_1v_2,1}$};
\node[fill=white] at (  2, -0.5) {\large $x_{v_2}$};
\node[fill=white] at (  3.5,  -0.5) {\large $y_{v_2v_3,2}$};
\node[fill=white] at (  5,  -0.5) {\large $x_{v_3}$};





\path [draw=black,postaction={on each segment={mid arrow=black, ultra thick}}]
(v1) -- (v2)
(v2) -- (v3)
;
\end{tikzpicture}%
 \\
$x_{v_1}\,y_{v_1v_2, 1}\,
\textcolor{red}{x_{v_2}}\,y_{v_2v_3,2}\,
x_{v_3}\,y_{v_3v_4,3}\,
x_{v_4}\,y_{v_4v_2,4}\,
\textcolor{red}{x_{v_2}}$ &
$x_{v_1}\,y_{v_1v_2,1}\,x_{v_2}\,
y_{v_2v_3,2}\,x_{v_3}$
\end{tabular}
\caption{\label{fig:encoding:walks}
A monomial encoding of a walk (left) and a path (right) in static graphs.
Arrows indicate the direction of the walk.
Variables highlighted in red are repeated if and only if a walk is not a path.}
\end{figure}

\begin{figure}
\centering
\begin{tabular}{c c}
\begin{tikzpicture}[scale=\tikzscale,every node/.style={scale=\tikzscale}]]

\input{tikz/tikz-defs}

\node[exnode] (v1) at ( -0.5, 0) {$v_1$};
\node[exnode] (v2) at (    2, 0) {$v_2$};
\node[exnode] (v3) at (    3, 1.5) {$v_3$};
\node[exnode] (v4) at (    1, 1.5) {$v_4$};

\node[fill=white] at ( -1, -0.5) {\large $x_{v_1}$};
\node[fill=white] at (  1, -0.5) {\large $y_{v_1v_2,1,j_1}$};
\node[fill=white] at (  2, -0.5) {\large $x_{v_2}$};
\node[fill=white] at (  3.25, 0.75) {\large $y_{v_2v_3,2,j_2}$};
\node[fill=white] at (  3.5,  2) {\large $x_{v_3}$};
\node[fill=white] at (  2,  2) {\large $y_{v_3v_4,3,j_3}$};
\node[fill=white] at (  0.5,  2) {\large $x_{v_4}$};
\node[fill=white] at (  0.75,  0.75) {\large $y_{v_4v_2,4,j_4}$};





\path [draw=black,postaction={on each segment={mid arrow=black, ultra thick}}]
(v1) -- (v2)
(v2) -- (v3)
(v3) -- (v4)
(v4) -- (v2)
;
\end{tikzpicture}%
 &
\begin{tikzpicture}[scale=\tikzscale,every node/.style={scale=\tikzscale}]]

\input{tikz/tikz-defs}

\node[exnode] (v1) at ( -0.5, 0) {$v_1$};
\node[exnode] (v2) at (    2, 0) {$v_2$};
\node[exnode] (v3) at (  4.5, 0) {$v_3$};

\node[fill=white] at ( -1, -0.5) {\large $x_{v_1}$};
\node[fill=white] at (  1, -0.5) {\large $y_{v_1v_2,1,j_1}$};
\node[fill=white] at (  2, -0.5) {\large $x_{v_2}$};
\node[fill=white] at (  3.5,  -0.5) {\large $y_{v_2v_3,2,j_2}$};
\node[fill=white] at (  5,  -0.5) {\large $x_{v_3}$};





\path [draw=black,postaction={on each segment={mid arrow=black, ultra thick}}]
(v1) -- (v2)
(v2) -- (v3)
;
\end{tikzpicture}%
 \\
$x_{v_1}\,y_{v_1v_2,1,j_1}\,
\textcolor{red}{x_{v_2}}\,y_{v_2v_3,2,j_2}\,
x_{v_3}\,y_{v_3v_4,3,j_3}\,
x_{v_4}\,y_{v_4v_2,4,j_4}\,
\textcolor{red}{x_{v_2}}$ &
$x_{v_1}\,y_{v_1v_2,1,j_1}\,x_{v_2}\,
y_{v_2v_3,2,j_2}\,x_{v_3}$
\end{tabular}
\caption{\label{fig:encoding:tempwalks}
A monomial encoding of a temporal walk (left) and a temporal path (right). Arrows
indicate the direction of the walk. Variables highlighted in red are repeated if and only if a temporal   walk
is not a temporal path.}
\end{figure}

\subsection{Generating temporal walks}
To generate a polynomial encoding of temporal walks, we turn to a dynamic-programming recursion.
That is, we make use of polynomial encoding of temporal walks
of length $\ell-2$ to generate the encoding of temporal walks of length
$\ell-1$ recursively.

Consider the example illustrated in Figure~\ref{fig:encoding:temppath:1}, where the vertex $u$ has
incoming edges from vertices $\{v_1,v_2\}$ at timestamp $i$, i.e.,
$N_i(u)=\{v_1,v_2\}$.
Let $\prepp{u}{\ell}{i}$ denote the polynomial encoding of all temporal walks of
length $\ell-1$ ending at vertex $u$ at latest time $i$. Following our notation,
$\prepp{v_1}{\ell-1}{i}$, $\prepp{v_2}{\ell-1}{i}$ represent the polynomial
encoding of walks ending at vertices $v_1, v_2$, respectively, such that all
walks have length $\ell -1$ and end at latest time $i$. From the definition of a
temporal walk, it is clear that we can walk from vertices $v_1, v_2$ to vertex $u$ at
time $i$ if we have reached $v_1$ and/or $v_2$ at latest time $i$. The
polynomial encoding of walks of length $\ell-1$ ending at vertex $u$ at latest
time $i$ can be written as
\[
\prepp{u}{\ell}{i} = x_u \sum_{j \leq i} y_{v_1u,\ell-1,j}\,\prepp{v_1}{\ell-1}{j}
                  + x_u \sum_{j \leq i} y_{v_2u,\ell-1,j}\,\prepp{v_2}{\ell-1}{j}.
\]

By generalizing the above intuition, a generating polynomial for
temporal walks can be written as
\[
\prepp{u}{1}{i} = x_u, \text{ for each } u \in V \text{ and } i \in [\tau], \text{ and}
\]
\begin{equation}
\prepp{u}{\ell}{i} = x_u \sum_{v \in N_i(u)} \sum_{j \leq i}
y_{vu,\ell-1,j}\,\prepp{v}{\ell-1}{j}
\end{equation}
for each $u \in V$, $\ell \in \{2,\dots,k\}$ and $i \in [\tau]$.


\begin{figure}
\centering
\begin{tikzpicture}[scale=\tikzscale,every node/.style={scale=\tikzscale}]]

\input{tikz/tikz-defs}

\node[exnode] (v1) at ( 0, 2) {$v_1$};
\node[exnode] (v2) at ( 0, 0) {$v_2$};
\node[exnode] (u)  at ( 3, 1) {$u$};

\node[fill=white] at (  1.6, 0.2) {\large $i$};
\node[fill=white] at (  1.6, 1.9) {\large $i$};

\node[fill=white] at ( 0.5, -0.75) {\large $\mathcal{P}_{v_1,\ell-1,j}(\vec{x},\vec{y})$};
\node[fill=white] at ( 0.5, 2.75) {\large $\mathcal{P}_{v_2,\ell-1,j}(\vec{x},\vec{y})$};

\node[fill=white, text width=8cm] at ( 7.5, 0.4) {\large
$\mathcal{P}_{u,\ell,i}(\vec{x},\vec{y})  =  x_{u} \displaystyle{\sum_{j \leq i}} y_{v_1u,\ell-1,j} \, \mathcal{P}_{v_1,\ell-1,j}(\vec{x},\vec{y})\,+ $\\
        $\,\,\,\,\,\,\,\,\,\,\,\,\,\,\,\,\,\,\,\,\,\,\,\,\,\,\,\,\,\,\,x_{u} \displaystyle{\sum_{j \leq i}} y_{v_2 u,\ell-1,j} \, \mathcal{P}_{v_2,\ell-1,j}(\vec{x},\vec{y})$};




\path [draw=black,postaction={on each segment={mid arrow=black, ultra thick}}]
(v1) -- (u)
(v2) -- (u)
;
\end{tikzpicture}%
\caption{\label{fig:encoding:temppath:1}
Polynomial encoding of temporal walks.}
\end{figure}

\section{Dataset statistics}\label{appendix:dataset-statistics}
For each real-world dataset we report the number of vertices $n$, the maximum timestamp $\tau$, the number of temporal edges $m$, and the average temporal degree $d_{\text{avg}} = \frac{m}{n \tau}$.
Furthermore, for each dataset we construct a static undirected graph or an {\em underlying graph} by considering the edges without the timestamp information ignoring multiple edges (if any) between the same set of vertices. For a temporal graph $G$, we denote the corresponding underlying (static) graph as $\ulg{G}$. For the underlying graph we report the number of edges $\ulg{m}$, the average degree $\ulg{d}_{\text{avg}}$, the length of the diameter $\ulg{D}_{\text{max}}$, the feedback edge number (FEN), and the feedback vertex number (FVN).
The dataset statistics are reported in Table~\ref{table:exp:datasets}.

The {\em average degree} $\ulg{d}_{\text{avg}}=\frac{\ulg{m}}{n}$ is the ratio of number of edges and the number of vertices in the static graph. The {\em diameter} of a static graph is the length of a longest    shortest path between any two vertices and it can be computed in time $\bigO(n \ulg{m})$, where $n$ is number of vertices and $\ulg{m}$ is number of edges, however, the run\-time can be further reduced to $\bigO(\ulg{m})$ in practice~\cite{crescenzi2013on}. The diameter of the underlying graph is neither a lower bound nor an upper bound on the maximum length of the restless path. We report this value only to show that usually the path length parameter is orders of magnitude less than the parameters FVN and FEN in most real-world graph datasets.

The {\em feedback edge number} of a static graph $G=(V,E)$ is the minimum number of edges that   need to be removed from $G$ in order to make $G$ acyclic. Formally, a set $F \subseteq E$ is a \emph{feedback edge set} if $(V,E \setminus F)$ does not    contain a cycle. In other words, the feedback edge number of $G$ is the size of a minimum feedback edge set for $G$.
The feedback edge number can be computed in polynomial-time by computing a maximum spanning      forest of $G$ and by taking its edge-complement.

The {\em feedback vertex number} of a static graph $G=(V,E)$ is the minimum number of vertices   that need to be removed from $G$ in order to make $G$ acyclic. Formally, a set $F \subseteq V$ is a \emph{feedback vertex set} if $(V \setminus F, E \setminus \{(u,v): u \in F~\text{or}~v \in F\})$ does not contain a cycle. The feedback vertex number of $G$ is the size of minimum feedback vertex set of $G$. In contrast to the feedback edge number, computing the feedback vertex number is \np-hard (see \eg,~\cite{karp1972reducibility}). For computing the feedback vertex number, we use the solvers from the work of Iwata et al.~\cite{iwata2016half,fvs-code}.

\begin{table}
\centering
\caption{\label{table:exp:datasets} Dataset statistics.
Here,
$n$ is the number of vertices,
$m$ is the number of temporal edges,
$\tau$ is the number of timestamps,
$d_{\text{avg}}$ is the average degree of the temporal graph; and
$\ulg{m}$ is number of edges,
$\ulg{d}_{\text{avg}}$ is the average degree,
$\ulg{D}_{\text{max}}$ is the diameter,
FEN is the feedback edge number,
FVN is the feedback vertex number of the underlying static graph.
See main text for a precise definition of these parameters.
}
\footnotesize
\begin{tabular}{l r r r r r r r r r}
\toprule
& \multicolumn{4}{c}{{\em  temporal graph}} &
\multicolumn{5}{c}{{\em  static undirected graph}}\\
\cmidrule(l{2em}){2-5}
\cmidrule(l{2em}){6-10}
Dataset & $n$ & $m$ & $\tau$ & $d_{\text{avg}}$ & $\ulg{m}$ &
$\ulg{d}_{\text{avg}}$ & $\ulg{D}_{\text{max}}$ &
FEN & FVN\\
\midrule
{\em Copenhagen}\\
\cmidrule(r{1em}){1-1}
  {\sf Calls} & 536 &  3\,600 & 4\,028 & 0.00 & 621 & 1.16 & 22 & 248 & 50 \\
    {\sf SMS} & 568 & 24\,333 & 4\,032 & 0.01 & 697 & 1.23 & 20 & 233 & 58 \\
\midrule
{\em Socio-patterns}\\
\cmidrule{1-1}
      {\sf LH10} &   73 &     300\,252 & 12\,960 & 0.31 &  1\,381 &  18.92 & 3 &  1\,309 &  50 \\
    {\sf InVS13} &   95 &     788\,494 & 49\,679 & 0.16 &  3\,915 &  41.21 & 2 &  3\,821 &  85 \\
    {\sf InVS15} &  219 &  2\,566\,388 & 49\,679 & 0.24 & 16\,725 &  76.37 & 3 & 16\,508 & 198 \\
      {\sf SFHH} &  403 &  2\,834\,970 &  5\,328 & 1.32 & 73\,557 & 182.52 & 2 & 73\,155 & 384 \\
{\sf LyonSchool} &  242 & 13\,188\,984 &  5\,887 & 9.26 & 26\,594 & 109.89 & 2 & 26\,353 & 235 \\
  {\sf Thiers13} &  328 & 37\,226\,078 & 19\,022 & 5.96 & 43\,496 & 132.61 & 2 & 43\,169 & 315 \\
\midrule
{\em Koblenz}\\
\cmidrule(r{1em}){1-1}
  {\sf sqwikibooks} &    6\,607 &    21\,709 &    906 &  0.00 &   7\,731 &  1.17 &   7 &   6\,693 &     22 \\
 {\sf pswiktionary} &   26\,595 &    66\,112 &    982 &  0.00 &  53\,167 &  2.00 &  10 &  39\,591 &     30 \\
 {\sf sawikisource} &   48\,960 &   106\,991 &    948 &  0.00 &  72\,608 &  1.48 &   4 &  60\,588 &     83 \\
       {\sf knwiki} &   78\,142 &   714\,594 & 1\,107 &  0.00 & 306\,077 &  3.92 &   9 & 254\,671 & 1\,965 \\
     {\sf epinions} &  131\,828 &   841\,372 &    189 &  0.01 & 711\,783 &  5.40 &  16 & 652\,338 &      - \\
\midrule
{\em Transport}\\
\cmidrule(r{1em}){1-1}
     {\sf Kuopio} &     549 &     32\,122 & 1\,232 &  0.05 &     699 & 1.27 &  29 &    219 &   43 \\
     {\sf Rennes} &  1\,407 &    109\,075 & 1\,223 &  0.06 &  1\,670 & 1.19 &  57 &    329 &   95 \\
   {\sf Grenoble} &  1\,547 &    114\,492 & 1\,314 &  0.06 &  1\,679 & 1.09 &  82 &    489 &   50 \\
     {\sf Venice} &  1\,874 &    118\,519 & 1\,474 &  0.04 &  2\,647 & 1.41 &  61 &    875 &  222 \\
    {\sf Belfast} &  1\,917 &    122\,693 & 1\,132 &  0.06 &  2\,180 & 1.14 &  64 &    291 &   95 \\
   {\sf Canberra} &  2\,764 &    124\,305 & 1\,095 &  0.04 &  3\,206 & 1.16 &  51 &    475 &  135 \\
      {\sf Turku} &  1\,850 &    133\,512 & 1\,260 &  0.06 &  2\,335 & 1.26 &  67 &    665 &  157 \\
 {\sf Luxembourg} &  1\,367 &    186\,752 & 1\,211 &  0.11 &  1\,903 & 1.39 &  34 &    687 &  160 \\
     {\sf Nantes} &  2\,353 &    196\,421 & 1\,280 &  0.07 &  2\,743 & 1.17 &  82 &    524 &  145 \\
   {\sf Toulouse} &  3\,329 &    224\,516 & 1\,233 &  0.05 &  3\,734 & 1.12 & 111 &    542 &  147 \\
    {\sf Palermo} &  2\,176 &    226\,215 & 1\,270 &  0.08 &  2\,559 & 1.18 &  90 &    384 &  123 \\
   {\sf Bordeaux} &  3\,435 &    236\,595 & 1\,307 &  0.05 &  4\,026 & 1.17 &  98 &    668 &  211 \\
{\sf Antofagasta} &     650 &    293\,921 & 1\,097 &  0.41 &     963 & 1.48 &  51 &    327 &   94 \\
    {\sf Detroit} &  5\,683 &    214\,863 & 1\,510 &  0.03 &  5\,946 & 1.05 & 206 &    647 &  100 \\
   {\sf Winnipeg} &  5\,079 &    333\,882 & 1\,296 &  0.05 &  5\,846 & 1.15 &  84 &    818 &  259 \\
   {\sf Brisbane} &  9\,645 &    392\,805 & 1\,283 &  0.03 & 11\,681 & 1.21 & 164 & 2\,306 &  607 \\
   {\sf Adelaide} &  7\,548 &    404\,300 & 1\,270 &  0.04 &  9\,234 & 1.22 &  89 & 1\,992 &  524 \\
     {\sf Dublin} &  4\,571 &    407\,240 & 1\,256 &  0.07 &  5\,537 & 1.21 &  81 & 1\,559 &  319 \\
     {\sf Lisbon} &  7\,073 &    526\,179 & 1\,457 &  0.05 &  8\,817 & 1.25 & 101 & 2\,158 &    - \\
     {\sf Prague} &  5\,147 &    670\,423 & 1\,510 &  0.09 &  6\,714 & 1.30 &  67 & 2\,446 &  470 \\
   {\sf Helsinki} &  6\,986 &    686\,457 & 1\,465 &  0.07 &  9\,022 & 1.29 &  74 & 2\,401 &  655 \\
     {\sf Athens} &  6\,768 &    724\,851 & 1\,506 &  0.07 &  7\,978 & 1.18 & 131 & 1\,375 &  399 \\
     {\sf Berlin} &  4\,601 & 1\,048\,218 & 1\,520 &  0.15 &  6\,600 & 1.43 &  46 & 2\,245 &  575 \\
       {\sf Rome} &  7\,869 & 1\,051\,211 & 1\,506 &  0.09 & 10\,068 & 1.28 & 100 & 2\,519 &  686 \\
  {\sf Melbourne} & 19\,493 & 1\,098\,227 & 1\,441 &  0.04 & 21\,434 & 1.10 & 254 & 2\,531 &  665 \\
     {\sf Sydney} & 24\,063 & 1\,265\,135 & 1\,519 &  0.03 & 28\,695 & 1.19 & 115 & 5\,176 & 1511 \\
      {\sf Paris} & 11\,950 & 1\,823\,872 & 1\,359 &  0.11 & 13\,726 & 1.15 & 159 & 3\,736 &  613 \\
\bottomrule
\end{tabular}
\end{table}

\newpage
\section{Baselines} \label{appendix:baselines}
In this section we discuss the choice of baselines considered for comparison.

\para{Random restless walks.}
Algorithmic approaches based on random walks can be used to estimate reachability in temporal graphs. The approach works by performing a random walk in a temporal graph and update reachability using the transitive closure property by respecting time constraints. More precisely, let $u, v$ and $w$ be distinct vertices.  Now, if there exists a temporal walk from $u$ to $v$ ending at time $i$ and a temporal walk starting at time $j$ from $v$ to $w$ such that $i \leq j$, then there is a temporal walk from $u$ to $w$. Most reachability methods using temporal walks can be extended to estimate reachability with temporal paths, since a temporal walk can be transformed into a temporal path by removing loops in the walk. However, a similar approach of transforming reachability using restless walks to restless paths is not straightforward, since a restless walk cannot be transformed into a restless path by simply removing loops, as it may not satisfy waiting-time constraints (see Figure~\ref{fig:random-walk}).

\begin{figure}
\centering
\tikzset{every picture/.style={scale=0.7}}
\setlength{\tabcolsep}{25pt}
\begin{tabular}{c c}
\begin{tikzpicture}[scale=\tikzscale,every node/.style={scale=\tikzscale}]

\input{tikz/tikz-defs}

\node[exnode, draw=black] (v1) at (   0, 0) {$v_1$};
\node[exnode, draw=black] (v2) at (   3, 0) {$v_2$};
\node[exnode, draw=black] (v3) at ( 4.5, 2) {$v_3$};
\node[exnode, draw=black] (v4) at ( 1.5, 2) {$v_4$};
\node[exnode, draw=black] (v5) at (   6, 0) {$v_5$};

\node[fill=white] at ( 1.5, -0.5) {\large $1$};
\node[fill=white] at ( 4.2,    1) {\large $2$};
\node[fill=white] at ( 3, 2.5) {\large $3$};
\node[fill=white] at ( 1.8, 1) {\large $4$};
\node[fill=white] at ( 4.5, -0.5) {\large $5$};

\path [draw=black, postaction={on each segment={mid arrow=black, ultra thick}}]
(v1) -- (v2)
(v2) -- (v3)
(v3) -- (v4)
(v4) -- (v2)
(v2) -- (v5)
;

\end{tikzpicture}%
 &
\begin{tikzpicture}[scale=\tikzscale,every node/.style={scale=\tikzscale}]

\input{tikz/tikz-defs}

\node[exnode, ultra thick, draw=blue] (v1) at (   0, 0) {$v_1$};
\node[exnode, ultra thick, draw=blue] (v2) at (   3, 0) {$v_2$};
\node[exnode, ultra thick, draw=red] (v3) at ( 4.5, 2) {$v_3$};
\node[exnode, ultra thick, draw=red] (v4) at ( 1.5, 2) {$v_4$};
\node[exnode, ultra thick, draw=blue] (v5) at (   6, 0) {$v_5$};

\node[fill=white] at ( 1.5, -0.5) {\large $1$};
\node[fill=white] at ( 4.2,    1) {\large $2$};
\node[fill=white] at (   3,  2.5) {\large $3$};
\node[fill=white] at ( 1.8,    1) {\large $4$};
\node[fill=white] at ( 4.5, -0.5) {\large $5$};

\path [draw=black, postaction={on each segment={mid arrow=black, ultra thick}}]
(v1) -- (v2)
(v2) -- (v3)
(v3) -- (v4)
(v4) -- (v2)
(v2) -- (v5)
;

\path [draw=black,ultra thick, color=blue, postaction={on each segment={mid arrow=blue, ultra thick}}]
(v1) -- (v2)
(v2) -- (v5)
;

\path [draw=black,ultra thick, color=red, postaction={on each segment={mid arrow=red, ultra thick}}]
(v2) -- (v3)
(v3) -- (v4)
(v4) -- (v2)
;
\end{tikzpicture}%
 \\
A restless walk &  A temporal path (highlighted in blue)
\end{tabular}
\caption{\label{fig:random-walk}
An illustration of infeasibility to transform a restless walk to a restless path. On the left, a restless (temporal) walk $v_1~e_{v_1,v_2,1}~v_2~e_{v_2,v_3,2}~v_3~e_{v_3,v_4,3}~v_4~e_{v_4,v_2,4}~v_2~e_{v_2,v_5,5}~v_5$ from vertex $v_1$ to $v_5$, where resting time of vertices is $\delta(v_1)=\dots=\delta(v_5)=2$.
On the right, a temporal path $v_1~e_{v_1,v_2,1}~v_2~e_{v_2,v_5,5}~v_5$ from $v_1$ to $v_5$ (highlighted in blue) obtained by removing the loop (highlighted in red). Observe that the obtained temporal path is not a restless path since the difference in edge timestamps entering and leaving $v_2$ is $4>\delta(v_2)=2$. In conclusion, even though there is a restless walk from $v_1$ to $v_5$, there exists no restless path from $v_1$ to $v_5$.}
\end{figure}

\para{Temporal graph expansion.}
A time-expansion of a temporal graph to a static directed graph has been applied to reduce temporal reachability to reachability questions in static directed graphs~\cite{mertzios2019temporal,wu2016efficient,zschoche2020complexity,sengupta2019arrow, casteigts2021finding}. Again, the approach can be used to solve reachability problems when vertices need to be connected via a restless walk, however, it fails to solve reachability problems when vertices must be connected via a restless path. Recall that we described a transformation of a temporal graph $G$ to a static directed graph $\deltaexp{G}$ which respects waiting-time restrictions called $\delta$-expansion in Section~\ref{sec:algo:deterministic}. For an illustration, see Figure~\ref{fig:delta-expansion}.

From the construction of $\delta$-expansion, it is easy to see that there exists a restless walk from vertex $v_i$ to $v_j$ in $G$ if and only if there exist a directed path from vertex $v^\ell_i$ to $v^{\ell'}_j$ in $\deltaexp{G}$ for some $\ell, \ell' \in [\tau]$ such that $\ell \leq \ell'$. Using $\delta$-expansion, we can use existing algorithmic approaches for solving reachability problems in static directed graphs to solve reachability problems where vertices must be connected via a restless walk. Observe that even though there exists a directed path from $v^1_1$ to $v^5_5$ in $\deltaexp{G}$, there is no restless path from $v_1$ to $v_5$ in $G$, so it is not straightforward to employ this approach to solve reachability problems when vertices must be connected via a restless path. Even though we presented a deterministic algorithm to solve \restlesspath via a transformation to \rainbowpath, the algorithm is not practical for graphs with large number of vertices.

\para{Index construction.}
In a $2$-hop labeling of static graphs, each vertex $v \in V$ is assigned a label-set pair $(\lin{v}, \lout{v})$ such that $v$ is reachable from each $u \in \lin{v} \subseteq V$ and each $w \in \lout{v} \subseteq V$ is reachable from $v$. From the transitive closure property, it follows that a vertex $v$ is reachable from a vertex $u$ if and only if $\lout{u} \cap \lin{v} \neq \emptyset$. Computing a $2$-hop labeling of minimum size is \np-hard~\cite[Theorem~4.1]{cohen2003reachability}, while approximable within a logarithmic factor~\cite[Theorem~4.2]{cohen2003reachability}.

The approach is extended to solve temporal reachability using a transformation to a directed acyclic graph (DAG)~\cite[Section~3]{wu2016reachability}. Here, the authors try to      answer reachability questions in temporal graphs constrained by time intervals. More specifically, given two vertices $u,v \in V$ and timestamps $t_1, t_2 \in [\tau]$, the goal is to decide whether $v$ is reachable from $u$ via a temporal path (or a temporal walk) such that the timestamps of the edges in the path are in range $\{t_1,t_1+1,\dots,t_1+t_2\}$.
A key difference in our reachability model is that the transition time of an edge can be zero which makes the $\delta$-expansion described in Figure~\ref{fig:delta-expansion} a static directed graph but not a DAG. Additionally, we established that even though two vertices are reachable via a directed path in the $\delta$-expansion, it does not imply that there exists a restless path connecting the vertices in the temporal graph.

We would like to note that \restlesspath is \np-complete for all integers $\Delta \geq 1$ and $\tau \geq \Delta + 2$, even if there exists at most one temporal edge between any two vertices~\cite[Theorem~5]{casteigts2021finding}. So it is highly unlikely that there exists a straightforward extension of $2$-hop cover to solve \restlesspath even if we constrain time intervals.

\para{Parameterized (exact) algorithms.}
Casteigts et al.~\cite{casteigts2021finding} presented algorithms for solving \restlesspath parameterized by the feedback edge number of the underlying graph and the timed feedback vertex number (TFVN) of the temporal graph.
Additionally, they showed that the TFVN of the temporal graph is lower bounded by the FVN of the underlying graph.
For a temporal graph $G=(V,E)$ with maximum timestamp $\tau$, the \emph{timed feedback vertex set} of $G$ is a set $F \subseteq V \times [\tau]$ of vertex appearances such that $\ulg{(G-F)}$ is acyclic, where $G-F = (V,E')$ such that $E'=E \setminus \{(u,v,i) \in E: (v,i) \in F~\text{or}~(v,i)~\in F\}$ and $\ulg{(G-F)}$ is the underlying graph of $(G-F)$. The {\em timed feedback vertex number} of a temporal graph $G$ is the minimum cardinality of a timed feedback vertex set of $G$.

From the statistics in Table~\ref{table:exp:datasets}, it is clear that the values of FVN and FEN are orders of magnitude greater than the diameter of the graph, for most datasets.  Most importantly, it appears difficult to change the value of the parameters FEN and FVN, whereas the length of the restless path $k$ to be found can be readily varied.  Indeed, as we will see, such an approach leads to a solution scaling to graphs with millions of temporal edges provided that the parameter $k$ remains small enough. 
Also note that the deterministic algorithm presented in Section~\ref{sec:algo:deterministic} is not practical for graphs with large value of $n$ considered for experiments.

We implemented the $\textsc{FPT}(k)$-algorithm by Casteigts et al.~\cite{casteigts2021finding}, which uses a different polynomial construction, with running time $\bigO(2^k (kn + km \Delta))$ and space complexity $\bigO(kn\tau)$.\footnote{The exact polynomial factors and space complexity are not detailed in their paper. Instead, the reported complexities are based on our calculations.} While the runtime for a single execution of the decision oracle to decide the existence of a \krestlesspath between a source and a target is comparable to our algorithm, however, solving \krestlessreach requires $\bigO(n)$ queries to identify all reachable vertices, increasing the running time by a factor of $n$.

\begin{figure}
    \centering
    \includegraphics[width=1.0\linewidth]{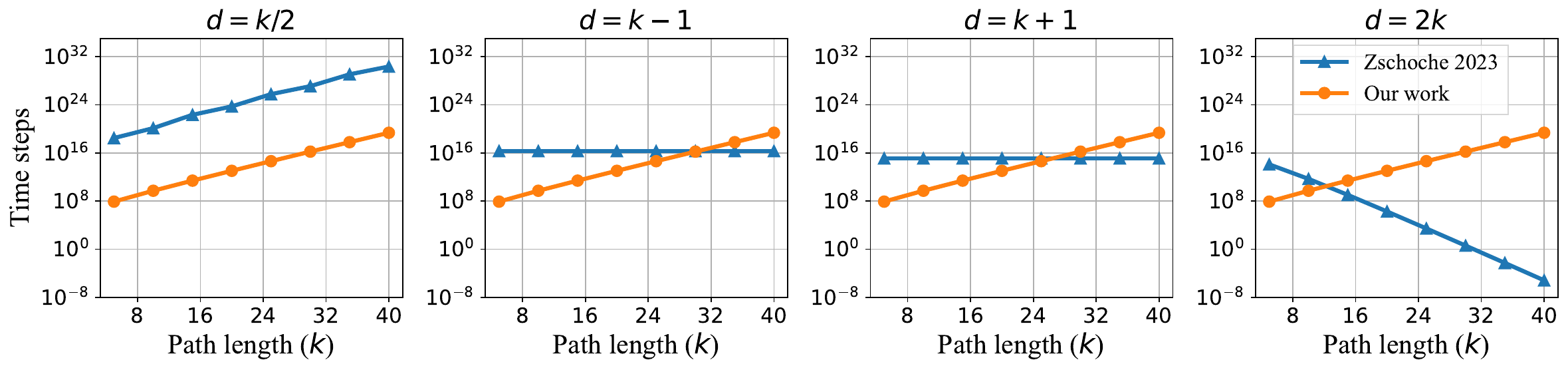}
    \caption{A comparison of number of time steps of our algorithm and \citet{zschoche2022restless} for solving \krestlesspath.}
    \label{app:fig:runtime-comp}
\end{figure}

In a recent work, Zschoche~\cite{zschoche2022restless} gave a randomized $\bigO(4^{k-d} (k-d)^2 m^3 \Delta)$-time algorithm to find a shortest restless path (\ie, one minimizing the length of the path) between a source $s$ and destination $z$, where $d$ is the length of a shortest temporal path from $s$ to $z$. For their algorithm to outperform ours, the values of $k$ must be large. In Figure~\ref{app:fig:runtime-comp}, we compare the time complexity of our algorithm---\ie, $\bigO(2^k k m \Delta)$---with that of Zschoche~\cite{zschoche2022restless} for various configurations: $d= \frac{k}{2}$ (left), $d = k-1$ (center-left), $d = k+1$ (center-right), $d = 2\,k$ (right), for $k=\{5, 10, \dots,40\}$, $n=10^4$, $m=10^6$, $\Delta=5$ and $\tau=100$. For $d=2\,k$, we observe that the runtime of both methods is similar for $k = 12$ and Zschoche's algorithm outperforming ours for $k > 12$. However, our implementation struggles to scale for $k > 10$, even for moderately sized graphs with $m=10^6$, making the practical runtime comparisons infeasible for large~$k$.\footnote{The results in Figure~\ref{app:fig:runtime-comp} represent the number of time steps calculated based on their respective theoretical time complexities. Although practical execution times may vary, the runtime scaling of both algorithms is expected to follow a similar pattern in practice.}

\end{document}